# The Influence of Adaptive Multicoding on Mutual Information and Channel Capacity for Uncertain Wideband CDMA Rayleigh Fading Channels


Richard J. Barton[1]

Electrical and Computer Engineering Department

University of Houston

Houston, TX 77204-4005

rbarton@uh.edu



**Abstract---We consider the problem of adaptive modulation for wideband DS-CDMA Rayleigh fading channels with imperfect channel state information (CSI). We assume a multidimensional signal subspace spanned by a collection of random spreading codes (multicoding) and study the effects of both the subspace dimension and the probability distribution of the transmitted symbols on the mutual information between the channel input and output in the presence of uncertainty regarding the true state of the channel. We develop approximations for the mutual information as well as both upper and lower bounds on the mutual information that are stated explicitly in terms of the dimension of the signal constellation, the number of resolvable fading paths on the channel, the current estimate of channel state, and the mean-squared-error of the channel estimate. We analyze these approximations and bounds in order to quantify the impact of signal dimension and symbol distribution on system performance. The results indicate that, as with space-time channels, higher signaling dimensions are always better, but the extent of the performance**



[1] This research was supported in part by NSF Grant Numbers CCR-9984462 and CCR-0196152.




**improvement depends on the relationship between the number of resolvable paths on the channel, the SNR, and the accuracy of the CSI estimates. Similarly, we show that the effect of the input symbol distribution on mutual information depends critically on the relationship between the SNR and the accuracy of the CSI estimates, and that adapting the distribution of the transmitted symbols in response to instantaneous changes in the channel state has the greatest impact when the CSI estimates are only partially accurate. Finally, we show that in the absence of any channel state information, the mutual information of the system for any constellation with a finite fourth moment decreases to zero at the same rate as the fourth moment of the fading process itself as the number of resolvable paths on the channel goes to infinity. Conversely, we show that if the distribution of the constellation is adapted in a particular manner so that both the dimension and the fourth moment of the constellation increase as the number of resolvable paths increases, the mutual information, and therefore channel capacity, can be made arbitrarily close to the capacity for a single-user AWGN channel with the same SNR.**

## 1. Introduction

In this paper, we study several aspects of information capacity on Rayleigh fading channels in the context of a single-user, wideband, direct-sequence code-division-multiple-access (DS-CDMA) environment. In particular, we study the effects of signal subspace dimension and input symbol distribution on the mutual information between channel input and output in the presence of uncertainty regarding the true state of the channel. Under reasonably general assumptions regarding the channel model, we develop approximations and bounds for the mutual information that are stated explicitly in terms of the dimension of the signal constellation, the number of resolvable fading paths on the channel, the current estimate of channel state, and the mean-



squared-error of the channel estimate. We analyze these approximations and bounds in order to estimate the effects of signal dimension and level of channel uncertainty on channel capacity and the structure of the capacity-achieving input symbol distributions. We also study the impact of adaptive modulation on channel capacity as a function of the level of channel uncertainty. Although we explicitly study mutual information in a single-user environment, our results are also applicable in certain cases to sum-capacity in a multiuser environment with only minor changes in the development. Similarly, the results presented in this paper are often applicable with only minor modification in a context more general than wideband DS-CDMA systems; however, the simplifying assumptions we have made regarding the channel model are perhaps most appropriate for wideband CDMA environments.

In this work, we consider an adaptive modulation technique in which the dimension and distribution of the signal constellation may be altered in response to characteristics of the transmission environment. In particular, we adopt *M*-ary quadrature amplitude modulation (MQAM) as the basic signaling format and study the effects of adapting not only the number of symbols in the constellation but also the coordinates and probability of transmission for each symbol with respect to a known set of basis signals. We refer to this approach as *adaptive multicoding*. Adaptive modulation techniques addressing variations of system performance have been studied by many different investigators [1-6]. In these previous studies, the effect on channel capacity of variable data rates, variable power levels, and variable coding strategies have all been investigated; however, the effect of factors such as channel estimation errors, constellation dimension, and symbol distribution have largely been ignored. Similarly, in previous work, multicoding techniques have been studied primarily as an expedient mechanism for implementing a variable-rate CDMA system.



We consider adaptive multicoding as a mechanism for adapting the long-term allocation of system resources in addition to adapting both the long- and short-term transmission characteristics of individual users. In this context, adaptive modulation becomes a system-level concept as well as a user-level concept. We show that allowing the dimension and distribution of the signal constellation to be adapted in response to variations in characteristics such as channel uncertainty, diversity level, and SNR can significantly improve the achievable data rate on the channel. Hence, in addition to improving performance for an individual user by adapting the instantaneous transmission power and constellation size in response to short-term variations in channel state, adaptive multicoding can be used to improve system performance by reallocating spreading codes among users and adapting the input symbol distribution for each user in response to long-term variations in channel behavior such as SNR, coherence time, delay spread, etc. among the individual users of the channel.

In relation to other previous work, the results presented in this paper are closely aligned with those presented in studies by Telatar and Tse [7], Medard and Gallager [8], and Subramanian and Hajek [9]. In [7], the mutual information and capacity associated with a wideband, spread-spectrum fading channel with a finite number of resolvable paths is studied explicitly in terms of the manner in which the energy is divided among the resolvable paths on the channel. In [8], the information capacity of wideband spread-spectrum systems is investigated instead by expressing the input signals in terms of an orthonormal set of functions each localized in time and frequency and studying the channel behavior in terms of the second and fourth moments of the coefficients of the expansion. Finally, the work in [9] investigates the relationship between mutual information and a fourth-moment quantity referred to as "fourthegy." The clear conclusion of all of these studies is that signals need to be bursty in time and/or frequency to achieve capacity on



very wideband fading channels. In particular, conventional DS-CDMA signaling schemes that spread power on the channel uniformly in time and frequency are very inefficient in the wideband regime. The problem stems from the fact that signals occupying very large bandwidths are transmitted on such channels over many independent fading paths, each with relatively little power, and the fading coefficients on these paths cannot be reliably estimated.

What is less clear from these earlier studies is the manner in which the distribution of power between the known (i.e., estimated) and unknown components of the fading coefficients influences the mutual information on the channel. Similarly, the extent to which constellation dimension and symbol distribution can be exploited to mitigate the loss in capacity caused by channel uncertainty on wideband spread-spectrum fading channels has not been well studied. The results presented in this paper provide some insight on these issues.

Several interesting conclusions, which are summarized below, can be drawn from the results presented in this paper. Each of these conclusions is discussed fully in the relevant sections of the paper.

- Assuming that a radially symmetric input symbol distribution maximizing mutual information on the channel always exists, arbitrarily good estimates of channel capacity and the associated capacity-achieving input distribution can be computed with relatively low complexity for small values of the constellation dimension $K$. This is true regardless of the level of channel uncertainty and the number of resolvable paths on the channel. While the computational procedure remains well defined for all values of $K$, the complexity increases rapidly as $K$ increases, making the procedure computationally intractable beyond some point. We conjecture that the capacity-achieving input distributions can always be chosen to be radially symmetric. This is certainly true for



both completely coherent and completely noncoherent channels, but the validity of the conjecture is less obvious for other levels of channel uncertainty. This result generalizes the results presented recently in [10] on the capacity of noncoherent Rayleigh fading channels.

- As with space-time channels, the dimension of the signal constellation on a wideband CDMA system has a significant impact on the mutual information on the channel. In fact, after normalization to account for performance improvement resulting from increase in antenna aperture size (and corresponding increase in SNR), the relationship between mutual information, constellation dimension, and number of resolvable paths on a multicoded CDMA channel is virtually identical to the relationship between mutual information, number of transmitting antennas and number of receiving antennas on a space-time channel. Further, on CDMA channels, the rate of increase in mutual information as a function of signal dimension depends critically on the relationship between the accuracy of the estimates of channel state information (CSI), the SNR, and the number of resolvable paths on the channel. In particular, the availability of additional signaling dimensions is generally of more importance on channels with relatively reliable CSI estimates than on channels with relatively unreliable CSI estimates and also of more importance on channels with relatively high SNR than on channels with relatively low SNR.

- Similarly, the impact of input symbol distribution on mutual information on a wideband CDMA channel depends greatly on the accuracy of the CSI estimates and the SNR. In particular, the proper choice of input symbol distribution is generally of more importance on channels with relatively unreliable CSI estimates than on channels with relatively



reliable CSI estimates and also of more importance on channels with relatively high SNR than on channels with relatively low SNR.

- In the absence of adaptive power control (which is not studied in this paper), adapting the size of the signal constellation and the distribution of the transmitted symbols in response to instantaneous changes in the channel state appears to result in only minor increases in mutual information compared to the best possible fixed constellation. This implies that channel capacity with or without channel side information is nearly the same on the channels studied in this paper. The extent of the increase in mutual information depends on the accuracy of the CSI estimates. In particular, the largest gains are associated with channels for which the CSI is only partially known, and there is no impact whatsoever on channels for which the CSI is known precisely. Further, regardless of the level of channel uncertainty, there is no performance advantage associated with adapting only the size of the constellation without simultaneously adapting either the average power or the distribution of the symbols.

- In the absence of any CSI estimates and subject to the usual constraint on average energy transmitted per symbol, the mutual information of the system for any constellation with finite fourth moment decreases to zero at the same rate as the fourth moment of the fading process itself as the number of fading paths on the channel goes to infinity. This result is analogous to results reported in [8, 9] and implies, in particular, that the mutual information for any fixed finite constellation goes to zero as the number of resolvable paths goes to infinity. Since the usual wide-sense-stationary-uncorrelated-scattering (WSSUS) assumption for a diffuse (non-specular) fading channel with a bounded delay power spectrum implies that the number of resolvable paths ultimately grows linearly



with bandwidth [11], this result also essentially implies that the infinite-bandwidth capacity for a CDMA system subject to a fourth-moment constraint is zero on a diffuse WSSUS Rayleigh fading channel. Conversely, if adaptive multicoding is employed so that both the dimension and the fourth moment of the constellation increase as the number of resolvable paths increases, the mutual information, and therefore the channel capacity, can be made arbitrarily close to $\mathcal{E}_s/N_0$ nats per symbol, which is identical to the capacity for an infinite-bandwidth, single-user AWGN channel with the same SNR. This result provides an $M$-ary CDMA analog of a result proved by Kennedy [12] for $M$-ary frequency-shift keying.

The remainder of the paper is organized as follows. Adaptive multicoding is introduced briefly in Section 2. In Section 3, we develop several approximations and bounds for mutual information that are used throughout the paper. In Section 4, we discuss the interpretation and implications of the technical results developed in Section 3. Conclusions are presented in Section 5, and several proofs are given in the Appendix.

## 2. Adaptive Multicoding

We assume that information is transmitted using $M$-ary CDMA modulation with $K$ randomly generated, unit-norm spreading codes assigned to each user. This corresponds to each user being assigned a $K$-dimensional signal subspace. In particular, we assume that the transmitted baseband signal for a single user in a single symbol interval takes the form

$$s(t) = \sum_{k=1}^{K} x_k \sum_{n=0}^{N-1} c_{kn} \psi(t - nT_c),$$

where $N$ is the length of the chip sequence, $T_c = T/N$ is the length of the chip interval, $\mathbf{x} = (x_1, x_2, \ldots, x_K)^T \in \mathcal{C}^K$ represents the value of the transmitted symbol, $\psi(t)$ is a chip waveform



of duration $T_c$, and $\mathbf{c}_k = \left( c_{k0}, c_{k1}, \ldots, c_{k,N-1} \right)^T$ is the $k$th chip sequence assigned to the user. We adopt a simple discrete-time representation for the channel output in which the observed output vector corresponding to a single symbol interval can be modeled as:

$$\mathbf{w} = \left( \sum_{k=1}^{K} x_k \mathbf{C}_k \right) (\hat{\boldsymbol{\alpha}} + \boldsymbol{\varepsilon}) + \boldsymbol{\gamma} = \mathbf{C} \cdot \mathbf{H_x} (\hat{\boldsymbol{\alpha}} + \boldsymbol{\varepsilon}) + \boldsymbol{\gamma},$$

where $\mathbf{w} \in \mathscr{C}^{N+L-1}$ represents the received vector, $\hat{\boldsymbol{\alpha}}$ represents a *known* estimate of the true channel-state information (CSI) vector $\boldsymbol{\alpha}$, $\boldsymbol{\varepsilon}$ represents the *unknown* vector of channel estimation errors, $\boldsymbol{\gamma}$ represents an additive noise vector including the effects of multiple-access interference, intersymbol interference, narrowband interference, and additive white Gaussian noise, $\mathbf{H_x} = \left[ x_1 \mathbf{I} \mid x_2 \mathbf{I} \mid \cdots \mid x_K \mathbf{I} \right]^T$, and the $(N+L-1) \times KL$ matrix $\mathbf{C} = \left[ \mathbf{C}_1 \mid \mathbf{C}_2 \mid \cdots \mid \mathbf{C}_K \right]$ is defined based on the circulant matrices

$$\mathbf{C}_k = \begin{bmatrix} c_{k,0} & 0 & \cdots & 0 \\ c_{k,1} & c_{k,0} & \cdots & 0 \\ \vdots & c_{k,1} & \cdots & \vdots \\ c_{k,N-1} & \vdots & \cdots & 0 \\ 0 & c_{k,N-1} & \cdots & c_{k,0} \\ 0 & 0 & \cdots & c_{k,1} \\ \vdots & \vdots & \vdots & \vdots \\ 0 & 0 & \cdots & c_{k,N-1} \end{bmatrix}, \quad k = 1,2,\ldots,K.$$

Note that the quantity $L$ is equivalent to the number of *resolvable paths* on the fading channel, and we have rather arbitrarily chosen an observation vector of length $N+L-1$, corresponding to the current symbol interval plus the delay spread of the channel.

Finally, we make the simplifying (and generally reasonable) assumption that the matrix $\mathbf{C} = \left[ \mathbf{C}_1 \mid \mathbf{C}_2 \mid \cdots \mid \mathbf{C}_K \right]$ is of full rank $KL \leq N$. Under this assumption, the singular value decomposition gives $\mathbf{C} = \mathbf{V}\mathbf{D}\mathbf{U}^*$, where $\mathbf{U}$ and $\mathbf{V}$ are unitary matrices, $\mathbf{D}$ is a "diagonal" matrix of rank $KL$, and the symbol "*" represents the conjugate-transpose operation. It then follows that



the original observed vector $\mathbf{w}$ can be transformed into an equivalent observed vector $\mathbf{y} \in \mathscr{C}^{KL}$ of the form

$$\mathbf{y} = \mathbf{U}^* \mathbf{H_x} (\hat{\boldsymbol{\alpha}} + \boldsymbol{\varepsilon}) + \boldsymbol{\gamma},$$

where (abusing notation slightly) $\boldsymbol{\gamma}$ again represents the additive noise vector. Note that the dimension of the transformed observation vector $\mathbf{y}$ has been reduced to $KL$ by simply ignoring the dimensions with zero power. Note also that when we wish to designate the channel response during a particular symbol interval $n$, we will attach subscripts to the appropriate vectors, viz., $\mathbf{y}_n$, $\hat{\boldsymbol{\alpha}}_n$, $\boldsymbol{\varepsilon}_n$, and $\boldsymbol{\gamma}_n$.

It should be noted that the discrete representation used here corresponds to the output from a chip-matched filter sampled at the chip rate, and it is implicitly assumed that the channel fading is fixed during a single detection interval. In fact, the results presented in this paper based on this simple model are actually quite general. For example, a very similar but completely general discrete representation for arbitrarily varying WSSUS fading channels and arbitrary continuous-time pseudo-noise signature waveforms could be developed by exploiting the sampling theorem and Bello's time-frequency channel model [13, 14]. All results and proofs would be essentially identical for the more general, but less intuitive, model.

### 3. Information Theoretic Bounds and Approximations

In order to evaluate the efficacy of the proposed adaptive modulation method, we wish to study the relationship between adaptive multicoding and the information capacity of the system given partial CSI. Toward that end, we present in this section several approximations and bounds on mutual information that will be used in subsequent sections of the paper. For these bounds to hold, we require several assumptions, which we adopt for the remainder of this paper:

1.  The effects of intersymbol interference are negligible; in particular $L << N$.



2. The vector-valued random processes $\{\hat{\boldsymbol{\alpha}}_n\}$, $\{\boldsymbol{\varepsilon}_n\}$, and $\{\boldsymbol{\gamma}_n\}$ are mutually uncorrelated, zero-mean, stationary, and ergodic proper complex Gaussian random processes with independent real and imaginary parts.

3. The multiplicative noise vector $\boldsymbol{\varepsilon}$ has diagonal covariance matrix $\Sigma_{\boldsymbol{\varepsilon}} = \text{diag}\{\delta_1^2, \delta_2^2, \ldots, \delta_L^2\}$.

4. The additive noise vector $\boldsymbol{\gamma}$ has covariance matrix $\Sigma_{\boldsymbol{\gamma}} = \sigma^2 \mathbf{I}$.[2]

5. $\mathcal{E}_s = E\{\mathcal{E}_{\mathbf{x}}\} = 1$, where $\mathcal{E}_{\mathbf{x}}$ represents the energy transmitted in an arbitrary symbol $\mathbf{x}$. That is, the average transmitted energy per symbol $\mathcal{E}_s$ is normalized to unity. Letting $\sigma^2 = N_0$, this implies that $1/\sigma^2$ is equivalent to the SNR $\mathcal{E}_s/N_0$.

Under these assumptions, given that symbol $\mathbf{x} = (x_1, x_2, \ldots, x_K)^T$ is transmitted and CSI estimate vector $\hat{\boldsymbol{\alpha}}$ is observed, the received vector $\mathbf{y}$ will be proper complex Gaussian with mean vector and covariance matrix given by

$$\boldsymbol{\mu}_{\mathbf{y}|\mathbf{x}} = \mathbf{U}^*\mathbf{H}_{\mathbf{x}}\hat{\boldsymbol{\alpha}}, \qquad (3.1)$$

and

---

[2] Although this is a typical simplifying assumption for this type of analysis, in our case it is equivalent to the assumption that $KL << N$, which may not be very realistic in all cases. The results presented in this section and the remainder of the paper can be extended to the more general case of $\Sigma_{\boldsymbol{\gamma}}$ diagonal in a straightforward manner but at a considerable cost in computational and analytical complexity. These extensions will be addressed in a later paper.



$$\boldsymbol{\Sigma}_{\mathbf{y}|\mathbf{x}} = \mathbf{U}^* \mathbf{B}_{\mathbf{x}} \mathbf{U},$$

$$\mathbf{B}_{\mathbf{x}} = \mathbf{H}_{\mathbf{x}} \boldsymbol{\Sigma}_{\boldsymbol{\varepsilon}} \mathbf{H}_{\mathbf{x}}^* + \sigma^2 \mathbf{I} \tag{3.2}$$

$$= \begin{bmatrix} |x_1|^2 \boldsymbol{\Sigma}_{\boldsymbol{\varepsilon}} + \sigma^2 \mathbf{I} & x_1 \bar{x}_2 \boldsymbol{\Sigma}_{\boldsymbol{\varepsilon}} & \cdots & x_1 \bar{x}_K \boldsymbol{\Sigma}_{\boldsymbol{\varepsilon}} \\ x_2 \bar{x}_1 \boldsymbol{\Sigma}_{\boldsymbol{\varepsilon}} & |x_2|^2 \boldsymbol{\Sigma}_{\boldsymbol{\varepsilon}} + \sigma^2 \mathbf{I} & \cdots & x_2 \bar{x}_K \boldsymbol{\Sigma}_{\boldsymbol{\varepsilon}} \\ \vdots & \vdots & \cdots & \vdots \\ x_K \bar{x}_1 \boldsymbol{\Sigma}_{\boldsymbol{\varepsilon}} & x_K \bar{x}_2 \boldsymbol{\Sigma}_{\boldsymbol{\varepsilon}} & \cdots & |x_K|^2 \boldsymbol{\Sigma}_{\boldsymbol{\varepsilon}} + \sigma^2 \mathbf{I} \end{bmatrix},$$

respectively.

We also assume that both the receiver and transmitter may have access to the estimated CSI vector $\hat{\boldsymbol{\alpha}}$; however, for purposes of this paper, we will restrict attention to adaptive modulation schemes that require very little feedback and are memoryless, that is, those for which the conditional distribution of the input symbol sequence satisfies

$$p\left(\bar{\mathbf{x}}_n | \bar{\boldsymbol{\alpha}}_n\right) = \prod_{i=1}^{n} p\left(\mathbf{x}_i | \hat{\boldsymbol{\alpha}}_i\right) \tag{3.3}$$

where $\bar{\mathbf{x}}_n = \{\mathbf{x}_1, \mathbf{x}_2, \ldots, \mathbf{x}_n\}$, $\bar{\boldsymbol{\alpha}}_n = \{\hat{\boldsymbol{\alpha}}_1, \hat{\boldsymbol{\alpha}}_2, \ldots, \hat{\boldsymbol{\alpha}}_n\}$, and $p(\cdot)$ represents the density function for a distribution with respect to an appropriate dominating measure $\mu$. Similarly, we assume that the conditional distribution for the sequence of output vectors given the sequence of input vectors and CSI estimates satisfies

$$p\left(\bar{\mathbf{y}}_n | \bar{\mathbf{x}}_n, \bar{\boldsymbol{\alpha}}_n\right) = \prod_{i=1}^{n} p\left(\mathbf{y}_i | \mathbf{x}_i, \hat{\boldsymbol{\alpha}}_i\right), \tag{3.4}$$

where $\bar{\mathbf{y}} = \{\mathbf{y}_1, \mathbf{y}_2, \ldots, \mathbf{y}_n\}$. This is essentially equivalent to the assumption that the predictability of the fading process on the channel is completely captured in the sequence of CSI estimate vectors $\{\hat{\boldsymbol{\alpha}}_n\}$ and that the sequence of corresponding CSI error vectors $\{\boldsymbol{\varepsilon}_n\}$ is i.i.d.; that is, the sequence $\{\boldsymbol{\varepsilon}_n\}$ constitutes an "innovations sequence" for the fading process.



Ideally then, the purpose of the proposed adaptive modulation scheme is to choose the sequence of conditional densities $p(\mathbf{x}_n | \hat{\boldsymbol{\alpha}}_n)$ in order to maximize the quantity $\lim_{n \to \infty} \frac{1}{n} I(\bar{\mathbf{x}}_n; \bar{\mathbf{y}}_n | \bar{\boldsymbol{\alpha}}_n)$ subject to the average energy constraint

$$E\{\mathcal{E}_\mathbf{x}\} = \int_{\hat{\boldsymbol{\alpha}}} p(\hat{\boldsymbol{\alpha}}) \left[ \int_\mathbf{x} \|\mathbf{x}\|^2 \, p(\mathbf{x} | \hat{\boldsymbol{\alpha}}) \mu(d\mathbf{x}) \right] \mu(d\hat{\boldsymbol{\alpha}}) = 1.$$

Note that assumptions (3.3) and (3.4) imply that

$$I(\bar{\mathbf{x}}_n; \bar{\mathbf{y}}_n | \bar{\boldsymbol{\alpha}}_n) = \sum_{i=1}^n I(\mathbf{x}_i; \mathbf{y}_i | \hat{\boldsymbol{\alpha}}_i) = n I(\mathbf{x}; \mathbf{y} | \hat{\boldsymbol{\alpha}}),$$

so we wish to solve the following optimization problem

$$\max_{p(\mathbf{x} | \hat{\boldsymbol{\alpha}})} \; I(\mathbf{x}; \mathbf{y} | \hat{\boldsymbol{\alpha}})$$
$$\text{subject to: } E\{\mathcal{E}_\mathbf{x}\} = 1. \tag{3.5}$$

To solve Problem (3.5), we take the following approach. Let $I_{\hat{\boldsymbol{\alpha}}}(\mathbf{x}; \mathbf{y})$ represent $I(\mathbf{x}; \mathbf{y} | \hat{\boldsymbol{\alpha}})$ with $\hat{\boldsymbol{\alpha}}$ treated as a constant. We first seek a solution to the problem

$$\max_{p(\mathbf{x} | \hat{\boldsymbol{\alpha}})} \; I_{\hat{\boldsymbol{\alpha}}}(\mathbf{x}; \mathbf{y})$$
$$\text{subject to: } E\{\mathcal{E}_\mathbf{x} | \hat{\boldsymbol{\alpha}}\} = 1, \tag{3.6}$$

for arbitrary noise power $\sigma^2$ in order to characterize

$$f(\hat{\boldsymbol{\alpha}}, \gamma) = \max_{p(\mathbf{x} | \hat{\boldsymbol{\alpha}})} \left\{ I(\mathbf{x}; \mathbf{y} | \hat{\boldsymbol{\alpha}}) \; \middle| \; E\{\mathcal{E}_\mathbf{x} | \hat{\boldsymbol{\alpha}}\} = \gamma, \sigma^2 = 1 \right\},$$

as a function of the instantaneous CSI estimate $\hat{\boldsymbol{\alpha}}$ and the instantaneous SNR $\gamma$. We then seek to characterize the optimal power control strategy $\gamma^*(\hat{\boldsymbol{\alpha}})$ by solving

$$\max_{\gamma(\hat{\boldsymbol{\alpha}})} \int_{\hat{\boldsymbol{\alpha}}} p(\hat{\boldsymbol{\alpha}}) f\left( \hat{\boldsymbol{\alpha}}, \frac{\gamma(\hat{\boldsymbol{\alpha}})}{\sigma^2} \right) \mu(d\hat{\boldsymbol{\alpha}})$$
$$\text{subject to } \int_{\hat{\boldsymbol{\alpha}}} p(\hat{\boldsymbol{\alpha}}) \gamma(\hat{\boldsymbol{\alpha}}) \mu(d\hat{\boldsymbol{\alpha}}) = 1.$$



Having identified the optimal power control strategy $\gamma^*(\hat{\boldsymbol{\alpha}})$, a useful adaptive modulation scheme that approximately solves Problem (3.5) can then be implemented by choosing the symbol distribution for each symbol interval in order to satisfy (3.6) subject to the instantaneous power constraint imposed by $\gamma^*(\hat{\boldsymbol{\alpha}})$.

In this paper, we study approximate solutions to Problem (3.6) and investigate the effects of signal dimension, symbol distribution, and channel uncertainty on mutual information. It is interesting to note that the general form of the solution for Problem (3.6) is known only for the completely coherent case and the completely noncoherent case. In the completely coherent case (perfect CSI estimation or $\hat{\boldsymbol{\alpha}} = \boldsymbol{\alpha}$), Problem (3.6) is equivalent to a standard parallel Gaussian channel problem, and the maximizing conditional input distribution is known to be Gaussian. Discussions of this case can be found in many places including, for example, [15]. In the completely noncoherent case (no CSI estimates or $\hat{\boldsymbol{\alpha}} = \boldsymbol{0}$), the maximizing input distribution for $K = 1$ has only recently been characterized in [10]. In the other cases, that is when $0 \neq \hat{\boldsymbol{\alpha}} \neq \boldsymbol{\alpha}$ or $\hat{\boldsymbol{\alpha}} = \boldsymbol{0}, K > 1$, the form of the maximizing input distribution for Problem (3.6) remains an open problem.

In related work, the problem of adaptive power control and coding for flat Rayleigh fading channels in one-dimension has been considered by Goldsmith and Varaiya in [3]. The optimal power control strategy in that case is determined by water-filling in the time domain. For the multidimensional case considered here, it is easy to see that time-domain water filling again gives the optimal power control strategy for the Rayleigh fading case with perfect channel estimates. Unfortunately, determining the optimal strategy in the presence of channel uncertainty is much more problematic. In this paper, we will restrict attention to fixed power transmission strategies.



In the remainder of this section, we present several lemmas and corollaries that will be used to provide estimates of $I_{\hat{\boldsymbol{\alpha}}}(\mathbf{x};\mathbf{y})$ for optimization and performance evaluation under the assumption that $\boldsymbol{\Sigma}_{\boldsymbol{\gamma}} = \sigma^2 \mathbf{I}$. Most proofs are omitted at this point and appear instead in the Appendix. The first lemma gives an upper bound for $I_{\hat{\boldsymbol{\alpha}}}(\mathbf{x};\mathbf{y})$ explicitly in terms of $\hat{\boldsymbol{\alpha}}$, $\boldsymbol{\Sigma}_{\boldsymbol{\varepsilon}}$, and the second-order statistics of the input symbol distribution.

**Lemma 1**. Let $\mathbf{x} = \left(x_1, x_2, \ldots, x_K\right)^T$ represent an arbitrary transmitted symbol and let

$$\mu_i = E\{x_i\},$$
$$s_i^2 = E\left\{|x_i|^2\right\},$$
$$\sigma_i^2 = Var\{x_i\} = s_i^2 - |\mu_i|^2,$$
$$s_{ij} = E\{x_i \bar{x}_j\},$$
$$\sigma_{ij} = Cov\left(x_i, x_j\right) = s_{ij} - \mu_i \bar{\mu}_j.$$

Let the matrix $\mathbf{F}$ be given by

$$\mathbf{F} = \begin{bmatrix} s_1^2 \boldsymbol{\Sigma}_{\boldsymbol{\varepsilon}} + \sigma_1^2 \hat{\boldsymbol{\alpha}}\hat{\boldsymbol{\alpha}}^* + \sigma^2 \mathbf{I} & s_{12}\boldsymbol{\Sigma}_{\boldsymbol{\varepsilon}} + \sigma_{12}\hat{\boldsymbol{\alpha}}\hat{\boldsymbol{\alpha}}^* & \cdots & s_{1K}\boldsymbol{\Sigma}_{\boldsymbol{\varepsilon}} + \sigma_{1K}\hat{\boldsymbol{\alpha}}\hat{\boldsymbol{\alpha}}^* \\ \hline s_{21}\boldsymbol{\Sigma}_{\boldsymbol{\varepsilon}} + \sigma_{21}\hat{\boldsymbol{\alpha}}\hat{\boldsymbol{\alpha}}^* & s_2^2 \boldsymbol{\Sigma}_{\boldsymbol{\varepsilon}} + \sigma_2^2 \hat{\boldsymbol{\alpha}}\hat{\boldsymbol{\alpha}}^* + \sigma^2 \mathbf{I} & \cdots & s_{2K}\boldsymbol{\Sigma}_{\boldsymbol{\varepsilon}} + \sigma_{2K}\hat{\boldsymbol{\alpha}}\hat{\boldsymbol{\alpha}}^* \\ \vdots & \vdots & & \vdots \\ \hline s_{K1}\boldsymbol{\Sigma}_{\boldsymbol{\varepsilon}} + \sigma_{K1}\hat{\boldsymbol{\alpha}}\hat{\boldsymbol{\alpha}}^* & s_{K2}\boldsymbol{\Sigma}_{\boldsymbol{\varepsilon}} + \sigma_{K2}\hat{\boldsymbol{\alpha}}\hat{\boldsymbol{\alpha}}^* & \cdots & s_K^2 \boldsymbol{\Sigma}_{\boldsymbol{\varepsilon}} + \sigma_K^2 \hat{\boldsymbol{\alpha}}\hat{\boldsymbol{\alpha}}^* + \sigma^2 \mathbf{I} \end{bmatrix}.$$

Then, under Assumptions 1-5, we have

$$I_{\hat{\boldsymbol{\alpha}}}(\mathbf{x};\mathbf{y}) \leq \ln|\mathbf{F}| - E\left\{\ln\left|\boldsymbol{\mathcal{E}}_{\mathbf{x}}\boldsymbol{\Sigma}_{\boldsymbol{\varepsilon}} + \sigma^2 \mathbf{I}\right|\right\} - (K-1)L\ln\sigma^2$$

$$\leq K\sum_{i=1}^{L}\ln\left(1 + \frac{\delta_i^2}{K\sigma^2}\right) + K\ln\left(1 + \sum_{i=1}^{L}\frac{|\hat{\alpha}_i|^2}{\delta_i^2 + K\sigma^2}\right) - E\left\{\sum_{i=1}^{L}\ln\left(1 + \boldsymbol{\mathcal{E}}_{\mathbf{x}}\frac{\delta_i^2}{\sigma^2}\right)\right\}.$$

**Corollary 1**. Under Assumptions 1-5, we have

$$I_{\hat{\boldsymbol{\alpha}}}(\mathbf{x};\mathbf{y}) \leq K\ln\left(1 + \sum_{i=1}^{L}\frac{|\hat{\alpha}_i|^2}{\delta_i^2 + K\sigma^2}\right) + \frac{1}{2}E\left\{\boldsymbol{\mathcal{E}}_{\mathbf{x}}^2\right\}\sum_{i=1}^{L}\frac{\delta_i^4}{\sigma^4}.$$

**Proof**. The result follows immediately from Lemma 1 and the inequality $x - \frac{1}{2}x^2 \leq \ln(1+x) \leq x$ for $x \geq 0$. ∎



To develop a lower bound for $I_{\hat{\boldsymbol{\alpha}}}(\mathbf{x};\mathbf{y})$, we adopt a technique used by Medard in [16].

**Lemma 2**. Let $H_{\hat{\boldsymbol{\alpha}}}(\mathbf{x})$ represent the conditional entropy given $\hat{\boldsymbol{\alpha}}$ of a discrete-distribution input source and let $h_{\hat{\boldsymbol{\alpha}}}(\mathbf{x})$ represent the conditional differential entropy given $\hat{\boldsymbol{\alpha}}$ of a continuous-distribution input source. Let $d$ be any number less than or equal to the minimum Euclidean distance between any two input symbols. Let $\boldsymbol{\mu_x} = (\mu_1, \mu_2, \ldots \mu_K)^T = E\{\mathbf{x}\}$ represent the expected value of the constellation of input symbols and let $\boldsymbol{\Sigma_x} = E\{(\mathbf{x} - \boldsymbol{\mu_x})(\mathbf{x} - \boldsymbol{\mu_x})^*\}$ represent the covariance matrix of the constellation. Finally, let the matrix $\mathbf{F}$ be as given in Lemma 1 and let the $KL \times K$ block diagonal matrix $\mathbf{A}$ be given by

$$\mathbf{A} = \begin{bmatrix} \hat{\boldsymbol{\alpha}} & \mathbf{0} & \mathbf{0} & \cdots & \mathbf{0} \\ \mathbf{0} & \hat{\boldsymbol{\alpha}} & \mathbf{0} & \cdots & \mathbf{0} \\ \mathbf{0} & \mathbf{0} & \hat{\boldsymbol{\alpha}} & \cdots & \mathbf{0} \\ \vdots & \vdots & \vdots & \cdots & \vdots \\ \mathbf{0} & \mathbf{0} & \mathbf{0} & \cdots & \hat{\boldsymbol{\alpha}} \end{bmatrix}.$$

Then, under Assumptions 1-5, we have

$$I_{\hat{\boldsymbol{\alpha}}}(\mathbf{x};\mathbf{y}) \geq H_{\hat{\boldsymbol{\alpha}}}(\mathbf{x}) + \ln\left(\frac{\pi^K d^{2K}}{4^K \cdot K!}\right) - \ln\left[(\pi e)^K \left|\frac{d^2}{4(K+1)}\mathbf{I} + \boldsymbol{\Sigma_x} - \boldsymbol{\Sigma_x}\mathbf{A}^*\mathbf{F}^{-1}\mathbf{A}\boldsymbol{\Sigma_x}\right|\right],$$

and

$$I_{\hat{\boldsymbol{\alpha}}}(\mathbf{x};\mathbf{y}) \geq h_{\hat{\boldsymbol{\alpha}}}(\mathbf{x}) - \ln\left[(\pi e)^K \left|\boldsymbol{\Sigma_x} - \boldsymbol{\Sigma_x}\mathbf{A}^*\mathbf{F}^{-1}\mathbf{A}\boldsymbol{\Sigma_x}\right|\right].$$

**Corollary 2**. If the input symbol constellation has mean zero and covariance matrix $\boldsymbol{\Sigma_x} = (1/K) \cdot \mathbf{I}$, then the bounds in Lemma 2 become

$$I_{\hat{\boldsymbol{\alpha}}}(\mathbf{x};\mathbf{y}) \geq H_{\hat{\boldsymbol{\alpha}}}(\mathbf{x}) + \ln\left(\frac{\pi^K d^{2K}}{4^K \cdot K!}\right) + K\ln\left(\frac{K}{\pi e}\right) - K\ln\left[\frac{d^2 K}{4(K+1)} + \left(1 + \sum_{i=1}^{L}\frac{|\hat{\alpha}_i|^2}{\delta_i^2 + K\sigma^2}\right)^{-1}\right],$$

and

$$I_{\hat{\boldsymbol{\alpha}}}(\mathbf{x};\mathbf{y}) \geq h_{\hat{\boldsymbol{\alpha}}}(\mathbf{x}) + K\ln\left(\frac{K}{\pi e}\right) + K\ln\left(1 + \sum_{i=1}^{L}\frac{|\hat{\alpha}_i|^2}{\delta_i^2 + K\sigma^2}\right).$$



**Proof**. In this case, it is straightforward to show that $\mathbf{F}$ is a block diagonal matrix of the form $\mathbf{F} = \mathrm{diag}\left[\frac{1}{K}\left(\boldsymbol{\Sigma}_{\boldsymbol{\varepsilon}} + \hat{\boldsymbol{\alpha}}\hat{\boldsymbol{\alpha}}^{*}\right) + \sigma^2 \mathbf{I}\right]$. It then follows that

$$\boldsymbol{\Sigma}_{\mathbf{x}} - \boldsymbol{\Sigma}_{\mathbf{x}}\mathbf{A}^{*}\mathbf{F}^{-1}\mathbf{A}\boldsymbol{\Sigma}_{\mathbf{x}} = \mathrm{diag}\left[\frac{1}{K}\left(1 + \sum_{i=1}^{K}\frac{|\hat{\alpha}_i|^2}{\delta_i^2 + K\sigma^2}\right)^{-1}\right],$$

and the result follows immediately by substitution into Lemma 2. ■

The following lemma provides a second lower bound for $I_{\hat{\boldsymbol{\alpha}}}(\mathbf{x};\mathbf{y})$ that is much tighter than the one given in Lemma 2 but restricted to the special case of uniform fading and orthogonal signal constellations.

**Lemma 3**. Let $\delta_i^2 = \beta/L$ for $0 \le \beta \le 1$, $i = 1,2,\ldots,L$, and $\mathcal{E} = 2mL\sigma^2$ for some $m > 1/\left(2L\sigma^2\right)$. Assume that the constellation consists of the zero symbol, which is transmitted with probability $p_0 = 1 - 1/\mathcal{E}$, and symbols $\left\{\mathbf{x}_1, \mathbf{x}_2, \ldots, \mathbf{x}_K\right\}$ of the form

$$\mathbf{x}_1 = \sqrt{\mathcal{E}}\left(1,0,\ldots,0\right)^T,$$
$$\mathbf{x}_2 = \sqrt{\mathcal{E}}\left(0,1,\ldots,0\right)^T,$$
$$\vdots$$
$$\mathbf{x}_K = \sqrt{\mathcal{E}}\left(0,0,\ldots,1\right)^T,$$

each transmitted independently with probability $p_i = 1/\left(K\mathcal{E}\right)$. Then, under Assumptions 1-5 and for $L$ sufficiently large, we have



$$I_{\hat{\boldsymbol{\alpha}}}(\mathbf{x};\mathbf{y}) \geq \frac{1}{2mL\sigma^2} \left( \begin{array}{l} \ln\left[K 2mL\sigma^2\right] \\[2mm] -\dfrac{1}{\sqrt{2\pi}}\displaystyle\int_{-\infty}^{\infty} e^{-x^2/2}\ln\left[\begin{array}{l}1 + K\left(2mL\sigma^2-1\right)\left(2m\beta+1\right)^L \\[1mm] \cdot e^{-2mL\left(\beta+\|\hat{\boldsymbol{\alpha}}\|^2\right)-2x\sqrt{m^2\beta^2 L+\left(2m^2\beta+m\right)L\|\hat{\boldsymbol{\alpha}}\|^2}}\end{array}\right]dx \\[4mm] -\dfrac{K-1}{\sqrt{2\pi}}\displaystyle\int_0^{\infty}\dfrac{1}{2y\sqrt{2\pi L\left[m^2\beta^2+\left(2m^2\beta+m\right)\|\hat{\boldsymbol{\alpha}}\|^2\right]}}\, e^{-\frac{\left[\ln y - 2mL\left(\beta+\|\hat{\boldsymbol{\alpha}}\|^2\right)\right]^2}{8mL\left[m\beta^2+\left(2m\beta+1\right)\|\hat{\boldsymbol{\alpha}}\|^2\right]}} \\[4mm] \cdot\displaystyle\int_{-\infty}^{\infty}\left[\Phi(x)\right]^{K-2} e^{-x^2/2}\ln\left[1+\dfrac{(K-1)e^{\frac{2x}{2m\beta+1}\sqrt{m^2\beta^2 L+mL\|\hat{\boldsymbol{\alpha}}\|^2}+\frac{2m\beta L}{2m\beta+1}}}{y+K\left(2mL\sigma^2-1\right)\left(2m\beta+1\right)^L}\right]dxdy \end{array} \right),$$

where $\Phi(x)$ represents the standard normal cumulative distribution function.

The final lemma gives an exact expression for the mutual information on the channels under consideration for one-dimensional signal constellations with radially symmetric symbol distributions under the assumption of uniform fading. In cases where the symbol distribution is only approximately radially symmetric (e.g., for finite constellations), Lemma 4 still provides a good approximation for the mutual information. The corollaries to Lemma 4 give simple necessary and sufficient conditions for an input distribution to maximize the mutual information on the channel. The integral equations given in Corollary 3 and Corollary 4 can be solved numerically in a straightforward manner resulting in a computationally tractable method for finding arbitrarily good estimates of the channel capacity for one-dimensional constellations in uniform fading. These results generalize the results reported recently in [10] on the capacity of noncoherent Rayleigh fading channels. Proofs of both Lemma 4 and Corollary 3 are given in the Appendix. The proof of Corollary 4 is straightforward and is omitted.

**Lemma 4**. Let $K=1$ and $\delta_i^2 = \beta/L$ for $0 \leq \beta \leq 1$, $i=1,2,\ldots,L$. Suppose that the input symbol constellation has a radially symmetric distribution and that the distribution of the amplitude



$|x| = r$ of an arbitrary input symbol $x$ is defined by the probability measure $\mu$. Then, under Assumptions 1-5, we have

$$I_{\hat{\boldsymbol{\alpha}}}(\mathbf{x};\mathbf{y}) = \begin{cases} -1 - \int_{\mathcal{R}^+}\left[\int_0^\infty \ln\left[\int_{\mathcal{R}^+} f(x,0,r,\rho,1,\beta,\sigma,\hat{\boldsymbol{\alpha}})\mu(d\rho)\right]\chi_2^2\left(\frac{2r^2\|\hat{\boldsymbol{\alpha}}\|^2}{\beta r^2 + \sigma^2},x\right)dx\right]\mu(dr), & L=1, \\[4mm] -L - \int_{\mathcal{R}^+}\left[\int_0^\infty\int_0^\infty \ln\left[\int_{\mathcal{R}^+} f(x,y,r,\rho,L,\beta,\sigma,\hat{\boldsymbol{\alpha}})\mu(d\rho)\right]\chi_2^2\left(\frac{2Lr^2\|\hat{\boldsymbol{\alpha}}\|^2}{\beta r^2 + L\sigma^2},x\right)\chi_{2L-2}^2(y)dxdy\right]\mu(dr), & L>1, \end{cases}$$

where

$$f(x,y,r,\rho,L,\beta,\sigma,\hat{\boldsymbol{\alpha}}) = \left(\frac{\beta r^2 + L\sigma^2}{\beta\rho^2 + L\sigma^2}\right)^L I_0\left(2\sqrt{\frac{x}{2}\left(\frac{L\rho^2\|\hat{\boldsymbol{\alpha}}\|^2}{\beta\rho^2 + L\sigma^2}\right)\left(\frac{\beta r^2 + L\sigma^2}{\beta\rho^2 + L\sigma^2}\right)}\right) e^{-\left[\frac{(x+y)(\beta r^2 + L\sigma^2)}{2(\beta\rho^2 + L\sigma^2)} + \frac{L\rho^2\|\hat{\boldsymbol{\alpha}}\|^2}{\beta\rho^2 + L\sigma^2}\right]},$$

$I_0(\cdot)$ is the modified Bessel function of the first kind with order zero, $\chi_2^2(\lambda,\cdot)$ is the pdf of a non-central $\chi^2$ random variable with two degrees of freedom and non-centrality parameter $\lambda$, and $\chi_{2L-2}^2(\cdot)$ is the pdf of a $\chi^2$ random variable with $2L-2$ degrees of freedom.

**Corollary 3.** Let $K=1$ and $\delta_i^2 = \beta/L$ for $0 \le \beta \le 1$, $i=1,2,\ldots,L$. Let the set of possible input symbol constellations be restricted to those with radially symmetric distributions and let the distribution of the amplitude $|x| = r$ of an arbitrary input symbol $x$ be defined by a probability measure $\mu$. Then for $L=1$ and under Assumptions 1-5, the distribution $\mu^*$ maximizes the quantity $I_{\hat{\boldsymbol{\alpha}}}(\mathbf{x};\mathbf{y})$ subject to the average energy constraint $E\{\mathcal{E}_{\mathbf{x}}|\hat{\boldsymbol{\alpha}}\} \le 1$ if and only if there exist nonnegative constants $\lambda_1$ and $\lambda_2$ such that

$$\lambda_1 + \lambda_2 r^2 + \ln(\beta r^2 + \sigma^2) + \frac{1}{2}\int_0^\infty f(x,0,r,1,\beta,\sigma,\hat{\boldsymbol{\alpha}})\left[1 + \ln\left(\int_{\mathcal{R}^+} f(x,0,\rho,1,\beta,\sigma,\hat{\boldsymbol{\alpha}})\mu^*(d\rho)\right)\right]dx \ge 0,$$

for all $r \ge 0$, where



$$f\left(x,y,r,L,\beta,\sigma,\hat{\boldsymbol{\alpha}}\right)=\left(\frac{1}{\beta r^2+L\sigma^2}\right)^L I_0\left(\sqrt{\frac{x}{\beta r^2+L\sigma^2}\left(\frac{2Lr^2\|\hat{\boldsymbol{\alpha}}\|^2}{\beta r^2+L\sigma^2}\right)}\right)e^{-\frac{1}{2}\left[\frac{x+y}{\beta r^2+L\sigma^2}+\frac{2Lr^2\|\hat{\boldsymbol{\alpha}}\|^2}{\beta r^2+L\sigma^2}\right]}.$$

Furthermore, except for a set of $\mu^*$-measure zero, we must have

$$\lambda_1+\lambda_2 r^2+\ln\left(\beta r^2+\sigma^2\right)+\frac{1}{2}\int_0^\infty f\left(x,0,r,1,\beta,\sigma,\hat{\boldsymbol{\alpha}}\right)\left[1+\ln\left(\int_{\mathscr{R}^+}f\left(x,0,\rho,1,\beta,\sigma,\hat{\boldsymbol{\alpha}}\right)\mu^*\left(d\rho\right)\right)\right]dx=0.$$

For $L>1$, the distribution $\mu^*$ maximizes the quantity $I_{\hat{\boldsymbol{\alpha}}}(\mathbf{x};\mathbf{y})$ subject to the average energy constraint $E\{\mathcal{E}_{\mathbf{x}}|\hat{\boldsymbol{\alpha}}\}\leq 1$ only if there exists $\lambda_1$ and $\lambda_2$ such that

$$0\leq\lambda_1+\lambda_2 r^2+\ln\left[\left(\beta r^2+L\sigma^2\right)^L\right]$$
$$+\frac{1}{2^L(L-2)!}\int_0^\infty\int_0^\infty f\left(u,v,r,L,\beta,\sigma,\hat{\boldsymbol{\alpha}}\right)\left(1+\ln\left[\int_{\mathscr{R}^+}f\left(u,v,\rho,L,\beta,\sigma,\hat{\boldsymbol{\alpha}}\right)\mu^*\left(d\rho\right)\right]\right)dudv,$$

for all $r\geq 0$, and

$$0=\lambda_1+\lambda_2 r^2+\ln\left[\left(\beta r^2+L\sigma^2\right)^L\right]$$
$$+\frac{1}{2^L(L-2)!}\int_0^\infty\int_0^\infty f\left(u,v,r,L,\beta,\sigma,\hat{\boldsymbol{\alpha}}\right)\left(1+\ln\left[\int_{\mathscr{R}^+}f\left(u,v,\rho,L,\beta,\sigma,\hat{\boldsymbol{\alpha}}\right)\mu^*\left(d\rho\right)\right]\right)dudv,$$

except for a set of $\mu^*$-measure zero.

**Corollary 4**. Let $K=1$ and $\delta_i^2=\beta/L$ for $0\leq\beta\leq 1$, $i=1,2,\ldots,L$. Let the set of possible input symbol constellations be restricted to those with radially symmetric distributions and let the distribution of the amplitude $|x|=r$ of an arbitrary input symbol $x$ be defined by a probability measure $\mu$. Suppose that $\mu^*$ takes the form

$$\mu^*(dr)=\sum_{i=0}^M p_i\delta\left(r-r_i\right)dr,$$



for some integer $M$ and some collection of probabilities $\{p_0, p_1, \ldots, p_M\}$ and radii $\{r_0, r_1, \ldots, r_M\}$, where $\delta(r)$ represents the Dirac delta function. For fixed $L$, $\beta$, $\sigma$, $\hat{\boldsymbol{\alpha}}$, let the function $g(r, \lambda_1, \lambda_2)$ be defined as

$$
\begin{aligned}
g(r, \lambda_1, \lambda_2) = {} & \lambda_1 + \lambda_2 r^2 + \ln\!\left(\beta r^2 + \sigma^2\right) \\
& + \frac{1}{2}\int_0^\infty f(x, 0, r, 1, \beta, \sigma, \hat{\boldsymbol{\alpha}})\left[1 + \ln\!\left(\int_{\mathcal{R}^+} f(x, 0, \rho, 1, \beta, \sigma, \hat{\boldsymbol{\alpha}})\mu^*(d\rho)\right)\right] dx,
\end{aligned}
$$

for $L = 1$ and

$$
\begin{aligned}
g(r, \lambda_1, \lambda_2) = {} & \lambda_1 + \lambda_2 r^2 + \ln\!\left[\left(\beta r^2 + L\sigma^2\right)^L\right] \\
& + \frac{1}{2^L(L-2)!}\int_0^\infty\int_0^\infty f(u, v, r, L, \beta, \sigma, \hat{\boldsymbol{\alpha}})\left(1 + \ln\!\left[\int_{\mathcal{R}^+} f(u, v, \rho, L, \beta, \sigma, \hat{\boldsymbol{\alpha}})\mu^*(d\rho)\right]\right) du\, dv,
\end{aligned}
$$

for $L > 1$. Then under Assumptions 1-5, the distribution $\mu^*$ maximizes the quantity $I_{\hat{\boldsymbol{\alpha}}}(\mathbf{x}; \mathbf{y})$ subject to the average energy constraint $E\{\mathcal{E}_{\mathbf{x}} | \hat{\boldsymbol{\alpha}}\} \leq 1$ if and only if there exist nonnegative $\lambda_1$ and $\lambda_2$ such that

$$
0 = g(r_i, \lambda_1, \lambda_2) = \int_{r_i}^{r_{i+1}} g^-(r, \lambda_1, \lambda_2)\, dr,
$$

for all $i = 0, 1, \ldots, M$, where $r_{M+1} = \infty$ and $g^-(\cdot) = \min\{0, g(\cdot)\}$.

## 4. Discussion

In this section, we discuss the implications of the theoretical results established in Section 3. We start by considering methods for numerically computing channel capacity for one-dimensional constellations using Lemma 4 and Corollaries 3 and 4.



## 4.1. Computation of Channel Capacity in One Dimension

The results of Lemma 4 and Corollaries 3 and 4 imply that it should be possible to compute channel capacity numerically for channels with small signaling dimension, an arbitrary number of resolvable paths $L$, and an arbitrary uncertainty level $0 \leq \beta \leq 1$. This has in fact already been done for the case $K = 1$, $L = 1$ (flat fading), and $\beta = 1$ (non-coherent detection) in [10]. In [10], the authors start by establishing necessary and sufficient conditions for the capacity achieving distribution, which are a special case of those given in Corollary 3[3]. They then analyze the resulting integral equations and demonstrate that the probability distribution of the symbol magnitude that maximizes mutual information on the channel must be a discrete distribution with a finite number of mass points, including a positive mass at zero. Finally, they give numerical results for the capacity of one-dimensional, noncoherent, flat Rayleigh fading channels over a range of SNR values.

Although we have not yet completed a similar analysis of the integral equations given in Corollary 3, it appears that very similar arguments can be used to show that the same results hold for the more general case considered here. That is, for any $K$, $L$, and $0 < \beta \leq 1$, the probability distribution of the symbol magnitude that maximizes mutual information on the channel must be a discrete distribution with a finite number of mass points, including a positive mass at zero. Our numerical analysis supports this contention and indicates that the number of components in the discrete distribution is quite small over a wide range of SNR values. Further, as we discuss a bit later in the paper, when $\beta = 0$ (coherent detection), mutual information is maximized by a radially symmetric proper complex Gaussian input symbol distribution for all values of $K$ and $L$. In this section, we compute estimates of channel capacity for $K = 1$, $L = 1$, and $0 < \beta \le 1$ under the assumption that the input symbol distribution that maximizes mutual information on the channel is radially symmetric with at most four mass points in the distribution of the symbol magnitude, including a positive mass at the origin. Rather than solving the integral equations given in Corollaries 3 and 4, we adopt the less elegant, but computationally more straightforward approach of numerically maximizing the expression for $I_{\hat{\boldsymbol{\alpha}}}(\mathbf{x}; \mathbf{y})$ given in Lemma 4 subject to the necessary constraints. That is, we solve the following optimization problem numerically.

---

[3] For the case $\beta = 1$, the output symbol distribution is independent of the phase of the input symbol, so it is clear that the capacity-achieving distributions can be assumed radially symmetric. In the more general case, this remains to be proven, but as discussed in Section 3, for the purposes of this paper, we merely assume that this is true and restrict attention to radially symmetric input symbol distributions.

$$C\left(\|\hat{\boldsymbol{\alpha}}\|^2, \beta\right) = \max_{\substack{\{p_0, p_1, p_2, p_3, \\ d_1, d_2, d_3\}}} \left[ -1 - \int_0^\infty \left( \sum_{n=0}^3 p_n \left[ \ln\left[ \sum_{m=0}^3 p_m f\left(x, d_n, d_m, \beta, \sigma, \|\hat{\boldsymbol{\alpha}}\|^2\right) \right] \chi_2^2\left( \frac{2 d_n^2 \|\hat{\boldsymbol{\alpha}}\|^2}{\beta d_n^2 + \sigma^2}, x \right) \right] \right) dx \right]$$

$$\text{subject to } \sum_{n=0}^3 p_n = 1, \quad \sum_{n=0}^3 p_n d_n^2 = 1, \quad d_0 = 0, \quad p_n \ge 0, d_n \ge 0, n = 1, 2, 3,$$

where

$$f\left(x, r, \rho, \beta, \sigma, \|\hat{\boldsymbol{\alpha}}\|^2\right) = \left( \frac{\beta r^2 + \sigma^2}{\beta \rho^2 + \sigma^2} \right) I_0\left( 2\sqrt{\frac{x}{2}\left( \frac{\rho^2 \|\hat{\boldsymbol{\alpha}}\|^2}{\beta \rho^2 + \sigma^2} \right)\left( \frac{\beta r^2 + \sigma^2}{\beta \rho^2 + \sigma^2} \right)} \right) e^{-\left[ \frac{x\left(\beta r^2 + \sigma^2\right)}{2\left(\beta \rho^2 + \sigma^2\right)} + \frac{2 \rho^2 \|\hat{\boldsymbol{\alpha}}\|^2}{\beta \rho^2 + \sigma^2} \right]}.$$

We then derive estimates of channel capacity $C_{T,R}(\beta)$ for the case in which CSI estimates are available at both the transmitter and receiver (adaptive modulation) by averaging $C\left(\|\hat{\boldsymbol{\alpha}}\|^2, \beta\right)$ over the distribution of $\|\hat{\boldsymbol{\alpha}}\|^2$ for a range of values in $0 < \beta \le 1$. Under Assumptions 1-5, it follows that $\|\hat{\boldsymbol{\alpha}}\|^2 \sim \frac{1-\beta}{2} \cdot \chi_2^2$ for $L = 1$. Hence, we have

$$C_{T,R}(\beta) = \int_0^\infty C\left( \frac{1-\beta}{2} u, \beta \right) \chi_2^2(u) du,$$



where $\chi_2^2(u)$ represents the pdf of a $\chi^2$ random variable with 2 degrees of freedom.

Similarly, to derive estimates of the channel capacity under the assumption that the CSI estimates are available only at the receiver, we solve the following optimization problem numerically.

$$C_R(\beta) = \max_{\substack{\{p_0,p_1,p_2,p_3,\} \\ \{d_1,d_2,d_3\}}} \int_0^\infty \left[ -1 - \int_0^\infty \left( \sum_{n=0}^3 p_n \left[ \ln \left[ \sum_{m=0}^3 p_m f\left(x,d_n,d_m,\beta,\sigma,\frac{1-\beta}{2}u\right) \right] \chi_2^2 \left( \frac{2d_n^2 \frac{1-\beta}{2} u}{\beta d_n^2 + \sigma^2}, x \right) \right] \right) dx \right] \chi_2^2(u) du$$

$$\text{subject to } \sum_{n=0}^3 p_n = 1, \quad \sum_{n=0}^3 p_n d_n^2 = 1, \quad d_0 = 0, \quad p_n \geq 0, d_n \geq 0, n = 1,2,3.$$

To compute the capacity for the case $K = 1$, $L = 1$, $\beta = 0$ (coherent detection), we make use of the fact that with perfect CSI estimates, the capacity achieving distribution is radially symmetric Gaussian regardless of the values of $K$ and $L$ and does not depend on the instantaneous value of $\|\hat{\boldsymbol{\alpha}}\|^2$. To see this, recall that for a radially symmetric $K$-dimensional proper complex Gaussian distribution with unit energy, the differential entropy is given by (see Fact 3 in the Appendix)

$$h(\mathbf{v}) = \ln\left[ (\pi e)^K \left| \frac{1}{K} \mathbf{I} \right| \right] = K \ln\left( \frac{\pi e}{K} \right).$$

Hence, the lower bound for $I_{\hat{\boldsymbol{\alpha}}}(\mathbf{x}; \mathbf{y})$ given by Corollary 2 becomes

$$I_{\hat{\boldsymbol{\alpha}}}(\mathbf{x}; \mathbf{y}) \geq K \ln\left( 1 + \frac{\|\hat{\boldsymbol{\alpha}}\|^2}{K\sigma^2} \right).$$

Further, for $\beta = 0$, the upper bound on $I_{\hat{\boldsymbol{\alpha}}}(\mathbf{x}; \mathbf{y})$ given by Corollary 1 becomes

$$I_{\hat{\boldsymbol{\alpha}}}(\mathbf{x}; \mathbf{y}) \leq K \ln\left( 1 + \frac{\|\hat{\boldsymbol{\alpha}}\|^2}{K\sigma^2} \right),$$

which is independent of input symbol distribution. It follows that a unit-energy radially symmetric Gaussian distribution maximizes mutual information when $\beta = 0$ for all values of $K$ and $L$ regardless of the instantaneous value of $\|\hat{\boldsymbol{\alpha}}\|^2$, and



$$C_{T,R}(0) = C_R(0) = E\left\{K\ln\left(1 + \frac{\|\hat{\boldsymbol{\alpha}}\|^2}{K\sigma^2}\right)\right\}.$$

In particular, for $K = 1$, $L = 1$, we have

$$C_{T,R}(0) = C_R(0) = \int_0^\infty \ln\left(1 + \frac{u}{2\sigma^2}\right)\chi_2^2(u)\,du.$$

In this section, we compute the value of $I_{\hat{\boldsymbol{\alpha}}}(\mathbf{x};\mathbf{y})$ under a radially symmetric Gaussian distribution for uncertainty values in the range $0 < \beta \le 1$ as well as for $\beta = 0$. For values in the range $0 < \beta \le 1$, the mutual information is computed using Lemma 4 with $\mu(dr)$ given by the appropriate Gaussian pdf. This approach provides not only the channel capacity for $\beta = 0$, but also a point of reference that indicates the performance advantage associated with other signal constellations for other uncertainty levels.

Finally, for comparison purposes, we also compute the value of $I_{\hat{\boldsymbol{\alpha}}}(\mathbf{x};\mathbf{y})$ for three additional constellations. The first is a constellation with symbols uniformly distributed on a circle of radius 1 in the complex plane. Since this is approximately equal to an MPSK constellation for large values of $M$, we refer to this constellation simply as the *PSK* constellation. The second is a constellation uniformly distributed over the disk of radius $r = \sqrt{2}$ in the complex plan. We refer to this constellation as the *uniform* constellation. The third is an adaptive constellation in which the symbols are uniformly distributed over $M$ circles of radii

$$r_m = m\sqrt{\frac{6M}{(M+1)(3M^2 + M - 1)}}, \quad m = 1,2,\ldots M,$$

with radial probabilities

$$p_m = \frac{2m-1}{M^2}, \quad m = 1,2,\ldots M,$$



where the integer $1 \le M \le M_{\max}$ is chosen adaptively in order to maximize the value of $I_{\hat{\alpha}}(\mathbf{x}; \mathbf{y})$.

We refer to this adaptive constellation as the *adaptive MQAM* (AMQAM) constellation. Note

that with $M_{\max} = 1$, the adaptive constellation reduces to the PSK constellation. On the other

hand, as $M_{\max} \to \infty$, the AMQAM constellation maximizes over all possible unit-energy

constellations with equally likely symbols uniformly distributed over the disk of radius $r = \sqrt{2}$.

Hence, the AMQAM constellation behaves something like a standard adaptive MQAM

constellation [6] and maximizes the mutual information between the two extremes given by the

PSK constellation and the uniform constellation. For purposes of this paper, we let $M_{\max} = 10$.

The capacity estimates for $K = 1$, $L = 1$, and $0 \le \beta \le 1$ are presented in Figures 1-7.

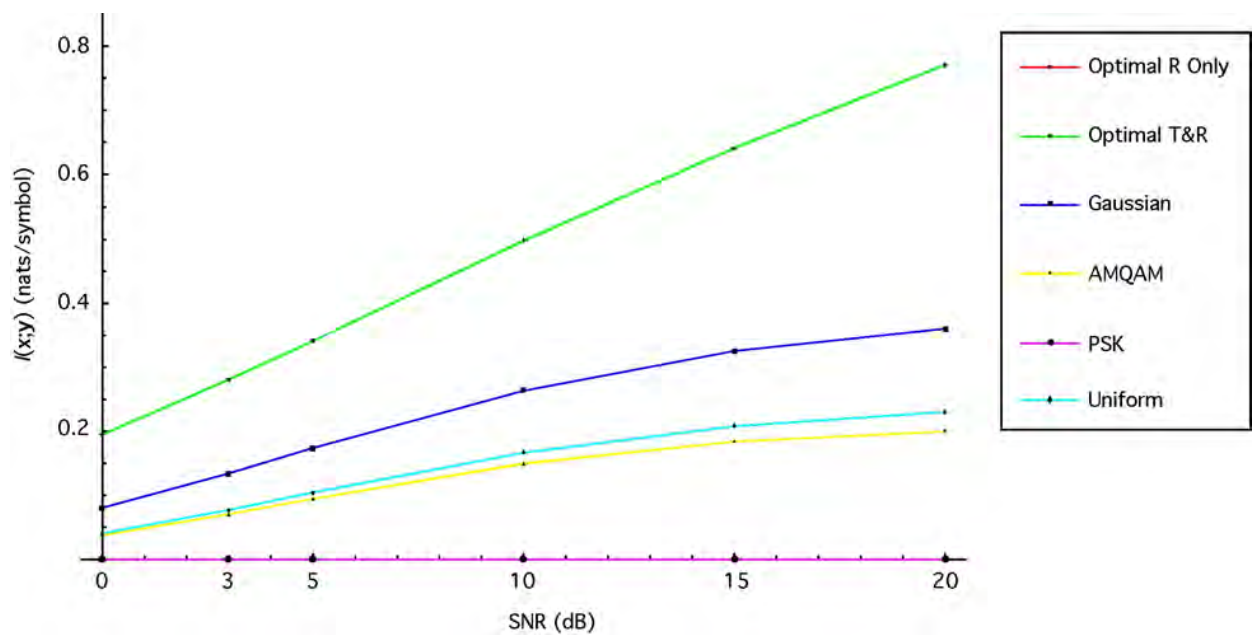

Figure 1. Estimated Channel Capacity for $K = 1$, $L = 1$, and $\beta = 1$ (Noncoherent Channel)



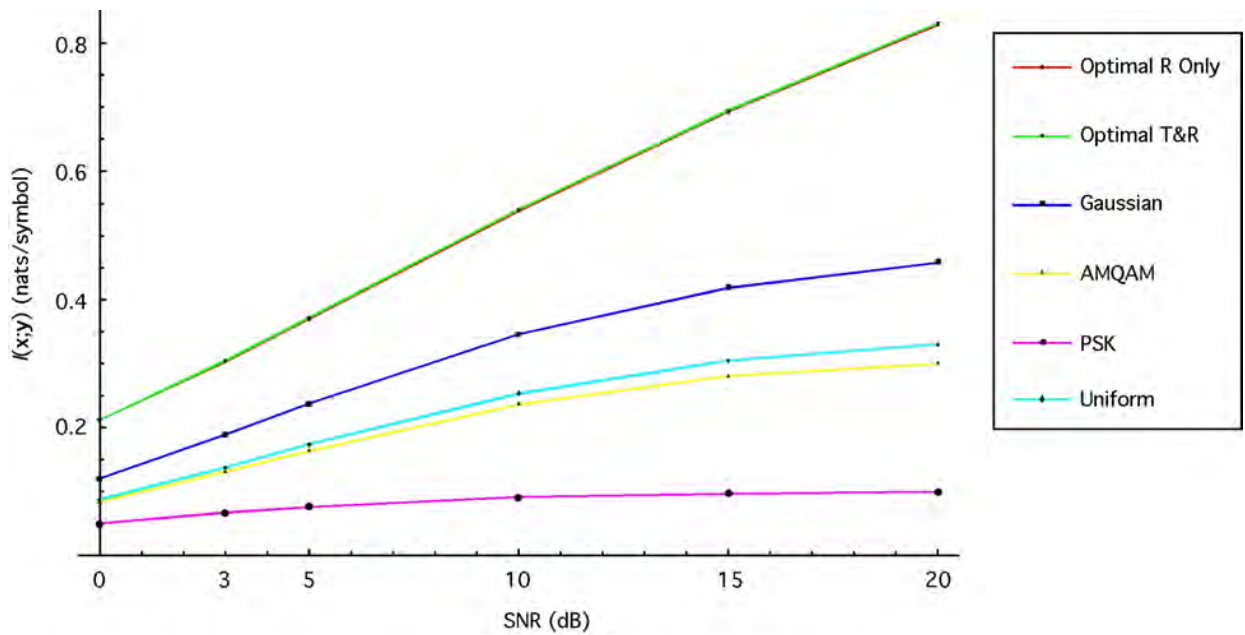

Figure 2. Estimated Channel Capacity for $K=1$, $L=1$, and $\beta=0.9$.

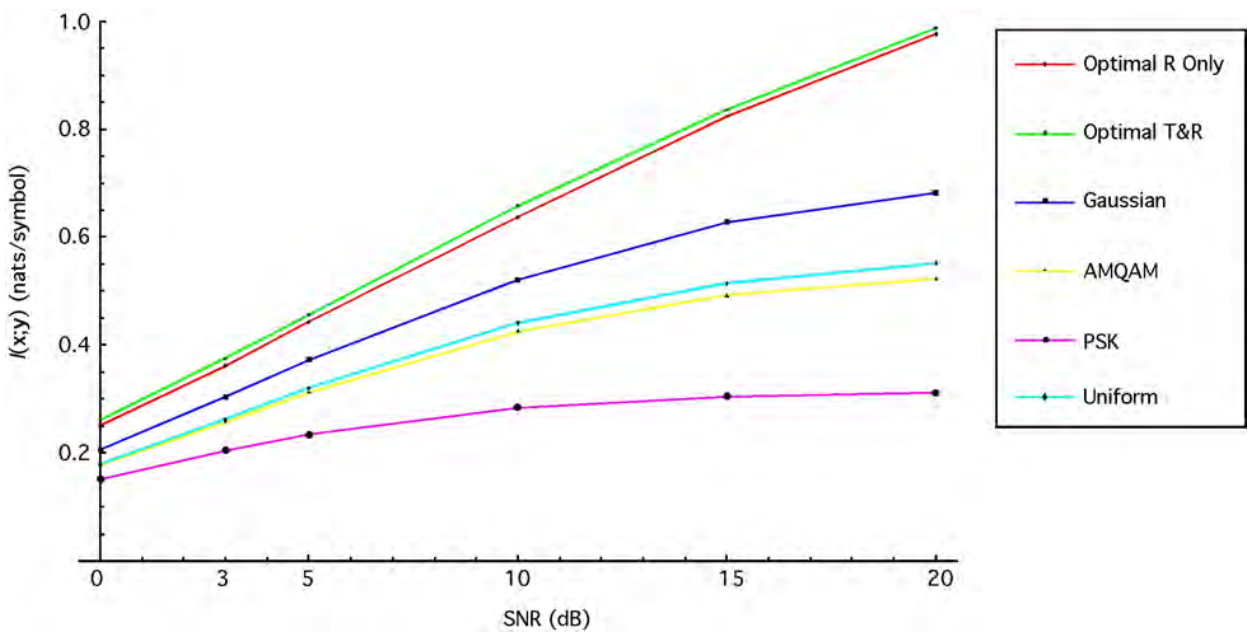

Figure 3. Estimated Channel Capacity for $K=1$, $L=1$, and $\beta=0.7$.



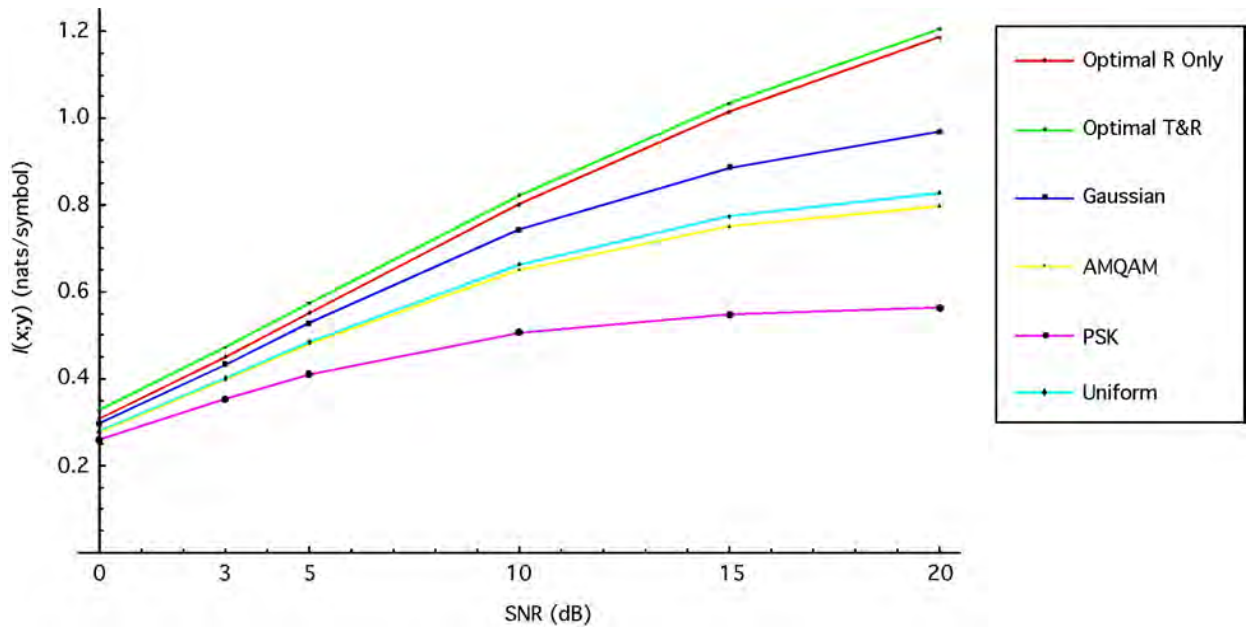

Figure 4. Estimated Channel Capacity for $K = 1$, $L = 1$, and $\beta = 0.5$.

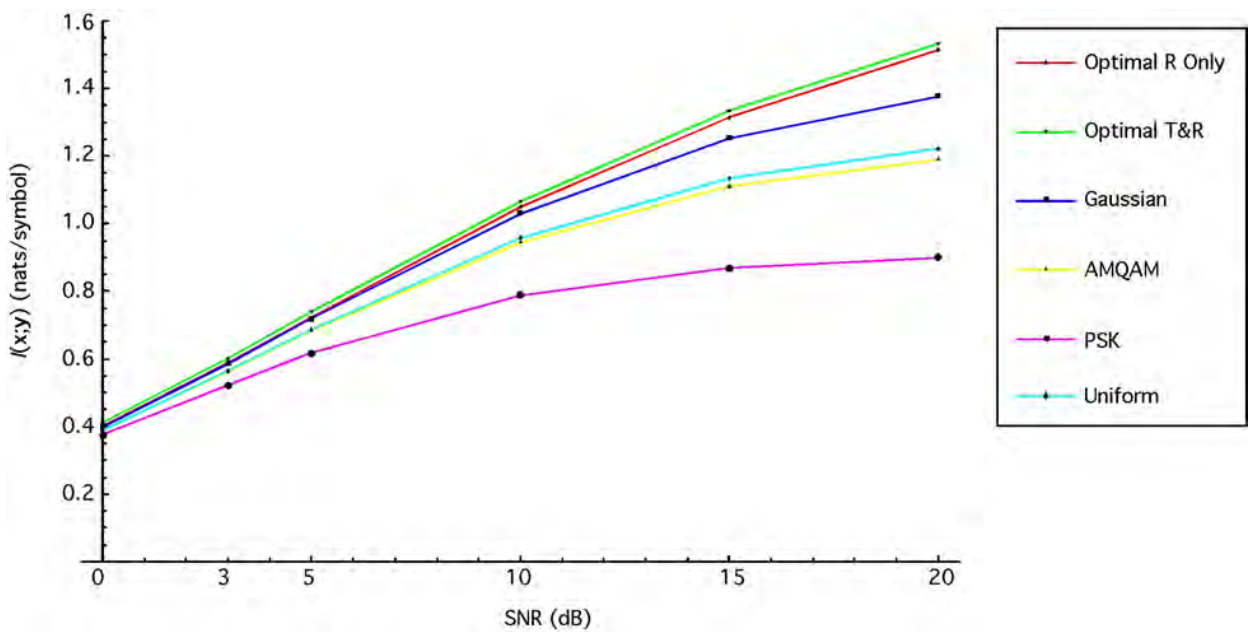

Figure 5. Estimated Channel Capacity for $K = 1$, $L = 1$, and $\beta = 0.3$.



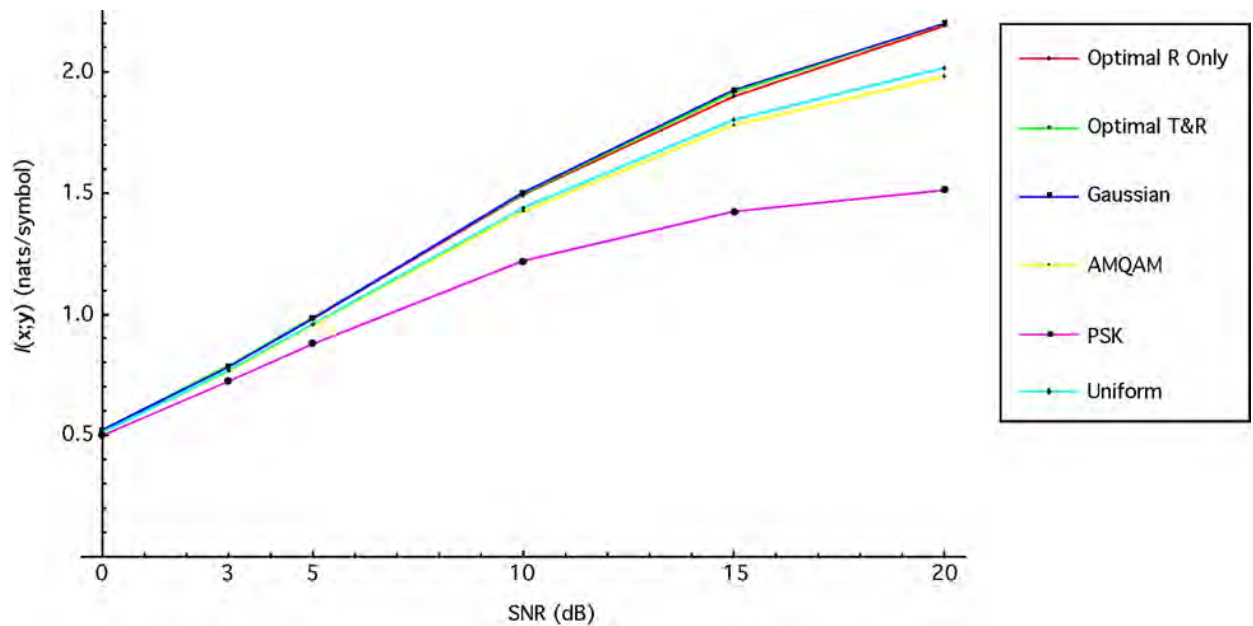

Figure 6. Estimated Channel Capacity for $K = 1$, $L = 1$, and $\beta = 0.1$.

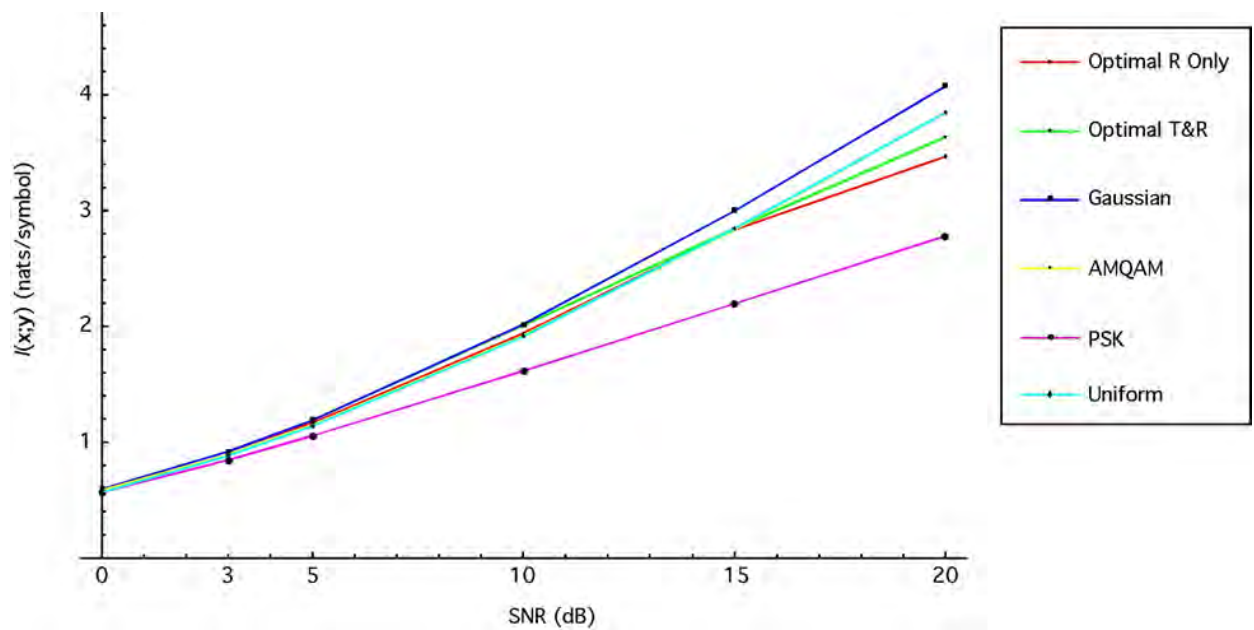

Figure 7. Estimated Channel Capacity for $K = 1$, $L = 1$, and $\beta = 0$ (Coherent Channel).

Each of these figures illustrates achievable data rates at a particular uncertainty level over the range of SNRs from 0 to 20 dB. For each uncertainty level greater than zero, the estimated capacity of the channel with CSI feedback to the transmitter is given by the curve labeled



"Optimal T&R." Similarly the estimated capacity of the channel with no feedback is given by the curve labeled "Optimal R Only." As discussed above, the capacity estimates for the range $0 < \beta \leq 1$ are computed as the maximum mutual information over the class of radially symmetric input symbol distributions with at most four mass points in the distribution of the symbol magnitude, including a positive mass at the origin. For $\beta = 0$, the capacity is achieved by the Gaussian constellation

Several conclusions can be drawn from the results presented in Figures 1-7. First notice that the variability in achievable data rates among the various constellations is much greater for channels with a high level of uncertainty than for channels with very little uncertainty. Similarly, the variability is much more pronounced at high SNR than low SNR. Hence, the proper choice of input symbol distribution is generally of more importance on channels with relatively unreliable CSI estimates than on channels with relatively reliable CSI estimates and also of more importance on channels with relatively high SNR than on channels with relatively low SNR.

Secondly, notice that the estimated channel capacities for nearly coherent channels are understated, but probably only slightly. In particular, for $\beta = 0.1$, the data rate achievable with a Gaussian input symbol distribution exceeds the estimated capacity of the channel with and without feedback, which implies that the estimated capacities are understated and more than four mass points are needed in the distribution of the symbol magnitude in order to achieve capacity for nearly coherent channels. However, even for $\beta = 0$ (coherent channel), the achievable data rate attainable within our constrained set of "optimal" constellations is only slightly less than capacity, which implies that the estimated capacities are not understated by much. Indeed, for large uncertainty values, say $\beta \geq 0.5$, the estimated capacities are almost certainly accurate. This is not obvious from Figures 1-7, but follows from the fact that the estimated capacity results for



$\beta = 1$ (noncoherent channel) presented in Figure 1 are attained with at most three levels of symbol magnitude and exactly match the capacity results for the noncoherent channel presented in [10]. Hence, for some range of uncertainty values near $\beta = 1$, only four mass points are necessary to achieve capacity and the estimated capacities are accurate; however, it is not clear at what uncertainty level more than four mass points are required.

Thirdly, notice that there is very little difference between the estimated capacity curves for the channel with and without CSI feedback over the entire uncertainty range and for all values of SNR. In particular, when $\beta = 0$ or $\beta = 1$, the two capacity values are identical. This follows trivially for $\beta = 1$, since there is really no feedback in that case, and holds also for $\beta = 0$ due to the fact that the capacity-achieving distribution is the same (Gaussian) for all values of $\hat{\boldsymbol{\alpha}}$ when the CSI is known precisely. As discussed in [3], this result continues to hold in the presence of adaptive power control but only for memoryless channels. For $0 < \beta < 1$, the estimated capacity with feedback always exceeds the capacity without feedback, but the increase is limited to about 0.5 dB, which is attained only for channels with uncertainty values near $\beta = 0.5$. This implies that the potential benefit of adaptive modulation will be greatest on channels for which the CSI estimates are "about 50% accurate," but the same benefit can be obtained with very slight increases in transmitted power.

Finally, notice that the maximum data rate achievable with the AMQAM constellation (with $M_{\max} = 10$) is substantially greater than the achievable data rate of the PSK constellation but slightly less than the achievable data rate of the uniform constellation. What is not obvious from the figures is that for the AMQAM constellation, $I_{\hat{\boldsymbol{\alpha}}}(\mathbf{x}; \mathbf{y})$ is always maximized by the constellation with the largest possible number of different symbol energy levels ($M = M_{\max}$) regardless of the value of $\hat{\boldsymbol{\alpha}}$. This implies that if only the size of the constellation is adapted



subject to the constraint that all transmitted symbols are equally likely and roughly uniformly distributed over the largest possible region compatible with the energy constraint, mutual information is always maximized by transmitting from the largest possible constellation regardless of the value of $\hat{\boldsymbol{\alpha}}$. Hence, in the absence of adaptive power control, no increase in channel capacity can be achieved by adapting the size of a conventional MQAM constellation.

## 4.2. Multidimensional Capacity Results

In this section, we study the effects of increasing the dimension of the signal constellation by employing adaptive multicoding. We start by computing capacities for the case $\beta = 0$ for various values of $K \geq 1$ and $L \geq 1$, and comparing these results with analogous results given in [17] for coherent space-time Rayleigh fading channels.

Recall from the discussion given above that for $\beta = 0$, $I_{\hat{\boldsymbol{\alpha}}}(\mathbf{x};\mathbf{y})$ will always be maximized by the same Gaussian constellation and the channel capacity with or without feedback is given by

$$K \int_0^\infty \ln\left(1 + \frac{u}{2KL\sigma^2}\right) \chi_{2L}^2(u) du\,.$$

Similarly, for an analogous space-time channel with $K$ transmitting and $L$ receiving antennas, the capacity is given in [17] by

$$\int_0^\infty \ln\left(1 + \frac{u}{K\sigma^2}\right)\left(\sum_{k=0}^{m-1} \frac{k!}{(k+n-m)!}\left[L_k^{n-m}(u)\right]^2\right) u^{n-m} e^{-u} du\,,$$

where $m = \min\{K,L\}$, $n = \max\{K,L\}$, and $L_k^{n-m}(u)$ is the $k$th order Laguerre polynomial with respect to the weighting function $w(u) = u^{n-m} e^{-u}$. In order to compare these two capacity results on an equal footing, we first normalize the space-time capacity so that the average received power does not depend on the number of receiving antennas. This gives



$$\int_0^\infty \ln\left(1 + \frac{u}{KL\sigma^2}\right)\left(\sum_{k=0}^{m-1} \frac{k!}{(k+n-m)!}\left[L_k^{n-m}(u)\right]^2\right)u^{n-m}e^{-u}du.$$

The capacity results corresponding to $\beta = 0$ for the space-time channel are presented in Figures 8-

10 and the results for the multicoded CDMA channel are presented in Figures 11-13.

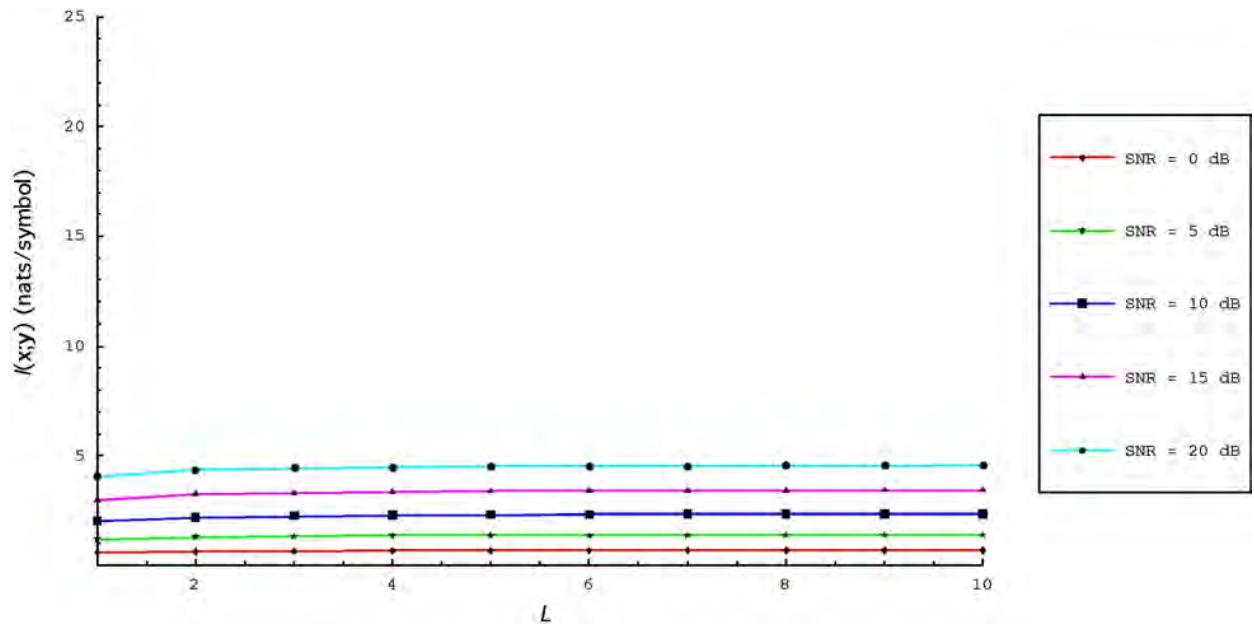

Figure 8. Capacity for Normalized Space-Time Channel With $K = 1$ Transmitting Antennas and

$1 \le L \le 10$ Receiving Antennas ($\beta = 0$).



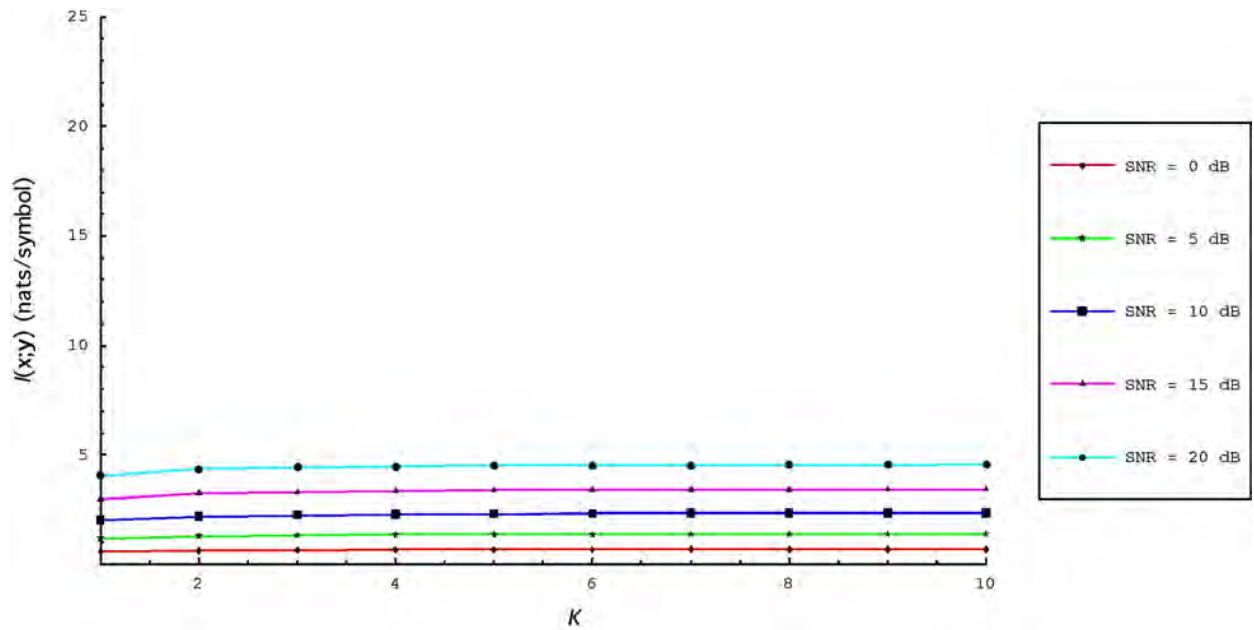

Figure 9. Capacity for Normalized Space-Time Channel With $L = 1$ Receiving Antennas and

$1 \leq K \leq 10$ Transmitting Antennas $(\beta = 0)$.

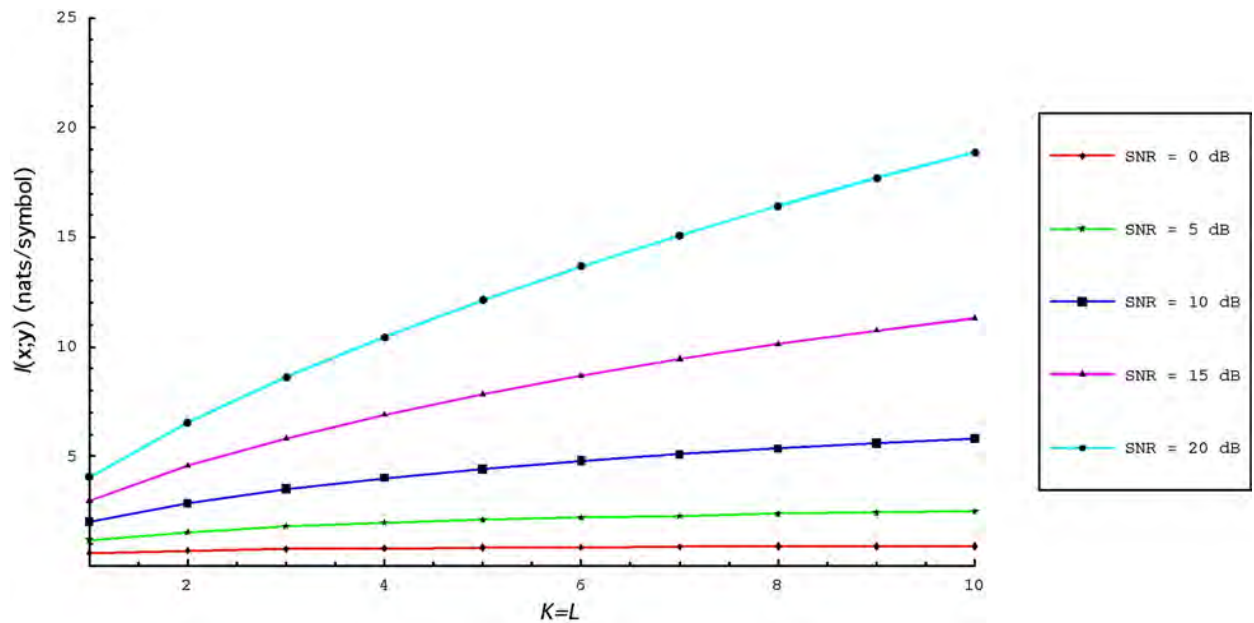

Figure 10. Capacity for Normalized Space-Time Channel With $1 \leq K = L \leq 10$ Receiving and

Transmitting Antennas $(\beta = 0)$.



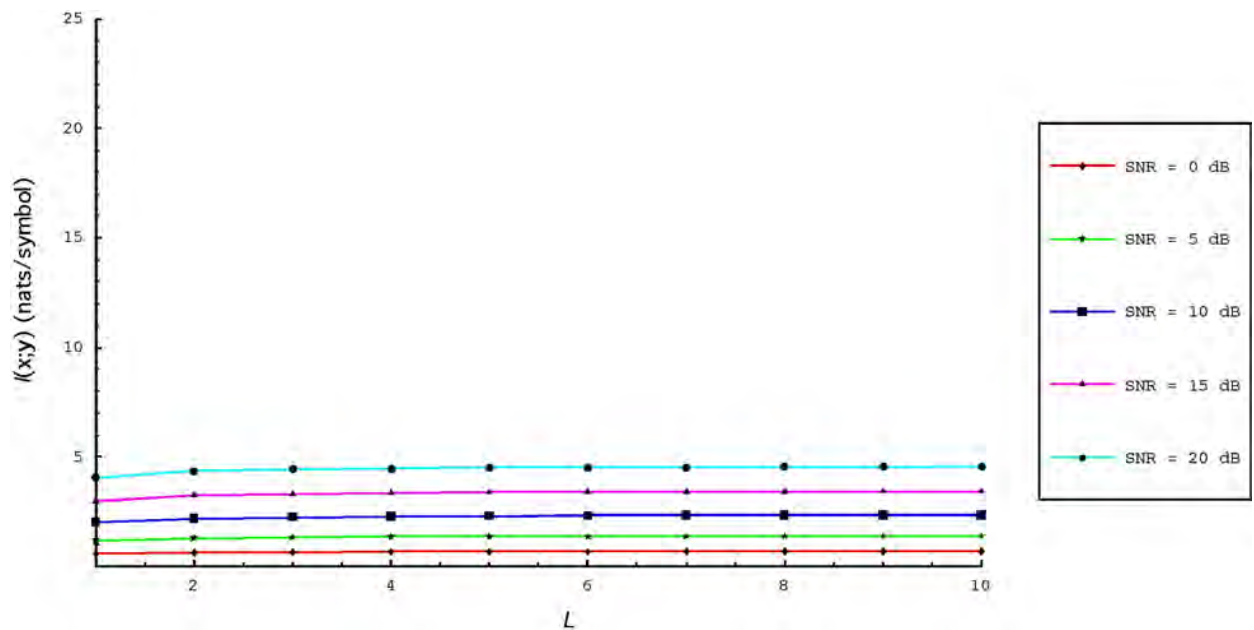

Figure 11. Capacity for CDMA Channel With $K=1$ Spreading Code and $1 \leq L \leq 10$ Resolvable Paths ($\beta = 0$).

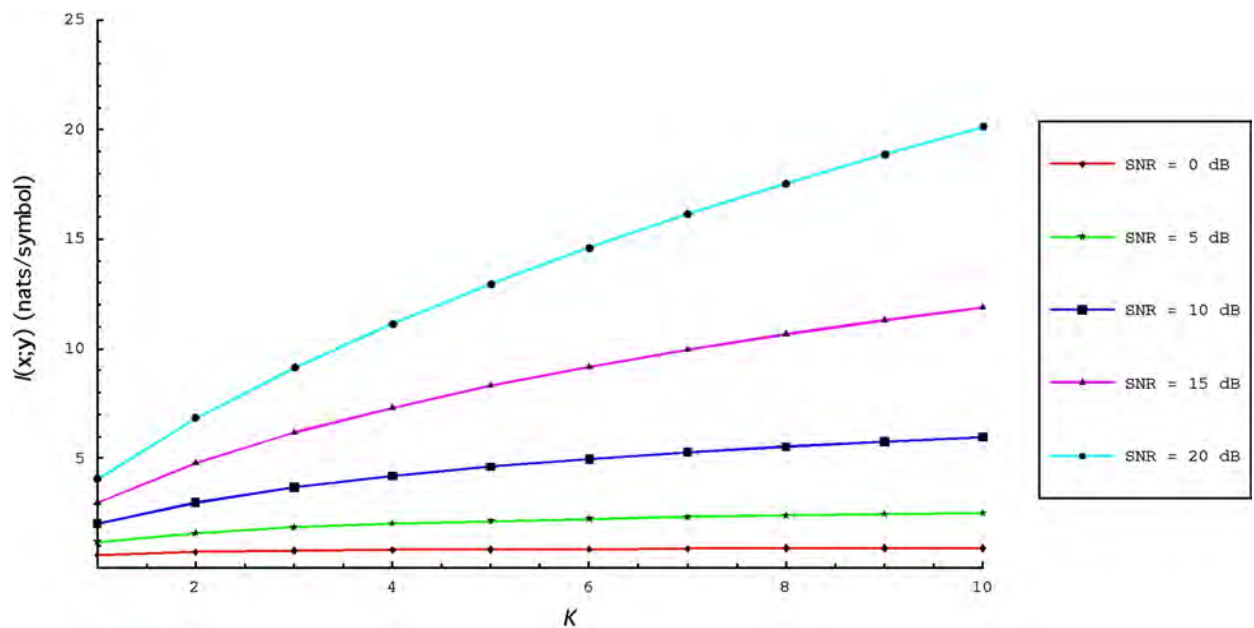

Figure 12. Capacity for CDMA Channel With $L=1$ Resolvable Path and $1 \leq K \leq 10$ Spreading Codes ($\beta = 0$).



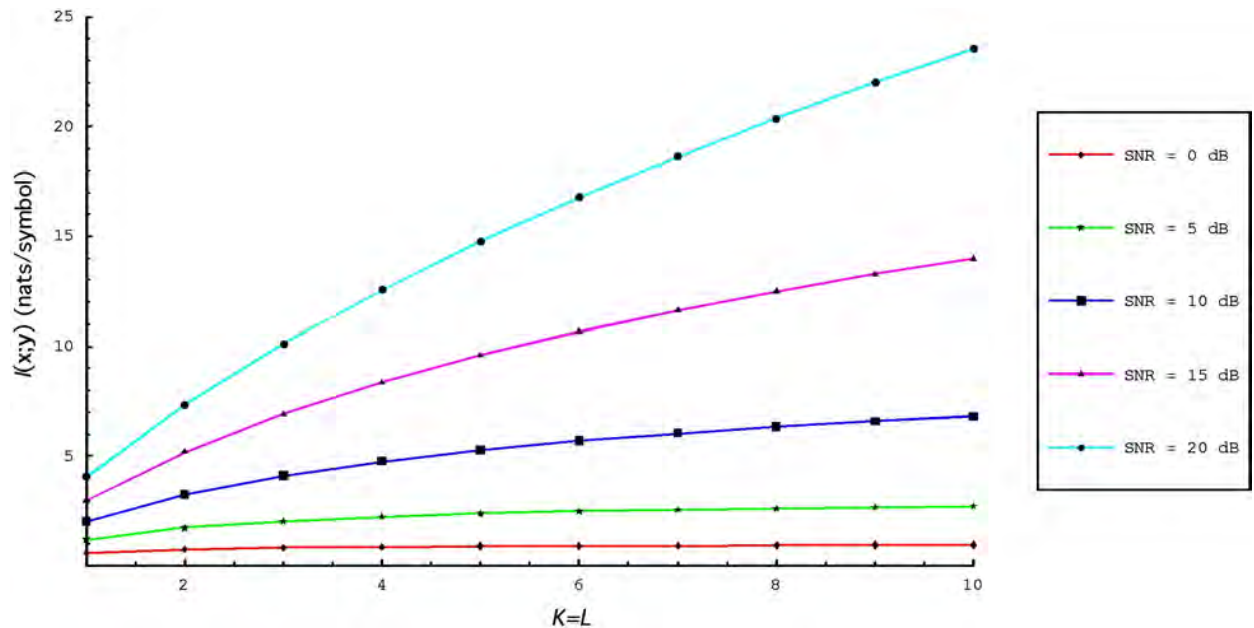

Figure 13. Capacity for CDMA Channel With $1 \leq K = L \leq 10$ Spreading Codes and Resolvable

Paths ($\beta = 0$).

As Figures 8 and 11 indicate, the capacity curves for the normalized space-time channel and the CDMA channel for the case $K = 1$, $1 \leq L \leq 10$ are virtually identical and very flat for the entire range of values $1 \leq L \leq 10$. This is to be expected since there is no increase in signaling dimension in either case and we have normalized out the SNR gain associated with multiple receiving antennas. Hence, the only improvement in performance comes from a diversity gain as $L$ increases, which is the same in both cases.

On the other hand, for the case $L = 1$, $1 \leq K \leq 10$ (Figures 8 and 12) the capacity on the CDMA channel increases with signaling dimension while the capacity on the normalized space-time channel does not change significantly. In fact, the performance on the normalized space-time channel is exactly the same for the two scenarios $K = 1$, $1 \leq L \leq 10$ and $L = 1$, $1 \leq K \leq 10$. This is again to be expected, since the effective signaling dimension on the space-time channel is constrained to unity with only one receiving antenna, and the diversity gain is the same for



multiple transmitting antennas as it is for multiple receiving antennas. In contrast, the signaling dimension increases linearly on the CDMA channel, and the capacity increases monotonically with dimension.

Finally, for the case $1 \leq K = L \leq 10$, which is depicted in Figures 10 and 13, the capacity increases monotonically with dimension for both the normalized space-time channel and the CDMA channel. It is somewhat surprising to note, however, that the capacity curves for the normalized space-time channel in this case are strictly dominated by the capacity curves for the CDMA channel. In fact, the space-time curves are very similar to the curves for the CDMA channel for the more restrictive case $L = 1$, $1 \leq K \leq 10$. The difference in behavior appears to result from our assumption that $\mathbf{\Sigma_\gamma} = \sigma^2 \mathbf{I}$ or equivalently that $N >> KL$ (see footnote 2 above). In fact, for finite values of $N \geq KL$ and fixed spreading codes, the actual capacity of the CDMA channel will be strictly less than depicted in Figures 16 and 17. This is a result of the fact that we will not truly have $\mathbf{\Sigma_\gamma} = \sigma^2 \mathbf{I}$ in this case, and a procedure analogous to water filling across dimension will be required to compute capacity. We conjecture that the performance in both cases would be identical under the assumption that $N = KL$ if capacity is averaged over all possible collections of random spreading codes.

To determine the effect of increasing the dimension of the signal constellation when $0 < \beta \leq 1$, we must resort to estimating channel capacity using the bounds developed in Section 3. As discussed earlier, this is due to the fact that numerical solutions to the integral equations given in Corollaries 3 and 4 become computationally burdensome for values of $K$ greater than 2 or 3. We have chosen to use only lower bounds for the capacity estimates here, since they appear to be tighter than the corresponding upper bounds and naturally provide more conservative estimates. In particular, we estimate channel capacity as the maximum of the following two lower bounds:



1. the lower bound given by Lemma 2 maximized over the AMQAM constellation discussed in Section 4.1 above, and

2. the lower bound given by Lemma 3 maximized over the set of orthogonal constellations defined in the lemma.

To compute each of these two lower bounds, we assume that the number of resolvable paths is fixed at $L = 10$. This is a modest figure for a wideband CDMA channel and is also large enough that the bounds given by Lemma 3 are reasonably tight. As a point of reference, we present in Figures 14 and 15, respectively, the true capacity for the coherent CDMA channel with $L = 10$ resolvable paths and $1 \leq K \leq 10$ spreading codes and the capacity estimate for the same channel computed using the lower bounds. The estimated capacities for values of $0 < \beta \leq 1$ are presented in Figures 16-21.

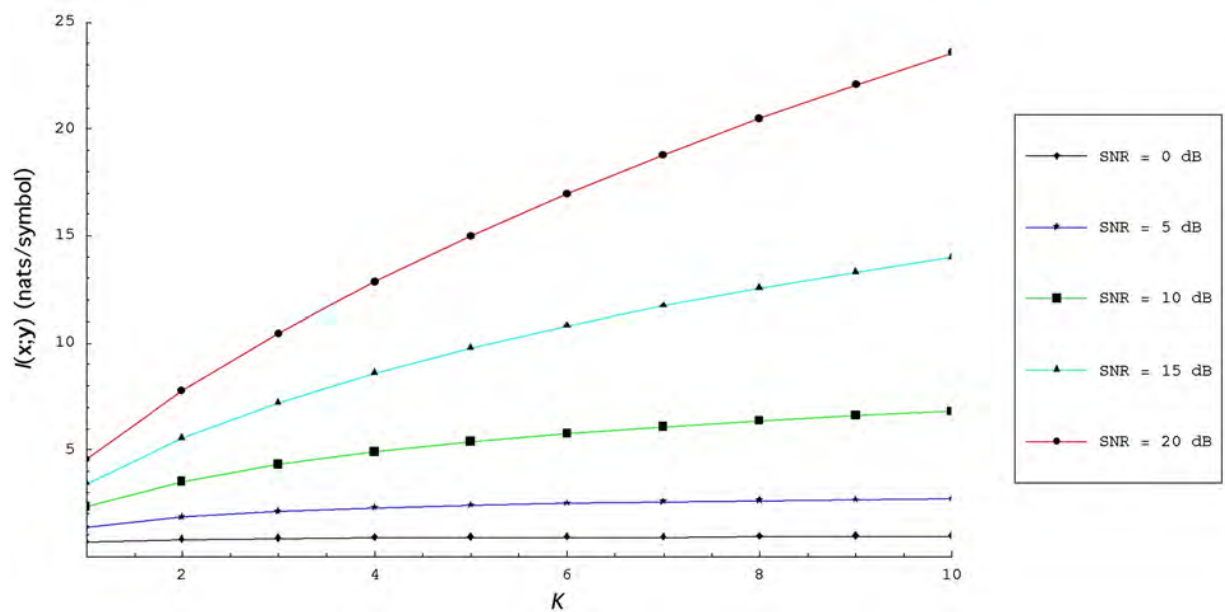

Figure 14. Capacity for CDMA Channel With $L = 10$ Resolvable Paths and $1 \leq K \leq 10$ Spreading Codes ($\beta = 0$).



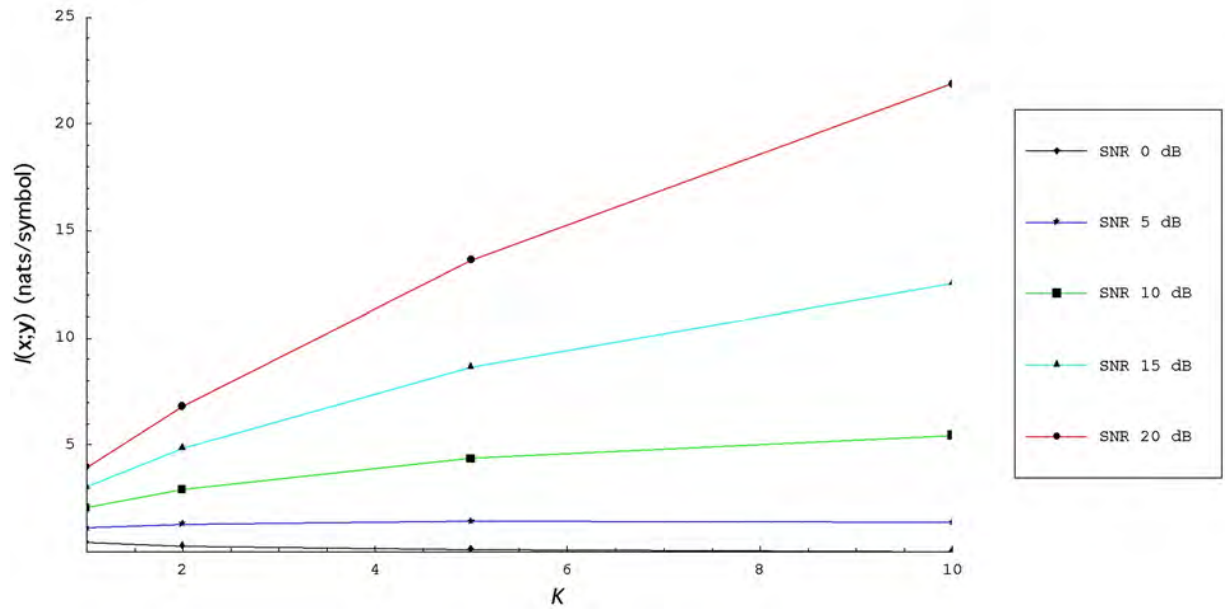

Figure 15. Estimated Capacity as a Function of Constellation Dimension $K$ for multicoded

CDMA Channel with $L = 10$ and $\beta = 0$ (Coherent Channel).

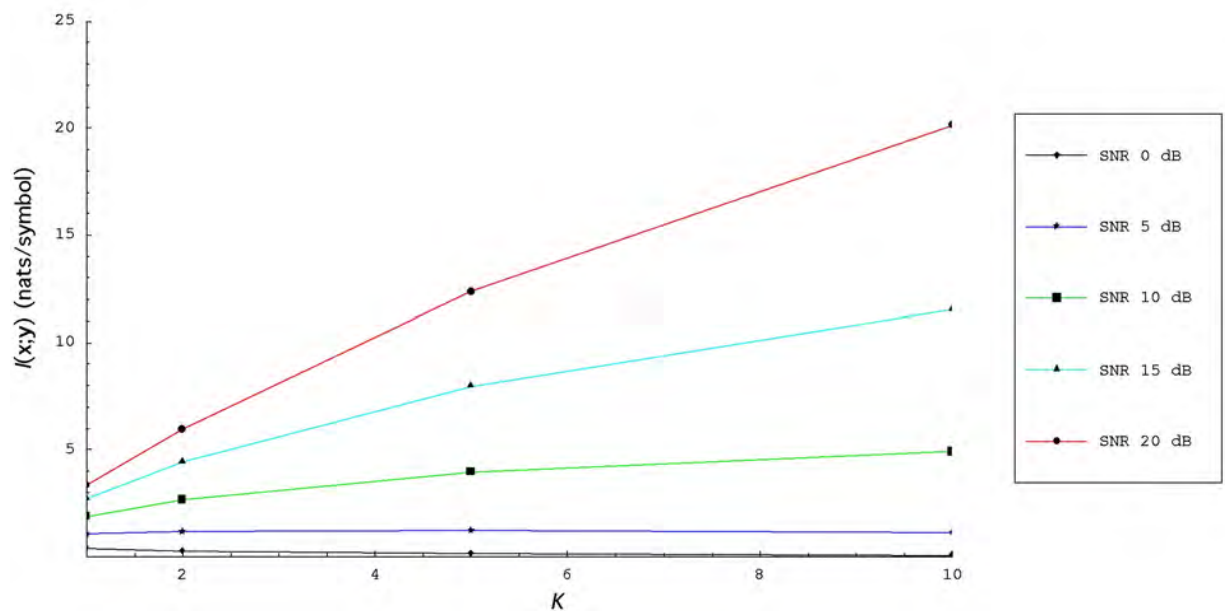

Figure 16. Estimated Capacity as a Function of Constellation Dimension $K$ for multicoded

CDMA Channel with $L = 10$ and $\beta = 0.1$.



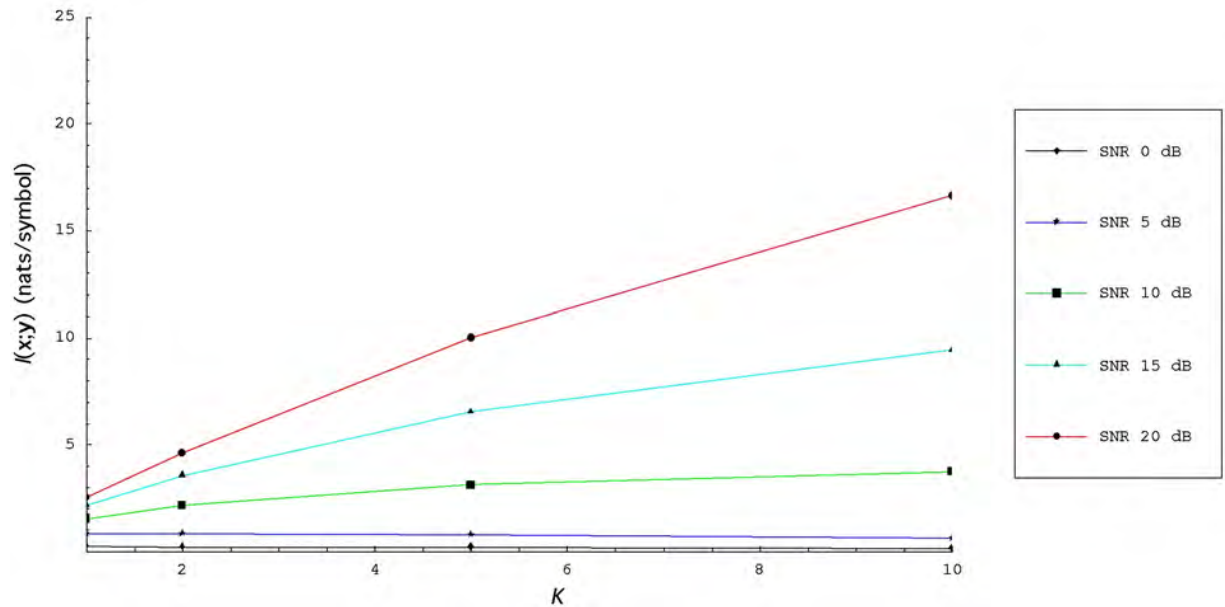

Figure 17. Estimated Capacity as a Function of Constellation Dimension $K$ for multicoded

CDMA Channel with $L = 10$ and $\beta = 0.3$.

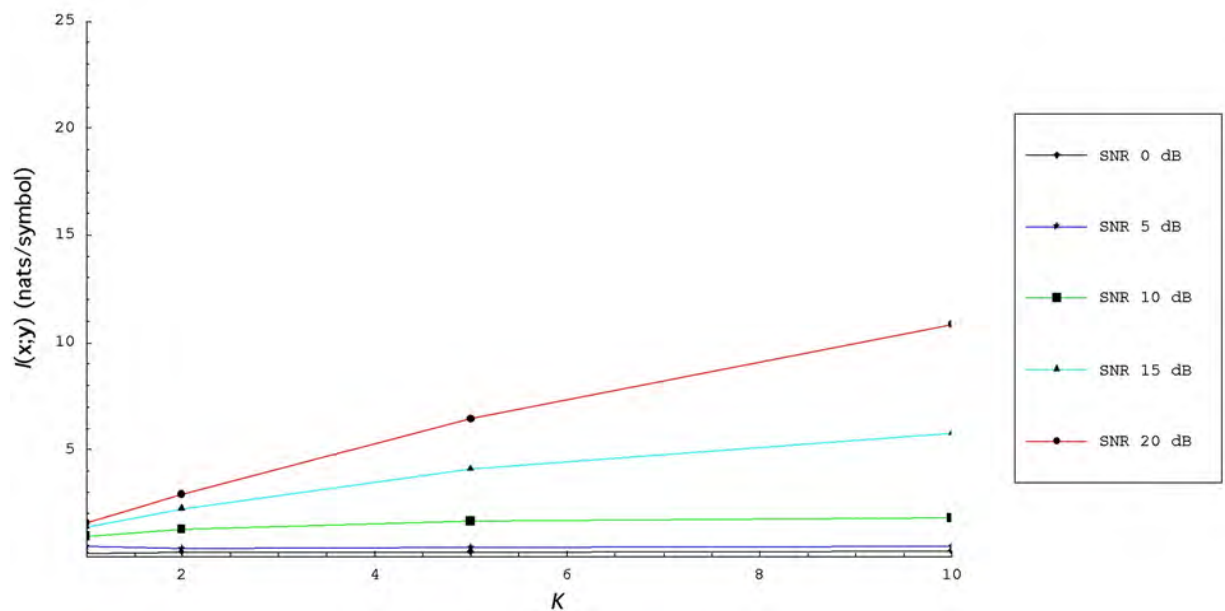

Figure 18. Estimated Capacity as a Function of Constellation Dimension $K$ for multicoded

CDMA Channel with $L = 10$ and $\beta = 0.5$.



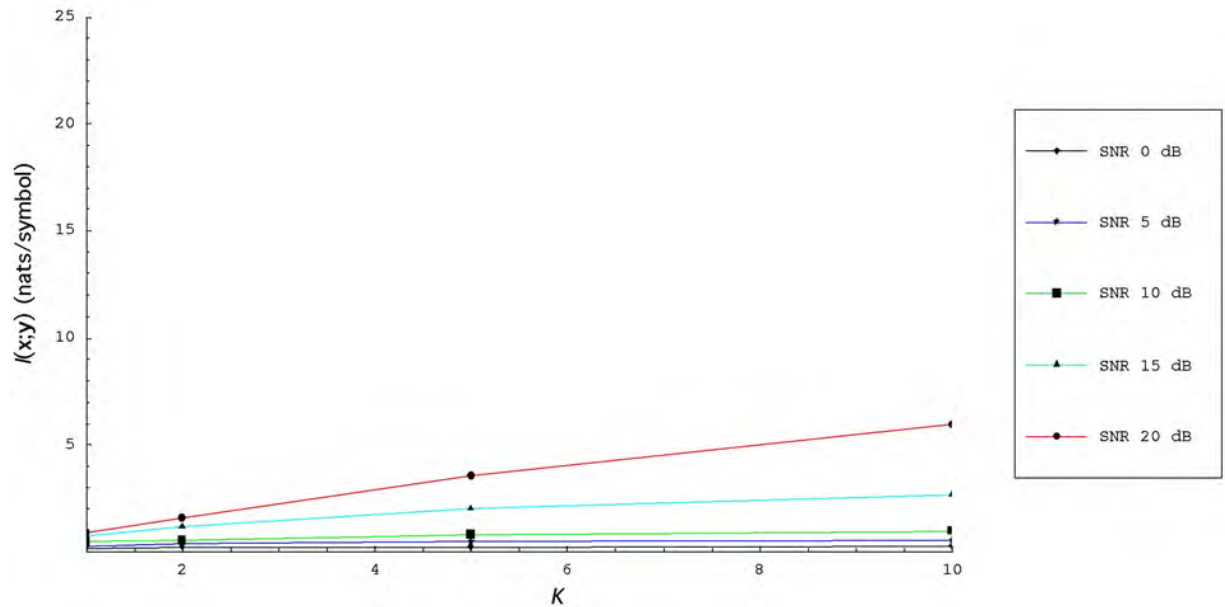

Figure 19. Estimated Capacity as a Function of Constellation Dimension $K$ for multicoded

CDMA Channel with $L = 10$ and $\beta = 0.7$.

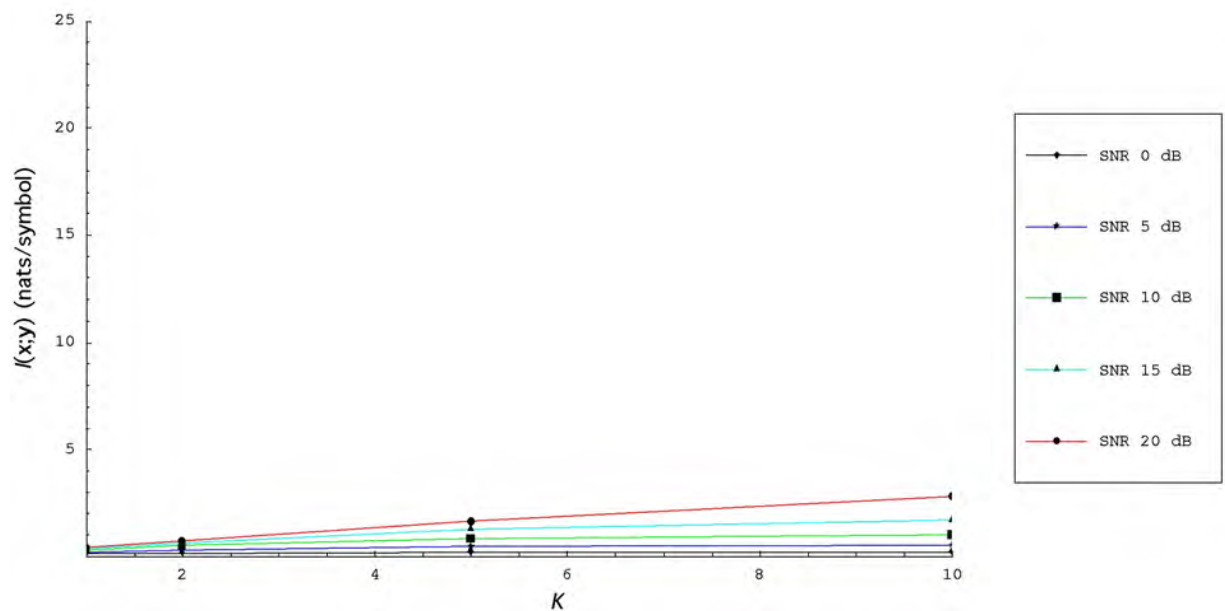

Figure 20. Estimated Capacity as a Function of Constellation Dimension $K$ for multicoded

CDMA Channel with $L = 10$ and $\beta = 0.9$.



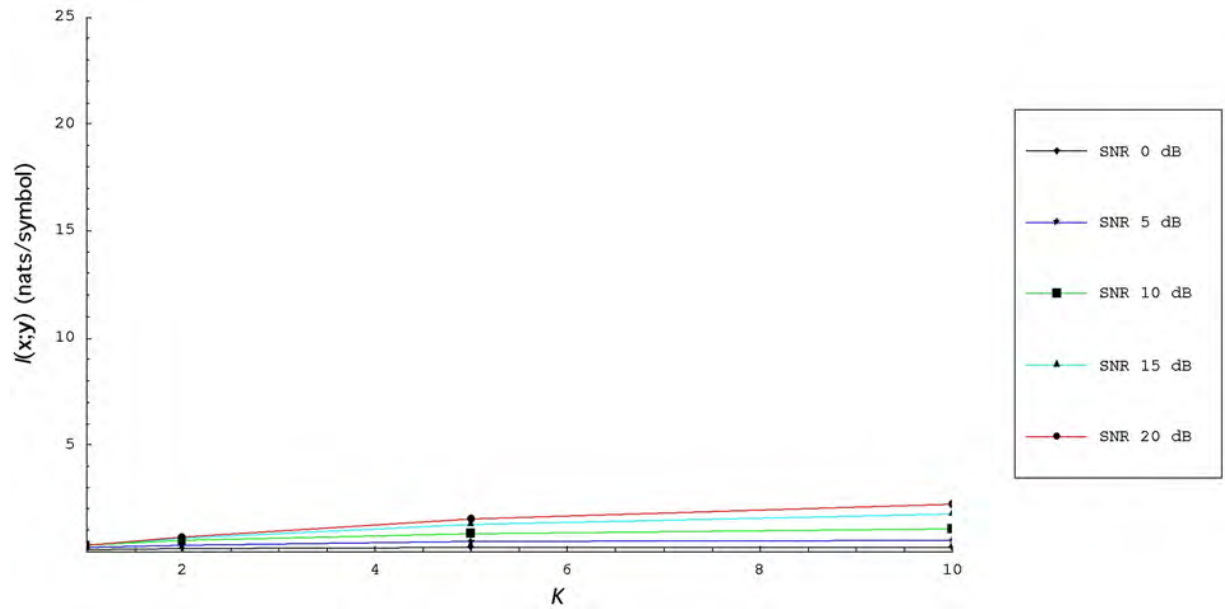

Figure 21. Estimated Capacity as a Function of Constellation Dimension $K$ for multicoded

CDMA Channel with $L = 10$ and $\beta = 1$ (Noncoherent Channel).

As illustrated in Figures 14 and 15, the estimated capacity based on the lower bounds is a

reasonable approximation for the true capacity for the coherent channel. For the higher SNR

values in particular, the estimated capacity seems to be within 1-2 dB of the true capacity when

the CSI is known precisely. Assuming a similar relationship between the estimated and true

capacity for values of $0 < \beta \leq 1$, it is clear that the impact of increasing signal dimension is much

greater on channels with low uncertainty than on channels with high uncertainty. Similarly, the

capacity increases much faster as a function of signal dimension when the SNR is high than

when it is low. Hence, the availability of additional signaling dimensions is generally of more

importance on channels with relatively reliable CSI estimates than on channels with relatively

unreliable CSI estimates and also of more importance on channels with relatively high SNR than

on cannels with relatively low SNR.



### 4.3. Asymptotic Results

In this final subsection, we discuss the capacity of the CDMA Rayleigh fading channel as both the spreading gain $N$ and the number of resolvable paths $L$ on the channel go to infinity. Such behavior would be observed in particular in a diffuse fading environment if the symbol duration $T = NT_c$ remains fixed as the bandwidth $B \propto 1/T_c$ of the signal is increased without bound. In order to ensure that Assumptions 1 and 4 hold asymptotically, we make the additional assumption that $KL/N \to 0$ as $N \to \infty$. Note that in a wideband diffuse fading environment, it would generally be expected that $L = \varepsilon/T_c$ for some $0 < \varepsilon \ll 1$, so the assumption $KL/N \to 0$ implies that $\varepsilon K/T \to 0$ and in particular that $T \to \infty$ as $B \to \infty$.

Notice first that, in the absence of any CSI estimates, Corollary 1 implies that

$$I_{\hat{\boldsymbol{\alpha}}}(\mathbf{x};\mathbf{y}) \le \frac{1}{2} E\left\{ \mathcal{E}_{\mathbf{x}}^2 \right\} \sum_{i=1}^{L} \frac{\delta_i^4}{\sigma^4}.$$

Hence, for any fading scenario (such as uniform fading) for which the fourth moment of the fading process goes to zero; that is,

$$\lim_{L \to \infty} \sum_{i=1}^{L} \delta_i^4 = 0,$$

we have $\lim_{L \to \infty} I_{\hat{\boldsymbol{\alpha}}}(\mathbf{x};\mathbf{y}) = 0$, as long as the fourth moment

$$E\left\{ \mathcal{E}_{\mathbf{x}}^2 \right\} = E\left\{ \sum_{i=1}^{K} \sum_{j=1}^{K} |x_i|^2 |x_j|^2 \right\}$$

of the constellation is finite. This immediately implies that, if we consider only constellations with a finite bound on the fourth moment, the infinite-bandwidth capacity of a diffuse CDMA Rayleigh fading channel with no side information is zero. This is consistent with a number of related results reported recently in the literature [7-9, 18].



On the other hand, it follows from Lemma 3 (see Appendix) that in the special case of uniform fading, if $K = e^{2mL}$ for $m \geq 1$ and $L$ large, the mutual information on the noncoherent channel for a unit energy orthogonal constellation with $E\{\mathcal{E}_\mathbf{x}^2\} = 2mL\sigma^2$ satisfies

$$I_{\hat{\boldsymbol{\alpha}}}(\mathbf{x};\mathbf{y}) \geq \frac{1}{2mL\sigma^2}\left[2mL - L\ln(2m+1) - \frac{4mL}{2m+1}\sqrt{m}e^{-1} - m\sqrt{\frac{2L}{\pi}}\right] \rightarrow \frac{1}{\sigma^2} \text{ as } m,L \rightarrow \infty.$$

This implies that the infinite-bandwidth single-user capacity for a diffuse CDMA Rayleigh fading channel with no side information is identical to the single-user capacity of an AWGN channel with the same SNR. Note, however, that to achieve capacity in this fashion, it is necessary to transmit a zero symbol a large percentage of the time, which means we need to adopt an impulsive on-off-keying strategy in the channel coding. This result provides an *M*-ary CDMA analog for a result proven by Kennedy [12] for *M*-ary frequency-shift keying and is consistent with results given in [8] and [9] which imply that some type of bursty and impulsive transmission strategy must always be adopted to achieve capacity in the wideband regime.

Note also that it is not clear from Lemma 4 that an increase in signal dimension along with the fourth moment of the constellation is necessary to achieve the indicated lower bound on mutual information, only that it is sufficient. However, we conjecture that the increase in dimension is also necessary. This conjecture is based on the behavior of the mutual information in the special case of perfect CSI estimates. In this case, we have $\hat{\boldsymbol{\alpha}} = \boldsymbol{\alpha}$, $\boldsymbol{\delta} = \mathbf{0}$, and

$$I_{\hat{\boldsymbol{\alpha}}}(\mathbf{x};\mathbf{y}) = K\ln\left(1 + \frac{1}{K\sigma^2}\sum_{i=1}^{L}|\alpha_i|^2\right) \leq \frac{1}{\sigma^2}\sum_{i=1}^{L}|\alpha_i|^2,$$

where the inequality in this relationship is strict unless $K\sigma^2 >> \sum_{i=1}^{L}|\alpha_i|^2$. It follows that the maximum conditional mutual information on a perfectly known channel is achieved only by letting the dimension of the constellation increase faster than the received SNR.



A couple of comments are worth making at this point. First, to attain capacity in the wideband noncoherent case using the multidimensional impulsive on-off-keying scenario that we have described would require that the symbol duration increase exponentially with the signal bandwidth. This follows from the facts that asymptotically, the number $L$ of resolvable paths can be expected to grow linearly with the bandwidth, and the number of required signal dimensions (hence also the spreading gain and symbol duration) is $K = e^{2mL}$. This is not a disadvantage from the point of view of Shannon capacity, but certainly has an undesirable impact on the delay-limited capacity of the system. Secondly, although such an impulsive on-off-keying transmission scheme can be viewed as a type of impulsive, time-hopping CDMA, it is not equivalent to the so-called time-hopping CDMA schemes being proposed currently for some impulse radio systems [19]. Indeed, the proposed impulse radio systems can be viewed as conventional CDMA schemes with {0,1} rather than {-1,1} spreading codes and very large bandwidth. As such, they can be expected to suffer severe degradation in capacity on channels with a large number of resolvable paths. In order to avoid such degradation, it is necessary to introduce the impulsive signal structure within the symbol stream rather than within each individual symbol and also to transmit a large number of bits per symbol.

## 5. Conclusion

In this paper, we have investigated the influence of signal subspace dimension and input symbol distribution on channel capacity in the presence of uncertainty regarding available channel state information for a wideband CDMA Rayleigh fading channel. The channel model adopted incorporates adaptive multicoding as a means of altering both the dimension and probability distribution of the transmitted signal constellation. Several conclusions can be drawn from the



results presented in the paper, which are of considerable importance for wideband CDMA systems operating in a fading environment.

First, it is possible to compute arbitrarily good estimates of channel capacity and the associated capacity-achieving input distribution with relatively low complexity for small values of the constellation dimension $K$. Formulas for mutual information and necessary and sufficient conditions for an input symbol distribution to maximize mutual information have been developed for the case $K = 1$ explicitly in terms of SNR, the number of resolvable paths on the channel, the observed estimate of CSI, and the MSE of the CSI estimates.

Second, the dimension of the signal constellation on a wideband CDMA system has a significant impact on the mutual information on the channel, but the rate of increase in mutual information as a function of signal dimension depends critically on the relationship between the accuracy of the CSI estimates, the SNR, and the number of resolvable paths on the channel. In particular, the availability of additional signaling dimensions is generally of more importance on channels with relatively reliable CSI estimates than on channels with relatively unreliable CSI estimates and also of more importance on channels with relatively high SNR than on channels with relatively low SNR.

Similarly, the impact of input symbol distribution on mutual information on a wideband CDMA channel depends greatly on the accuracy of the CSI estimates and the SNR. In particular, the proper choice of input symbol distribution is generally of more importance on channels with relatively unreliable CSI estimates than on channels with relatively reliable CSI estimates and also of more importance on channels with relatively high SNR than on channels with relatively low SNR. In contrast to this, it is somewhat surprising that in the absence of adaptive power control, adapting the size of the signal constellation and the distribution of the transmitted



symbols in response to instantaneous changes in the channel state leads to only minor increases in mutual information compared to the best possible fixed constellation. Hence, channel capacity with or without channel side information is nearly the same on the channels studied in this paper. The largest gains in mutual information are associated with channels for which the CSI is only partially known, and there is no impact whatsoever on channels for which the CSI is known precisely. Further, regardless of the level of channel uncertainty, there is no performance advantage associated with adapting only the size of the constellation without simultaneously adapting either the average power or the distribution of the symbols.

Finally, in the absence of any CSI estimates and subject to the usual constraint on average energy transmitted per symbol, the mutual information of the system for any constellation with finite fourth moment decreases to zero at the same rate as the fourth moment of the fading process itself as the number of fading paths on the channel goes to infinity. This essentially implies that the infinite-bandwidth capacity for a CDMA system subject to a fourth-moment constraint is zero on a diffuse WSSUS Rayleigh fading channel. Conversely, if adaptive multicoding is employed so that both the dimension and the fourth moment of the constellation increase as the number of resolvable paths increases, the mutual information, and therefore the channel capacity, can be made arbitrarily close  to the capacity for an infinite-bandwidth, single-user AWGN channel with the same SNR.

Many of the results presented in this paper can be extended in a straightforward manner to a more general channel model than the one adopted herein. In particular, Lemma 4 and its corollaries can be extended to provide expressions for mutual information and capacity on multiple-input, multiple-output (MIMO) channels and multiuser channels. These and other extensions will be addressed in a future paper.



**Appendix.**

In this Appendix, we prove Lemmas 1-5 and Corollary 3. Throughout the Appendix, we adopt the notation introduced in Section 3 of this paper. We will need the following facts in the proof of the Lemma 1. Fact 1 is obvious, Fact 2 follows immediately from the Schur determinant formula [20], and proofs of Facts 3 and 4 can be found in [21].

**Fact 1**. The mean vector and covariance matrix for the output vector $\mathbf{y}$ are given by

$$\boldsymbol{\mu}_{\mathbf{y}} = E\left\{\boldsymbol{\mu}_{\mathbf{y}|\mathbf{x}}\right\} \text{ and } \boldsymbol{\Sigma}_{\mathbf{y}} = E\left\{\boldsymbol{\Sigma}_{\mathbf{y}|\mathbf{x}} + \boldsymbol{\mu}_{\mathbf{y}|\mathbf{x}}\boldsymbol{\mu}_{\mathbf{y}|\mathbf{x}}^{*}\right\} - \boldsymbol{\mu}_{\mathbf{y}}\boldsymbol{\mu}_{\mathbf{y}}^{*}, \text{ respectively.}$$

**Fact 2**. If $\mathbf{A}$ is an $n \times m$ matrix and $\mathbf{B}$ is an $m \times n$ matrix, then $\det(\mathbf{I} + \mathbf{AB}) = \det(\mathbf{I} + \mathbf{BA})$.

**Fact 3**. The (differential) entropy of an $N$-dimensional proper complex Gaussian random vector $\mathbf{v}$ with arbitrary mean and nonsingular covariance matrix $\boldsymbol{\Sigma}$ is given by

$$h(\mathbf{v}) = \ln\left[(\pi e)^{N}|\boldsymbol{\Sigma}|\right] \text{ nats },$$

where $|\boldsymbol{\Sigma}|$ denotes the determinant of $\boldsymbol{\Sigma}$.

**Fact 4**. The entropy of an arbitrary complex-valued continuous random vector with nonsingular covariance matrix $\boldsymbol{\Sigma}$ is maximized by the corresponding Gaussian random vector.

**Proof of Lemma 1**. Using Facts 3 and 4 and the additional fact that the conditional distribution of the channel output is Gaussian given the value of the input symbol, we see that

$$I_{\hat{\alpha}}(\mathbf{x};\mathbf{y}) = h(\mathbf{y}) - h(\mathbf{y}|\mathbf{x}) \le \ln|\boldsymbol{\Sigma}_{\mathbf{y}}| - E\left\{\ln|\boldsymbol{\Sigma}_{\mathbf{y}|\mathbf{x}}|\right\}.$$

Then using Fact 2 and the expressions for $\boldsymbol{\mu}_{\mathbf{y}|\mathbf{x}}$ and $\boldsymbol{\Sigma}_{\mathbf{y}|\mathbf{x}}$ given by (3.1) and (3.2), we get

$$\begin{aligned}
I_{\hat{\alpha}}(\mathbf{x};\mathbf{y}) &\le \ln|\boldsymbol{\Sigma}_{\mathbf{y}}| - E\left\{\ln|\mathbf{B}_{\mathbf{x}}|\right\} \\
&= \ln|\boldsymbol{\Sigma}_{\mathbf{y}}| - E\left\{\ln|\mathbf{H}_{\mathbf{x}}\boldsymbol{\Sigma}_{\boldsymbol{\varepsilon}}\mathbf{H}_{\mathbf{x}}^{*} + \sigma^{2}\mathbf{I}|\right\} \\
&= \ln|\boldsymbol{\Sigma}_{\mathbf{y}}| - E\left\{\ln|\mathcal{E}_{\mathbf{x}}\boldsymbol{\Sigma}_{\boldsymbol{\varepsilon}} + \sigma^{2}\mathbf{I}|\right\} - (K-1)L\ln\sigma^{2}.
\end{aligned}$$



Now, using Fact 1 and the expressions for $\boldsymbol{\mu}_{\mathbf{y}|\mathbf{x}}$ and $\boldsymbol{\Sigma}_{\mathbf{y}|\mathbf{x}}$, it is straightforward to show that $|\boldsymbol{\Sigma}_{\mathbf{y}}| = |\mathbf{F}|$. Hence,

$$I_{\hat{\boldsymbol{\alpha}}}(\mathbf{x};\mathbf{y}) \leq \ln|\mathbf{F}| - E\left\{\ln\left|\boldsymbol{\mathcal{E}}_{\mathbf{x}}\boldsymbol{\Sigma}_{\boldsymbol{\epsilon}} + \sigma^2\mathbf{I}\right|\right\} - (K-1)L\ln\sigma^2,$$

which establishes the first inequality. Fischer's inequality [20] then immediately implies that

$$I_{\hat{\boldsymbol{\alpha}}}(\mathbf{x};\mathbf{y}) \leq \sum_{k=1}^{K}\ln\left|s_k^2\boldsymbol{\Sigma}_{\boldsymbol{\epsilon}} + \sigma_k^2\hat{\alpha}\hat{\alpha}^* + \sigma^2\mathbf{I}\right| - E\left\{\ln\left|\boldsymbol{\mathcal{E}}_{\mathbf{x}}\boldsymbol{\Sigma}_{\boldsymbol{\epsilon}} + \sigma^2\mathbf{I}\right|\right\} - (K-1)L\ln\sigma^2$$

$$= \sum_{k=1}^{K}\ln\left|s_k^2\boldsymbol{\Sigma}_{\boldsymbol{\epsilon}} + \sigma^2\mathbf{I}\right| + \sum_{k=1}^{K}\ln\left|\mathbf{I} + \sigma_k^2\hat{\alpha}^*\left(s_k^2\boldsymbol{\Sigma}_{\boldsymbol{\epsilon}} + \sigma^2\mathbf{I}\right)^{-1}\hat{\alpha}\right| - E\left\{\ln\left|\boldsymbol{\mathcal{E}}_{\mathbf{x}}\boldsymbol{\Sigma}_{\boldsymbol{\epsilon}} + \sigma^2\mathbf{I}\right|\right\} - (K-1)L\ln\sigma^2.$$

Maximizing this expression with respect to $\left\{s_k^2,\sigma_k^2\right\}_{k=1}^{K}$ leads to $s_k^2 \equiv \sigma_k^2 \equiv 1/K$. Making use of this result as well as the assumption that $\Sigma_{\boldsymbol{\epsilon}} = \mathrm{diag}\left\{\delta_1^2,\ldots,\delta_L^2\right\}$, it follows that

$$I_{\hat{\boldsymbol{\alpha}}}(\mathbf{x};\mathbf{y}) \leq \sum_{k=1}^{K}\ln\left(K^{-L}\left|\boldsymbol{\Sigma}_{\boldsymbol{\epsilon}} + K\sigma^2\mathbf{I}\right|\right) + \sum_{k=1}^{K}\ln\left|\mathbf{I} + \hat{\alpha}^*\left(\boldsymbol{\Sigma}_{\boldsymbol{\epsilon}} + K\sigma^2\mathbf{I}\right)^{-1}\hat{\alpha}\right| - E\left\{\ln\left|\boldsymbol{\mathcal{E}}_{\mathbf{x}}\boldsymbol{\Sigma}_{\boldsymbol{\epsilon}} + \sigma^2\mathbf{I}\right|\right\} - (K-1)L\ln\sigma^2$$

$$= -KL\ln K + K\sum_{l=1}^{L}\ln\left(\delta_l^2 + K\sigma^2\right) + K\ln\left(1 + \sum_{i=1}^{L}\frac{|\hat{\alpha}_i|^2}{\delta_i^2 + K\sigma^2}\right) - E\left\{\sum_{i=1}^{L}\ln\left(\boldsymbol{\mathcal{E}}_{\mathbf{x}}\delta_i^2 + \sigma^2\right)\right\} - (K-1)L\ln\sigma^2$$

$$= K\sum_{i=1}^{L}\ln\left(1 + \frac{\delta_i^2}{K\sigma^2}\right) + K\ln\left(1 + \sum_{i=1}^{L}\frac{|\hat{\alpha}_i|^2}{\delta_i^2 + K\sigma^2}\right) - E\left\{\sum_{i=1}^{L}\ln\left(1 + \boldsymbol{\mathcal{E}}_{\mathbf{x}}\frac{\delta_i^2}{\sigma^2}\right)\right\},$$

which completes the proof. ∎

**Proof of Lemma 2**. We prove the result for discrete-distribution input sources, which is the more difficult of the two results. The result for continuous-distribution input sources is proven in an analogous manner. We start by noting that

$$I_{\hat{\boldsymbol{\alpha}}}(\mathbf{x};\mathbf{y}) = H_{\hat{\boldsymbol{\alpha}}}(x) - H_{\hat{\boldsymbol{\alpha}}}(\mathbf{x}|\mathbf{y}) = H_{\hat{\boldsymbol{\alpha}}}(\mathbf{x}) - H_{\hat{\boldsymbol{\alpha}}}(\mathbf{x} - \hat{\mathbf{x}}(\mathbf{y})|\mathbf{y}),$$

where $H_{\hat{\boldsymbol{\alpha}}}(\mathbf{x}|\mathbf{y})$ is the conditional entropy of $\mathbf{x}$ given the observed output vector $\mathbf{y}$, $\hat{\mathbf{x}}(\mathbf{y})$ is an arbitrary estimate of $\mathbf{x}$ based on observation of the output vector $\mathbf{y}$, and $H_{\hat{\boldsymbol{\alpha}}}(\mathbf{x} - \hat{\mathbf{x}}(\mathbf{y})|\mathbf{y})$ is the conditional entropy of $\mathbf{e} = \mathbf{x} - \hat{\mathbf{x}}(\mathbf{y})$ given $\mathbf{y}$. Now, let $\tilde{\mathbf{e}} = \mathbf{e} + \mathbf{u}$, where $\mathbf{u}$ is a $K$-dimensional



complex-valued random vector, independent of both $\mathbf{x}$ and $\mathbf{y}$, uniformly distributed on the ball $\|\mathbf{u}\| \leq d/2$. Then, it is straightforward to show that

$$h_{\hat{\boldsymbol{\alpha}}}(\tilde{\mathbf{e}}|\mathbf{y}) = H_{\hat{\boldsymbol{\alpha}}}(\mathbf{e}|\mathbf{y}) + \log\left(\frac{\pi^K d^{2K}}{K! \, 4^K}\right),$$

where $h_{\hat{\boldsymbol{\alpha}}}(\tilde{\mathbf{e}}|\mathbf{y})$ represents the conditional (differential) entropy of $\tilde{\mathbf{e}}$ given $\mathbf{y}$. Hence,

$$I_{\hat{\boldsymbol{\alpha}}}(\mathbf{x};\mathbf{y}) = H_{\hat{\boldsymbol{\alpha}}}(\mathbf{x}) - H_{\hat{\boldsymbol{\alpha}}}(\mathbf{e}|\mathbf{y}) = H_{\hat{\boldsymbol{\alpha}}}(\mathbf{x}) - h_{\hat{\boldsymbol{\alpha}}}(\tilde{\mathbf{e}}|\mathbf{y}) + \ln\left(\frac{\pi^K d^{2K}}{K! \, 4^K}\right).$$

Since conditioning cannot increase entropy, we can write

$$I_{\hat{\boldsymbol{\alpha}}}(\mathbf{x};\mathbf{y}) \geq H_{\hat{\boldsymbol{\alpha}}}(\mathbf{x}) - h_{\hat{\boldsymbol{\alpha}}}(\tilde{\mathbf{e}}) + \ln\left(\frac{\pi^K d^{2K}}{K! \, 4^K}\right).$$

Using Fact 4 above and the relationship $\boldsymbol{\Sigma}_{\tilde{\mathbf{e}}} = \boldsymbol{\Sigma}_{\mathbf{e}} + \left[d^2/4(K+1)\right] \cdot \mathbf{I}$, we can then conclude that

$$I_{\hat{\boldsymbol{\alpha}}}(\mathbf{x};\mathbf{y}) \geq H_{\hat{\boldsymbol{\alpha}}}(\mathbf{x}) + \log\left(\frac{\pi^K d^{2K}}{K! \, 4^K}\right) - \ln\left[(\pi e)^K \left|\boldsymbol{\Sigma}_{\mathbf{e}} + \frac{d^2}{4(K+1)}\mathbf{I}\right|\right].$$

To establish the desired bound, we let $\hat{\mathbf{x}}(\mathbf{y}) = \mathbf{G}\mathbf{y} + \mathbf{m}$ be the minimum-mean-squared-error (MMSE) affine estimate of $\mathbf{x}$ based on $\mathbf{y}$. Using standard properties of affine MMSE estimates and the asymptotic expressions for $\boldsymbol{\mu}_{\mathbf{y}|\mathbf{x}}$ and $\boldsymbol{\Sigma}_{\mathbf{y}|\mathbf{x}}$ given by **(3.1)** and **(3.2)**, it is straightforward to show that

$$\mathbf{G} = \boldsymbol{\Sigma}_{\mathbf{x}}\mathbf{A}^*\mathbf{F}^{-1}\begin{bmatrix}\mathbf{C}_1^* \\ \mathbf{C}_2^* \\ \vdots \\ \mathbf{C}_K^*\end{bmatrix}, \quad \mathbf{m} = \boldsymbol{\mu}_{\mathbf{x}} - \sum_{i=1}^{K}\mu_i\mathbf{G}\mathbf{C}_i\hat{\boldsymbol{\alpha}}, \quad \boldsymbol{\Sigma}_{\mathbf{e}} = \boldsymbol{\Sigma}_{\mathbf{x}} - \boldsymbol{\Sigma}_{\mathbf{x}}\mathbf{A}^*\mathbf{F}^{-1}\mathbf{A}\boldsymbol{\Sigma}_{\mathbf{x}},$$

and the result follows immediately from the previous inequality. ∎

**Proof of Lemma 3**. Let $\mathbf{x} = (x_1, x_2, \ldots, x_K)^T$ represent an arbitrary channel input symbol, and $\mathbf{y} = (y_1, y_2, \ldots, y_{KL})^T$ represent the observed channel output. Recall that we are assuming $\mathbf{y}$ is



proper complex Gaussian with mean $\boldsymbol{\mu}_{\mathbf{y}|\mathbf{x}} = \mathbf{U}^*\mathbf{H}_{\mathbf{x}}\hat{\boldsymbol{\alpha}}$ and covariance $\boldsymbol{\Sigma}_{\mathbf{y}|\mathbf{x}} = \mathbf{U}^*\mathbf{B}_{\mathbf{x}}\mathbf{U}$, where

$\mathbf{H}_{\mathbf{x}} = \begin{bmatrix} x_1\mathbf{I} \mid x_2\mathbf{I} \mid \cdots \mid x_K\mathbf{I} \end{bmatrix}^T$ and $\mathbf{B}_{\mathbf{x}} = \mathbf{H}_{\mathbf{x}}\boldsymbol{\Sigma}_{\boldsymbol{\varepsilon}}\mathbf{H}_{\mathbf{x}}^* + \sigma^2\mathbf{I}$. Hence, the conditional output distribution

given that symbol $\mathbf{x}$ was transmitted is given by

$$p_{\hat{\boldsymbol{\alpha}}}(\mathbf{y}|\mathbf{x}) = \frac{1}{\pi^{KL}|\mathbf{B}_{\mathbf{x}}|}e^{-(\mathbf{y}-\mathbf{U}\mathbf{H}_{\mathbf{x}}\hat{\boldsymbol{\alpha}})^*\mathbf{U}\mathbf{B}_{\mathbf{x}}^{-1}\mathbf{U}^*(\mathbf{y}-\mathbf{U}\mathbf{H}_{\mathbf{x}}\hat{\boldsymbol{\alpha}})}.$$

Similarly, if we let

$$\bar{\mathbf{y}} = \begin{bmatrix} \tilde{\mathbf{y}}_1 \\ \hline \tilde{\mathbf{y}}_2 \\ \hline \vdots \\ \hline \tilde{\mathbf{y}}_K \end{bmatrix} = \mathbf{U}^*\mathbf{y}, \quad \tilde{\mathbf{y}}_i = \begin{bmatrix} \tilde{y}_{i1} \\ \tilde{y}_{i2} \\ \vdots \\ \tilde{y}_{iL} \end{bmatrix}, i = 1,2,\ldots,K,$$

represent an equivalent transformed channel output, then the conditional transformed output

distribution given that symbol $\mathbf{x}$ was transmitted is given by

$$p_{\hat{\boldsymbol{\alpha}}}(\bar{\mathbf{y}}|\mathbf{x}) = \frac{1}{\pi^{KL}\sigma^{2(K-1)L}\left|\boldsymbol{\Sigma}_{\mathbf{x}}\boldsymbol{\Sigma}_{\boldsymbol{\varepsilon}} + \sigma^2\mathbf{I}\right|}e^{-\bar{\mathbf{y}}^*\mathbf{B}_{\mathbf{x}}^{-1}\bar{\mathbf{y}} - \hat{\boldsymbol{\alpha}}^*\mathbf{H}_{\mathbf{x}}^*\mathbf{B}_{\mathbf{x}}^{-1}\mathbf{H}_{\mathbf{x}}\hat{\boldsymbol{\alpha}} + 2\mathrm{Re}\left(\bar{\mathbf{y}}^*\mathbf{B}_{\mathbf{x}}^{-1}\mathbf{H}_{\mathbf{x}}\hat{\boldsymbol{\alpha}}\right)}$$

Now, $\mathcal{E} = 2mL\sigma^2$ for $m > 1/2L\sigma^2$, $\mathbf{x}_0 = \mathbf{0}$, $p_0 = p_{\hat{\boldsymbol{\alpha}}}(\mathbf{x}_0) = 1 - 1/\mathcal{E}$,

$$\delta_i^2 \equiv \frac{\beta}{L}, \quad \mathbf{x}_i = \sqrt{\mathcal{E}}\,\mathbf{e}_i, \quad p_i = p_{\hat{\boldsymbol{\alpha}}}(\mathbf{x}_i) = \frac{1}{K\mathcal{E}}, \quad i = 1,2,\ldots,K,$$

where $\mathbf{e}_i$ represents the $i$th standard basis vector in $\mathcal{C}^K$. It follows that

$$p_{\hat{\boldsymbol{\alpha}}}(\bar{\mathbf{y}}|\mathbf{x}_0) = \frac{1}{\pi^{KL}\sigma^{2KL}}e^{-\|\bar{\mathbf{y}}\|^2/\sigma^2},$$

and

$$p_{\hat{\boldsymbol{\alpha}}}(\bar{\mathbf{y}}|\mathbf{x}_i) = \frac{1}{\pi^{KL}\sigma^{2KL}(2m+1)^L}e^{-\frac{1}{\sigma^2}\left(\|\bar{\mathbf{y}}\|^2 - \frac{2m\beta}{2m\beta+1}\left\|\tilde{\mathbf{y}}_i + \frac{L\sigma^2}{\sqrt{2mL\sigma^2}\beta}\hat{\boldsymbol{\alpha}}\right\|^2 + \frac{L\sigma^2}{\beta}\|\hat{\boldsymbol{\alpha}}\|^2\right)},$$

for $i = 1,2,\ldots,K$. Then $I_{\hat{\boldsymbol{\alpha}}}(\mathbf{x};\mathbf{y}) = H_{\hat{\boldsymbol{\alpha}}}(\mathbf{x}) - H_{\hat{\boldsymbol{\alpha}}}(\mathbf{x}|\mathbf{y})$, where

$$H_{\hat{\boldsymbol{\alpha}}}(\mathbf{x}) = \left[1 - \frac{1}{2mL\sigma^2}\right]\ln\left[\frac{2mL\sigma^2}{2mL\sigma^2 - 1}\right] + \frac{1}{2mL\sigma^2}\ln\left[K2mL\sigma^2\right],$$



and

$$H_{\hat{\boldsymbol{\alpha}}}(\mathbf{x}|\mathbf{y}) = \sum_{i=0}^{K} p_i \int_{\bar{\mathbf{y}}} p_{\hat{\boldsymbol{\alpha}}}(\bar{\mathbf{y}}|\mathbf{x}_i) \ln\left[1 + \sum_{\substack{j=0 \\ j \neq i}}^{K} \frac{p_j}{p_i} \cdot \frac{p_{\hat{\boldsymbol{\alpha}}}(\bar{\mathbf{y}}|\mathbf{x}_j)}{p_{\hat{\boldsymbol{\alpha}}}(\bar{\mathbf{y}}|\mathbf{x}_i)}\right]$$

$$\leq p_0 \ln\left(1 + \frac{1}{p_0}\sum_{i=1}^{K} p_i \int_{\bar{\mathbf{y}}} p_{\hat{\boldsymbol{\alpha}}}(\bar{\mathbf{y}}|\mathbf{x}_i)\right) + \sum_{i=1}^{K} p_i \int_{\bar{\mathbf{y}}} p_{\hat{\boldsymbol{\alpha}}}(\bar{\mathbf{y}}|\mathbf{x}_i) \ln\left[1 + \frac{p_0}{p_i} \cdot \frac{p_{\hat{\boldsymbol{\alpha}}}(\bar{\mathbf{y}}|\mathbf{x}_0)}{p_{\hat{\boldsymbol{\alpha}}}(\bar{\mathbf{y}}|\mathbf{x}_i)} + \sum_{\substack{j=1 \\ j \neq i}}^{K} \frac{p_j}{p_i} \cdot \frac{p_{\hat{\boldsymbol{\alpha}}}(\bar{\mathbf{y}}|\mathbf{x}_j)}{p_{\hat{\boldsymbol{\alpha}}}(\bar{\mathbf{y}}|\mathbf{x}_i)}\right]$$

$$= \left[1 - \frac{1}{2mL\sigma^2}\right]\ln\left[\frac{2mL\sigma^2}{2mL\sigma^2 - 1}\right]$$

$$+ \frac{1}{K2mL\sigma^2}\sum_{i=1}^{K}\int_{\bar{\mathbf{y}}} p_{\hat{\boldsymbol{\alpha}}}(\bar{\mathbf{y}}|\mathbf{x}_i)\ln\left[1 + p_0 K 2mL\sigma^2 \frac{p_{\hat{\boldsymbol{\alpha}}}(\bar{\mathbf{y}}|\mathbf{x}_0)}{p_{\hat{\boldsymbol{\alpha}}}(\bar{\mathbf{y}}|\mathbf{x}_i)} + \sum_{\substack{j=1 \\ j \neq i}}^{K} \frac{p_{\hat{\boldsymbol{\alpha}}}(\bar{\mathbf{y}}|\mathbf{x}_j)}{p_{\hat{\boldsymbol{\alpha}}}(\bar{\mathbf{y}}|\mathbf{x}_i)}\right].$$

Hence, we have

$$I_{\hat{\boldsymbol{\alpha}}}(\mathbf{x};\mathbf{y}) \geq \frac{1}{2mL\sigma^2}\left(\ln\left[K2mL\sigma^2\right] - \frac{1}{K}\sum_{i=1}^{K}\int_{\bar{\mathbf{y}}} p_{\hat{\boldsymbol{\alpha}}}(\bar{\mathbf{y}}|\mathbf{x}_i)\ln\left[1 + p_0 K 2mL\sigma^2 \frac{p_{\hat{\boldsymbol{\alpha}}}(\bar{\mathbf{y}}|\mathbf{x}_0)}{p_{\hat{\boldsymbol{\alpha}}}(\bar{\mathbf{y}}|\mathbf{x}_i)} + \sum_{\substack{j=1 \\ j \neq i}}^{K} \frac{p_{\hat{\boldsymbol{\alpha}}}(\bar{\mathbf{y}}|\mathbf{x}_j)}{p_{\hat{\boldsymbol{\alpha}}}(\bar{\mathbf{y}}|\mathbf{x}_i)}\right]\right).$$

Now,

$$\int_{\bar{\mathbf{y}}} p_{\hat{\boldsymbol{\alpha}}}(\bar{\mathbf{y}}|\mathbf{x}_i)\ln\left[1 + p_0 K 2mL\sigma^2 \frac{p_{\hat{\boldsymbol{\alpha}}}(\bar{\mathbf{y}}|\mathbf{x}_0)}{p_{\hat{\boldsymbol{\alpha}}}(\bar{\mathbf{y}}|\mathbf{x}_i)} + \sum_{\substack{j=1 \\ j \neq i}}^{K} \frac{p_{\hat{\boldsymbol{\alpha}}}(\bar{\mathbf{y}}|\mathbf{x}_j)}{p_{\hat{\boldsymbol{\alpha}}}(\bar{\mathbf{y}}|\mathbf{x}_i)}\right]$$

$$= E_{\mathbf{x}_i}\left\{\ln\left[\begin{array}{l}1 + K\left(2mL\sigma^2 - 1\right)\left(2m\beta + 1\right)^L e^{\frac{L}{\beta}\|\hat{\boldsymbol{\alpha}}\|^2 - \frac{2m\beta}{\sigma^2(2m\beta+1)}\left\|\bar{\mathbf{y}}_i + \frac{L\sigma^2}{\sqrt{2mL\sigma^2}\beta}\hat{\boldsymbol{\alpha}}\right\|^2} \\ + \sum_{\substack{j=1 \\ j \neq i}}^{K} e^{\frac{2m\beta}{\sigma^2(2m\beta+1)}\left\|\bar{\mathbf{y}}_j + \frac{L\sigma^2}{\sqrt{2mL\sigma^2}\beta}\hat{\boldsymbol{\alpha}}\right\|^2 - \frac{2m\beta}{\sigma^2(2m\beta+1)}\left\|\bar{\mathbf{y}}_i + \frac{L\sigma^2}{\sqrt{2mL\sigma^2}\beta}\hat{\boldsymbol{\alpha}}\right\|^2}\end{array}\right]\right\}$$

$$= E_{\mathbf{x}_i}\left\{\ln\left[\begin{array}{l}1 + e^{-\frac{2m\beta}{\sigma^2(2m\beta+1)}\left\|\bar{\mathbf{y}}_i + \frac{L\sigma^2}{\sqrt{2mL\sigma^2}\beta}\hat{\boldsymbol{\alpha}}\right\|^2} \\ \left(K\left(2mL\sigma^2 - 1\right)\left(2m\beta + 1\right)^L e^{\frac{L}{\beta}\|\hat{\boldsymbol{\alpha}}\|^2} + \sum_{\substack{j=1 \\ j \neq i}}^{K} e^{\frac{2m\beta}{\sigma^2(2m\beta+1)}\left\|\bar{\mathbf{y}}_j + \frac{L\sigma^2}{\sqrt{2mL\sigma^2}\beta}\hat{\boldsymbol{\alpha}}\right\|^2}\right)\end{array}\right]\right\},$$



where "$E_{\mathbf{x}_i}$" denotes expectation with respect to the pdf $p_{\hat{\boldsymbol{\alpha}}}(\mathbf{y}|\mathbf{x}_i)$. Note that under $p_{\hat{\boldsymbol{\alpha}}}(\mathbf{y}|\mathbf{x}_i)$, the random variables

$$\frac{2m\beta}{\sigma^2(2m\beta+1)}\left\|\tilde{\mathbf{y}}_i + \frac{L\sigma^2}{\sqrt{2mL\sigma^2\beta}}\hat{\boldsymbol{\alpha}}\right\|^2 = \frac{2m\beta}{\sigma^2(2m\beta+1)}\sum_{l=1}^{L}\left|\tilde{y}_{il} - \frac{L\sigma^2}{\sqrt{2mL\sigma^2\beta}}\alpha_l\right|^2,$$

and

$$\frac{2m\beta}{\sigma^2(2m\beta+1)}\left\|\tilde{\mathbf{y}}_j + \frac{L\sigma^2}{\sqrt{2mL\sigma^2\beta}}\hat{\boldsymbol{\alpha}}\right\|^2 = \frac{2m\beta}{\sigma^2(2m\beta+1)}\sum_{l=1}^{L}\left|\tilde{y}_{jl} - \frac{L\sigma^2}{\sqrt{2mL\sigma^2\beta}}\alpha_l\right|^2, \quad j=2,\ldots,K,$$

are independent with

$$E_{\mathbf{x}_i}\left\{\frac{2m\beta}{\sigma^2(2m\beta+1)}\left\|\tilde{\mathbf{y}}_i + \frac{L\sigma^2}{\sqrt{2mL\sigma^2\beta}}\hat{\boldsymbol{\alpha}}\right\|^2\right\} = 2m\beta L\left[1 + \frac{2m\beta+1}{2m\beta^2}\|\hat{\boldsymbol{\alpha}}\|^2\right],$$

$$Var_{\mathbf{x}_i}\left\{\frac{2m\beta}{\sigma^2(2m\beta+1)}\left\|\tilde{\mathbf{y}}_i + \frac{L\sigma^2}{\sqrt{2mL\sigma^2\beta}}\hat{\boldsymbol{\alpha}}\right\|^2\right\} = 4m^2\beta^2 L\left[1 + \frac{2m\beta+1}{m\beta^2}\|\hat{\boldsymbol{\alpha}}\|^2\right],$$

and

$$E_{\mathbf{x}_i}\left\{\frac{2m\beta}{\sigma^2(2m\beta+1)}\left\|\tilde{\mathbf{y}}_j + \frac{L\sigma^2}{\sqrt{2mL\sigma^2\beta}}\hat{\boldsymbol{\alpha}}\right\|^2\right\} = \frac{2m\beta L}{2m\beta+1}\left(1 + \frac{1}{2m\beta^2}\|\hat{\boldsymbol{\alpha}}\|^2\right),$$

$$Var_{\mathbf{x}_i}\left\{\frac{2m\beta}{\sigma^2(2m\beta+1)}\left\|\tilde{\mathbf{y}}_j + \frac{L\sigma^2}{\sqrt{2mL\sigma^2\beta}}\hat{\boldsymbol{\alpha}}\right\|^2\right\} = \frac{4m^2\beta^2 L}{(2m\beta+1)^2}\left(1 + \frac{1}{m\beta^2}\|\hat{\boldsymbol{\alpha}}\|^2\right), \qquad j=2,\ldots,K.$$

Further, under mild regularity conditions on the asymptotic behavior of the vector $\hat{\boldsymbol{\alpha}}$, the central limit theorem implies that, for large values of $L$,

$$\frac{2m\beta}{\sigma^2(2m\beta+1)}\left\|\tilde{\mathbf{y}}_i + \frac{L\sigma^2}{\sqrt{2mL\sigma^2\beta}}\hat{\boldsymbol{\alpha}}\right\|^2 \sim \mathcal{N}\left(2m\beta L\left[1 + \frac{2m\beta+1}{2m\beta^2}\|\hat{\boldsymbol{\alpha}}\|^2\right], 4m^2\beta^2 L\left[1 + \frac{2m\beta+1}{m\beta^2}\|\hat{\boldsymbol{\alpha}}\|^2\right]\right),$$

and



$$\frac{2m\beta}{\sigma^2(2m\beta+1)}\left\|\tilde{\mathbf{y}}_j + \frac{L\sigma^2}{\sqrt{2mL\sigma^2}\beta}\hat{\boldsymbol{\alpha}}\right\|^2 \sim \mathscr{N}\left(\frac{2m\beta L}{2m\beta+1}\left(1+\frac{1}{2m\beta^2}\|\hat{\boldsymbol{\alpha}}\|^2\right), \frac{4m^2\beta^2 L}{(2m\beta+1)^2}\left(1+\frac{1}{m\beta^2}\|\hat{\boldsymbol{\alpha}}\|^2\right)\right), \quad j=2,\ldots,K\,.$$

Hence, for $L$ sufficiently large, we can make the approximation

$$\int_{\tilde{\mathbf{y}}} p_{\hat{\boldsymbol{\alpha}}}(\tilde{\mathbf{y}}|\mathbf{x}_i)\ln\left[1+p_0 K 2mL\sigma^2\frac{p_{\hat{\boldsymbol{\alpha}}}(\tilde{\mathbf{y}}|\mathbf{x}_0)}{p_{\hat{\boldsymbol{\alpha}}}(\tilde{\mathbf{y}}|\mathbf{x}_i)}+\sum_{\substack{j=1\\j\neq i}}^{K}\frac{p_{\hat{\boldsymbol{\alpha}}}(\tilde{\mathbf{y}}|\mathbf{x}_j)}{p_{\hat{\boldsymbol{\alpha}}}(\tilde{\mathbf{y}}|\mathbf{x}_i)}\right]$$

$$=E\left\{\ln\left(1+\frac{1}{Y}\left[K\left(2mL\sigma^2-1\right)(2m\beta+1)^L e^{\frac{L}{\beta}\|\hat{\boldsymbol{\alpha}}\|^2}+W\right]\right)\right\},$$

where $Y$ and $W$ are independent random variables with

$$Y=e^X, \quad X\sim\mathscr{N}\left(2m\beta L\left[1+\frac{2m\beta+1}{2m\beta^2}\|\hat{\boldsymbol{\alpha}}\|^2\right], 4m^2\beta^2 L\left[1+\frac{2m\beta+1}{m\beta^2}\|\hat{\boldsymbol{\alpha}}\|^2\right]\right),$$

$$W=\sum_{j=2}^{K}e^{X_j}, \quad X_j \text{ i.i.d. } \mathscr{N}\left(\frac{2m\beta L}{2m\beta+1}\left(1+\frac{1}{2m\beta^2}\|\hat{\boldsymbol{\alpha}}\|^2\right), \frac{4m^2\beta^2 L}{(2m\beta+1)^2}\left(1+\frac{1}{m\beta^2}\|\hat{\boldsymbol{\alpha}}\|^2\right)\right), \quad j=2,\ldots,K\,.$$

Note that $Y$ is a lognormal random variable and $W$ is the sum of $K-1\geq 0$ i.i.d. lognormal random variables. Let $p_Y(y)$ and $p_W(w)$ represent the pdfs of $Y$ and $W$, respectively. Then



$$E\left\{\ln\left(1+\frac{1}{Y}\left[K\left(2mL\sigma^2-1\right)\left(2m\beta+1\right)^L e^{\frac{L}{\beta}\|\hat{\boldsymbol{\alpha}}\|^2}+W\right]\right)\right\}$$

$$=\int_0^\infty\int_0^\infty\ln\left(1+\frac{1}{y}\left[K\left(2mL\sigma^2-1\right)\left(2m\beta+1\right)^L e^{\frac{L}{\beta}\|\hat{\boldsymbol{\alpha}}\|^2}+w\right]\right)p_Y(y)p_W(w)\,dy\,dw$$

$$=\int_0^\infty\int_0^\infty\left(\int_0^{\ln\left[1+\frac{1}{y}\left(K\left(2mL\sigma^2-1\right)\left(2m\beta+1\right)^L e^{\frac{L}{\beta}\|\hat{\boldsymbol{\alpha}}\|^2}+w\right)\right]}dx\right)p_Y(y)p_W(w)\,dy\,dw$$

$$=\int_0^\infty\ln\left[1+\frac{1}{y}\left(K\left(2mL\sigma^2-1\right)\left(2m\beta+1\right)^L e^{\frac{L}{\beta}\|\hat{\boldsymbol{\alpha}}\|^2}\right)\right]P(W>0)p_Y(y)\,dy$$

$$+\int_0^\infty\left(\int_{\ln\left[1+\frac{1}{y}\left(K\left(2mL\sigma^2-1\right)\left(2m\beta+1\right)^L e^{\frac{L}{\beta}\|\hat{\boldsymbol{\alpha}}\|^2}\right)\right]}^\infty P\left[W\geq y\left(e^x-1\right)-K\left(2mL\sigma^2-1\right)\left(2m\beta+1\right)^L e^{\frac{L}{\beta}\|\hat{\boldsymbol{\alpha}}\|^2}\right]dx\right)p_Y(y)\,dy$$

$$\leq\int_0^\infty\left(\int_0^{\ln\left[1+\frac{1}{y}\left(K\left(2mL\sigma^2-1\right)\left(2m\beta+1\right)^L e^{\frac{L}{\beta}\|\hat{\boldsymbol{\alpha}}\|^2}\right)\right]}dx\right)p_Y(y)\,dy$$

$$+\int_0^\infty\left(\int_{\ln\left[1+\frac{1}{y}\left(K\left(2mL\sigma^2-1\right)\left(2m\beta+1\right)^L e^{\frac{L}{\beta}\|\hat{\boldsymbol{\alpha}}\|^2}\right)\right]}^\infty\left[1-\Phi\left(\frac{\ln\left[\frac{\frac{y}{K-1}\left(e^x-1\right)}{-\frac{K\left(2mL\sigma^2-1\right)\left(2m\beta+1\right)^L e^{\frac{L}{\beta}\|\hat{\boldsymbol{\alpha}}\|^2}}{K-1}}\right]}{\frac{2m\beta}{2m\beta+1}\sqrt{L+\frac{L}{m\beta^2}\|\hat{\boldsymbol{\alpha}}\|^2}}-\sqrt{L}\frac{1+\frac{1}{2m\beta^2}\|\hat{\boldsymbol{\alpha}}\|^2}{\sqrt{1+\frac{1}{m\beta^2}\|\hat{\boldsymbol{\alpha}}\|^2}}\right)\right]^{K-1}dx\right)p_Y(y)\,dy,$$

where the final inequality follows from the upper bound given in [22] for the complementary cdf of the sum of $K-1$ lognormal random variables. Continuing from this point, we have



$$E\left\{\ln\left(1+\frac{1}{Y}\left[K\left(2mL\sigma^2-1\right)\left(2m\beta+1\right)^L e^{\frac{L}{\beta}\|\hat{\boldsymbol{\alpha}}\|^2}+W\right]\right)\right\}$$

$$\leq\int_0^\infty P\left[Y\leq K\left(2mL\sigma^2-1\right)\left(2m\beta+1\right)^L e^{\frac{L}{\beta}\|\hat{\boldsymbol{\alpha}}\|^2}\middle/\left(e^x-1\right)\right]dx$$

$$+\int_0^\infty\int_{\ln\left[1+\frac{1}{y}\left(K\left(2mL\sigma^2-1\right)\left(2m\beta+1\right)^L e^{\frac{L}{\beta}\|\hat{\boldsymbol{\alpha}}\|^2}\right)\right]}^\infty\left(1-\Phi\left[\frac{\ln\left[\frac{\frac{y}{K-1}\left(e^x-1\right)}{-\frac{K\left(2mL\sigma^2-1\right)\left(2m\beta+1\right)^L e^{\frac{L}{\beta}\|\hat{\boldsymbol{\alpha}}\|^2}}{K-1}}\right]}{\frac{2m\beta}{2m\beta+1}\sqrt{L+\frac{L}{m\beta^2}\|\hat{\boldsymbol{\alpha}}\|^2}}-\sqrt{L}\frac{1+\frac{1}{2m\beta^2}\|\hat{\boldsymbol{\alpha}}\|^2}{\sqrt{1+\frac{1}{m\beta^2}\|\hat{\boldsymbol{\alpha}}\|^2}}\right]\right)^{K-1}dx\,p_Y(y)dy$$

$$=\int_0^\infty\Phi\left(\frac{\ln\left[\frac{K\left(2mL\sigma^2-1\right)\left(2m\beta+1\right)^L e^{\frac{L}{\beta}\|\hat{\boldsymbol{\alpha}}\|^2}}{e^x-1}\right]}{\left(2m\beta\sqrt{L+\frac{2m\beta+1}{m\beta^2}L\|\hat{\boldsymbol{\alpha}}\|^2}\right)}-\sqrt{L}\frac{1+\frac{2m\beta+1}{2m\beta^2}\|\hat{\boldsymbol{\alpha}}\|^2}{\sqrt{1+\frac{2m\beta+1}{m\beta^2}\|\hat{\boldsymbol{\alpha}}\|^2}}\right)dx$$

$$+\int_0^\infty\int_{\ln\left[1+\frac{1}{y}\left(K\left(2mL\sigma^2-1\right)\left(2m\beta+1\right)^L e^{\frac{L}{\beta}\|\hat{\boldsymbol{\alpha}}\|^2}\right)\right]}^\infty\left(1-\Phi\left[\frac{\ln\left[\frac{\frac{y}{K-1}\left(e^x-1\right)}{-\frac{K\left(2mL\sigma^2-1\right)\left(2m\beta+1\right)^L e^{\frac{L}{\beta}\|\hat{\boldsymbol{\alpha}}\|^2}}{K-1}}\right]}{\frac{2m\beta}{2m\beta+1}\sqrt{L+\frac{L}{m\beta^2}\|\hat{\boldsymbol{\alpha}}\|^2}}-\sqrt{L}\frac{1+\frac{1}{2m\beta^2}\|\hat{\boldsymbol{\alpha}}\|^2}{\sqrt{1+\frac{1}{m\beta^2}\|\hat{\boldsymbol{\alpha}}\|^2}}\right]\right)^{K-1}dx\,p_Y(y)dy.$$



$$= 2\sqrt{m^2\beta^2 L + (2m^2\beta + m)L\|\hat{\boldsymbol{\alpha}}\|^2} \int_{-\infty}^{\infty} \frac{K(2mL\sigma^2-1)(2m\beta+1)^L e^{-2mL(\beta+\|\hat{\boldsymbol{\alpha}}\|^2)-2u\sqrt{m^2\beta^2 L+(2m^2\beta+m)L\|\hat{\boldsymbol{\alpha}}\|^2}}}{1+K(2mL\sigma^2-1)(2m\beta+1)^L e^{-2mL(\beta+\|\hat{\boldsymbol{\alpha}}\|^2)-2u\sqrt{m^2\beta^2 L+(2m^2\beta+m)L\|\hat{\boldsymbol{\alpha}}\|^2}}} \Phi(u)\,du$$

$$+ \int_0^{\infty} \int_{\ln\left[1+\frac{1}{y}\left(K(2mL\sigma^2-1)(2m\beta+1)^L e^{\frac{L}{\beta}\|\hat{\boldsymbol{\alpha}}\|^2}\right)\right]}^{\infty} \left(1-\left[\Phi\left(\frac{\ln\left[\frac{\frac{y}{K-1}(e^x-1)}{-\frac{K(2mL\sigma^2-1)(2m\beta+1)^L e^{\frac{L}{\beta}\|\hat{\boldsymbol{\alpha}}\|^2}}{K-1}}\right]}{\frac{2m\beta}{2m\beta+1}\sqrt{L+\frac{L}{m\beta^2}\|\hat{\boldsymbol{\alpha}}\|^2}}-\sqrt{L}\frac{1+\frac{1}{2m\beta^2}\|\hat{\boldsymbol{\alpha}}\|^2}{\sqrt{1+\frac{1}{m\beta^2}\|\hat{\boldsymbol{\alpha}}\|^2}}\right)\right]^{K-1}\right) dx\, p_Y(y)\,dy$$

Now, using integration by parts on the first integral and making a change of variable in the second followed again by integration by parts, we get

$$E\left\{\ln\left[1+\frac{1}{Y}\left[K(2mL\sigma^2-1)(2m\beta+1)^L e^{\frac{L}{\beta}\|\hat{\boldsymbol{\alpha}}\|^2}+W\right]\right]\right\}$$

$$\leq \frac{1}{\sqrt{2\pi}}\int_{-\infty}^{\infty} e^{-x^2/2}\ln\left[1+K(2mL\sigma^2-1)(2m\beta+1)^L e^{-2mL(\beta+\|\hat{\boldsymbol{\alpha}}\|^2)-2x\sqrt{m^2\beta^2 L+(2m^2\beta+m)L\|\hat{\boldsymbol{\alpha}}\|^2}}\right]dx$$

$$+ \int_0^{\infty}\left[\int_{-\infty}^{\infty} \frac{(K-1)\frac{2}{2m\beta+1}\sqrt{m^2\beta^2 L+mL\|\hat{\boldsymbol{\alpha}}\|^2}\, e^{\frac{2u}{2m\beta+1}\sqrt{m^2\beta^2 L+mL\|\hat{\boldsymbol{\alpha}}\|^2}+\frac{2m\beta L}{2m\beta+1}}\left(1-[\Phi(u)]^{K-1}\right)}{ye^{-\frac{L}{\beta}\|\hat{\boldsymbol{\alpha}}\|^2}+K(2mL\sigma^2-1)(2m\beta+1)^L+(K-1)e^{\frac{2u}{2m\beta+1}\sqrt{m^2\beta^2 L+mL\|\hat{\boldsymbol{\alpha}}\|^2}+\frac{2m\beta L}{2m\beta+1}}}\,du\right] p_Y(y)\,dy$$

$$= \frac{1}{\sqrt{2\pi}}\int_{-\infty}^{\infty} e^{-x^2/2}\ln\left[1+K(2mL\sigma^2-1)(2m\beta+1)^L e^{-2mL(\beta+\|\hat{\boldsymbol{\alpha}}\|^2)-2x\sqrt{m^2\beta^2 L+(2m^2\beta+m)L\|\hat{\boldsymbol{\alpha}}\|^2}}\right]dx$$

$$+ \int_0^{\infty}\left(\frac{K-1}{\sqrt{2\pi}}\int_{-\infty}^{\infty} [\Phi(u)]^{K-2} e^{-u^2/2}\ln\left[1+\frac{(K-1)e^{\frac{2u}{2m\beta+1}\sqrt{m^2\beta^2 L+mL\|\hat{\boldsymbol{\alpha}}\|^2}+\frac{2m\beta L}{2m\beta+1}}}{ye^{-\frac{L}{\beta}\|\hat{\boldsymbol{\alpha}}\|^2}+K(2mL\sigma^2-1)(2m\beta+1)^L}\right]du\right) p_Y(y)\,dy,$$

where,



$$p_Y(y) = \frac{1}{2ym\beta\sqrt{2\pi L\left(1+\frac{2m\beta+1}{2m\beta^2}\|\hat{\boldsymbol{\alpha}}\|^2\right)}} e^{-\frac{\left[\ln y - 2m\beta L\left(1+\frac{2m\beta+1}{2m\beta^2}\|\hat{\boldsymbol{\alpha}}\|^2\right)\right]^2}{8m^2\beta^2 L\left(1+\frac{2m\beta+1}{m\beta^2}\|\hat{\boldsymbol{\alpha}}\|^2\right)}}.$$

Putting all of this together, we now have

$$I_{\hat{\boldsymbol{\alpha}}}(\mathbf{x};\mathbf{y}) \ge \frac{1}{2mL\sigma^2}\left(\begin{array}{l} \ln\left[K2mL\sigma^2\right] \\[2mm] -\dfrac{1}{\sqrt{2\pi}}\displaystyle\int_{-\infty}^{\infty} e^{-x^2/2} \ln\left[\begin{array}{l}1+K\left(2mL\sigma^2-1\right)(2m\beta+1)^L \\ \cdot e^{-2mL\left(\beta+\|\hat{\boldsymbol{\alpha}}\|^2\right)-2x\sqrt{m^2\beta^2 L+(2m^2\beta+m)L\|\hat{\boldsymbol{\alpha}}\|^2}}\end{array}\right]dx \\[4mm] -\dfrac{K-1}{\sqrt{2\pi}}\displaystyle\int_0^\infty \dfrac{1}{2ym\beta\sqrt{2\pi L\left(1+\frac{2m\beta+1}{2m\beta^2}\|\hat{\boldsymbol{\alpha}}\|^2\right)}} e^{-\frac{\left[\ln y+\frac{L}{\beta}\|\hat{\boldsymbol{\alpha}}\|^2-2m\beta L\left(1+\frac{2m\beta+1}{2m\beta^2}\|\hat{\boldsymbol{\alpha}}\|^2\right)\right]^2}{8m^2\beta^2 L\left(1+\frac{2m\beta+1}{m\beta^2}\|\hat{\boldsymbol{\alpha}}\|^2\right)}} \\[4mm] \cdot\displaystyle\int_{-\infty}^\infty \left[\Phi(x)\right]^{K-2} e^{-x^2/2}\ln\left[1+\dfrac{(K-1)e^{\frac{2x}{2m\beta+1}\sqrt{m^2\beta^2 L+mL\|\hat{\boldsymbol{\alpha}}\|^2}+\frac{2m\beta L}{2m\beta+1}}}{y+K\left(2mL\sigma^2-1\right)(2m\beta+1)^L}\right]dx\,dy \end{array}\right)$$

$$= \frac{1}{2mL\sigma^2}\left(\begin{array}{l} \ln\left[K2mL\sigma^2\right] \\[2mm] -\dfrac{1}{\sqrt{2\pi}}\displaystyle\int_{-\infty}^{\infty} e^{-x^2/2} \ln\left[\begin{array}{l}1+K\left(2mL\sigma^2-1\right)(2m\beta+1)^L \\ \cdot e^{-2mL\left(\beta+\|\hat{\boldsymbol{\alpha}}\|^2\right)-2x\sqrt{m^2\beta^2 L+(2m^2\beta+m)L\|\hat{\boldsymbol{\alpha}}\|^2}}\end{array}\right]dx \\[4mm] -\dfrac{K-1}{\sqrt{2\pi}}\displaystyle\int_0^\infty \dfrac{1}{2y\sqrt{2\pi L\left[m^2\beta^2+\left(2m^2\beta+m\right)\|\hat{\boldsymbol{\alpha}}\|^2\right]}} e^{-\frac{\left[\ln y-2mL\left(\beta+\|\hat{\boldsymbol{\alpha}}\|^2\right)\right]^2}{8mL\left[m\beta^2+(2m\beta+1)\|\hat{\boldsymbol{\alpha}}\|^2\right]}} \\[4mm] \cdot\displaystyle\int_{-\infty}^\infty \left[\Phi(x)\right]^{K-2} e^{-x^2/2}\ln\left[1+\dfrac{(K-1)e^{\frac{2x}{2m\beta+1}\sqrt{m^2\beta^2 L+mL\|\hat{\boldsymbol{\alpha}}\|^2}+\frac{2m\beta L}{2m\beta+1}}}{y+K\left(2mL\sigma^2-1\right)(2m\beta+1)^L}\right]dx\,dy \end{array}\right),$$

which establishes the bound given in Lemma 3. To establish a second bound, which is not as tight, but easier to analyze, we note that



$$\frac{1}{\sqrt{2\pi}}\int_{-\infty}^{\infty}e^{-x^2/2}\ln\left[1+K\left(2mL\sigma^2-1\right)\left(2m\beta+1\right)^L e^{-2mL\left(\beta+\|\hat{\boldsymbol{\alpha}}\|^2\right)-2x\sqrt{m^2\beta^2 L+\left(2m^2\beta+m\right)L\|\hat{\boldsymbol{\alpha}}\|^2}}\right]dx$$

$$\leq \ln\left[1+K\left(2mL\sigma^2-1\right)\left(2m\beta+1\right)^L e^{-2mL\left(\beta+\|\hat{\boldsymbol{\alpha}}\|^2\right)}\right]-\sqrt{2L\frac{m^2\beta^2+\left(2m^2\beta+m\right)\|\hat{\boldsymbol{\alpha}}\|^2}{\pi}}\int_{-\infty}^{0}xe^{-x^2/2}dx$$

$$= \ln\left[1+K\left(2mL\sigma^2-1\right)\left(2m\beta+1\right)^L e^{-2mL\left(\beta+\|\hat{\boldsymbol{\alpha}}\|^2\right)}\right]+\sqrt{2L\frac{m^2\beta^2+\left(2m^2\beta+m\right)\|\hat{\boldsymbol{\alpha}}\|^2}{\pi}},$$

and

$$\frac{K-1}{\sqrt{2\pi}}\int_{-\infty}^{\infty}\left[\Phi(x)\right]^{K-2}e^{-x^2/2}\ln\left[1+\frac{(K-1)e^{\frac{2x}{2m\beta+1}\sqrt{m^2\beta^2 L+mL\|\hat{\boldsymbol{\alpha}}\|^2}+\frac{2m\beta L}{2m\beta+1}}}{y+K\left(2mL\sigma^2-1\right)\left(2m\beta+1\right)^L}\right]dxdy$$

$$\leq \ln\left[1+\frac{(K-1)e^{\frac{2m\beta L}{2m\beta+1}}}{y+K\left(2mL\sigma^2-1\right)\left(2m\beta+1\right)^L}\right]+\frac{2}{2m\beta+1}\sqrt{m^2\beta^2 L+mL\|\hat{\boldsymbol{\alpha}}\|^2}\frac{(K-1)}{\sqrt{2\pi}}\int_{0}^{\infty}\left[\Phi(x)\right]^{K-2}xe^{-x^2/2}dx$$

$$\leq \ln\left[1+\frac{(K-1)e^{\frac{2m\beta L}{2m\beta+1}}}{K\left(2mL\sigma^2-1\right)\left(2m\beta+1\right)^L}\right]+\frac{2}{2m\beta+1}\sqrt{m^2\beta^2 L+mL\|\hat{\boldsymbol{\alpha}}\|^2}\left(a\left[\Phi(a)\right]^{K-1}-a2^{1-K}+\frac{K-1}{\sqrt{2\pi}}e^{-a^2/2}\right),$$

for all $a>0$. Letting $a=\sqrt{2\ln K}$ and making the necessary substitutions gives

$$I_{\hat{\boldsymbol{\alpha}}}(\mathbf{x};\mathbf{y})\geq\frac{1}{2mL\sigma^2}\left(\begin{array}{l}\ln\left[K2mL\sigma^2\right]-\ln\left[1+K\left(2mL\sigma^2-1\right)\left(2m\beta+1\right)^L e^{-2mL\left(\beta+\|\hat{\boldsymbol{\alpha}}\|^2\right)}\right]\\[2mm]-\ln\left[1+\frac{(K-1)}{K\left(2mL\sigma^2-1\right)\left(2m\beta+1\right)^L}e^{\frac{2m\beta}{2m\beta+1}L}\right]\\[2mm]-\frac{2}{2m\beta+1}\left(\left[\Phi\left(\sqrt{2\ln K}\right)\right]^{K-1}-\left[\frac{1}{2}\right]^{K-1}\right)\sqrt{2L\ln K\left(m^2\beta^2+m\|\hat{\boldsymbol{\alpha}}\|^2\right)}\\[2mm]-\sqrt{2L\frac{m^2\beta^2+m\|\hat{\boldsymbol{\alpha}}\|^2}{\pi}}\left(\sqrt{1+\frac{2\beta\|\hat{\boldsymbol{\alpha}}\|^2}{m\beta^2+\|\hat{\boldsymbol{\alpha}}\|^2}}+\frac{K-1}{K\left(2m\beta+1\right)}\right)\end{array}\right)$$

which establishes the desired second bound. ∎

**Proof of Lemma 4**. We adopt the same notation used in the proof of Lemma 3, and we let $p(\mathbf{y})$ represent the pdf of the observed vector $\mathbf{y}$. Assuming that the distribution of $\mathbf{x}$ is radially



symmetric and the distribution of $r = \|\mathbf{x}\|$ is represented by a probability measure $\mu$, we have

(with a slight abuse of standard surface integral notation)

$$p(\mathbf{y}) = \int_{\mathcal{R}^+} \frac{1}{S_K(r)} \left[ \int_{\|\mathbf{x}\| = r} p(\mathbf{y}|\mathbf{x}) d\mathbf{x} \right] \mu(dr),$$

where

$$p(\mathbf{y}|\mathbf{x}) = \frac{1}{\left(\pi\sigma^2\right)^{KL}} \left( \frac{L\sigma^2}{\beta\mathcal{E}_{\mathbf{x}} + L\sigma^2} \right)^L e^{-\frac{1}{\sigma^2}\left[ \|\tilde{\mathbf{y}}\|^2 + \left( \frac{\mathcal{E}_{\mathbf{x}}L\sigma^2}{\beta\mathcal{E}_{\mathbf{x}} + L\sigma^2} \right) \|\hat{\boldsymbol{\alpha}}\|^2 - \left( \frac{\beta}{\beta\mathcal{E}_{\mathbf{x}} + L\sigma^2} \right) \tilde{\mathbf{y}}^* \mathbf{H}_{\mathbf{x}} \mathbf{H}_{\mathbf{x}}^* \tilde{\mathbf{y}} - \left( \frac{2L\sigma^2}{\beta\mathcal{E}_{\mathbf{x}} + L\sigma^2} \right) \text{Re}\left[ \tilde{\mathbf{y}}^* \mathbf{H}_{\mathbf{x}} \hat{\boldsymbol{\alpha}} \right] \right]},$$

for $\mathbf{x} \neq 0$. Note that

$$S_K(r) = \frac{2\pi^K}{(K-1)!} r^{2K-1},$$

represents the surface area of a $2K$-dimensional sphere of radius $r$ ($K$ complex dimensions equals $2K$ real dimensions), and

$$p(\mathbf{y}|\mathbf{0}) = \lim_{r \to 0} \frac{1}{S_K(r)} \int_{\|\mathbf{x}\| = r} p(\mathbf{y}|\mathbf{x}) d\mathbf{x}.$$

Then

$$
\begin{aligned}
I_{\hat{\boldsymbol{\alpha}}}(\mathbf{x}; \mathbf{y}) &= h(\mathbf{y}) - h(\mathbf{y}|\mathbf{x}) \\
&= h(\mathbf{y}) - E\left\{ \ln(\pi e)^L \left| \boldsymbol{\Sigma}_{\mathbf{y}|\mathbf{x}} \right| \right\} \\
&= h(\mathbf{y}) - L\ln(\pi e) - E\left\{ \ln\left( \left| \sigma^2 I + \frac{\beta}{L} \mathbf{H}_{\mathbf{x}} \mathbf{H}_{\mathbf{x}}^* \right| \right) \right\} \\
&= h(\mathbf{y}) - L\ln(\pi e) - KL\ln(\sigma^2) - L \cdot E\left\{ \ln\left( \frac{\beta\mathcal{E}_{\mathbf{x}} + L\sigma^2}{L\sigma^2} \right) \right\} \\
&= -\int_{\mathbf{y}} p(\mathbf{y})\ln[p(\mathbf{y})]d\mathbf{y} - L\ln(\pi e) - KL\ln(\sigma^2) - L \cdot E\left\{ \ln\left( \frac{\beta\mathcal{E}_{\mathbf{x}} + L\sigma^2}{L\sigma^2} \right) \right\} \\
&= \int_{\mathcal{R}^+} \ln\left[ \left( \frac{L\sigma^2}{\beta r^2 + L\sigma^2} \right)^L \right] \mu(dr) - L\ln(\pi e) - KL\ln(\sigma^2) - \int_{\mathcal{R}^+} E\left\{ \ln[p(\mathbf{y})] \,\big|\, \|\mathbf{x}\| = r \right\} \mu(dr).
\end{aligned}
$$

Now, for $K = 1$, we have



$$\frac{1}{S_K(r)}\int_{\|\mathbf{x}\|=r}p(\mathbf{y}|\mathbf{x})d\mathbf{x}=\frac{1}{2\pi r}\oint_{|x|=r}\left[\frac{1}{\left(\pi\sigma^2\right)^L}\left(\frac{L\sigma^2}{\beta r^2+L\sigma^2}\right)^L e^{-\frac{1}{\sigma^2}\left(\frac{L\sigma^2}{\beta r^2+L\sigma^2}\right)\left[\|\tilde{\mathbf{y}}\|^2+\|\hat{\boldsymbol{\alpha}}\|^2 r^2-2\mathrm{Re}\left[x\left(\tilde{\mathbf{y}}^*\hat{\boldsymbol{\alpha}}\right)\right]\right]}\right]dx$$

$$=\frac{1}{2\pi r\left(\pi\sigma^2\right)^L}\left(\frac{L\sigma^2}{\beta r^2+L\sigma^2}\right)^L e^{-\frac{1}{\sigma^2}\left(\frac{L\sigma^2}{\beta r^2+L\sigma^2}\right)\left[\|\tilde{\mathbf{y}}\|^2+\|\hat{\boldsymbol{\alpha}}\|^2 r^2\right]}$$

$$\cdot\oint_{|x|=r}e^{\frac{2}{\sigma^2}\left(\frac{L\sigma^2}{\beta r^2+L\sigma^2}\right)\mathrm{Re}\left[x\left(\tilde{\mathbf{y}}^*\hat{\boldsymbol{\alpha}}\right)\right]}dx$$

$$=\frac{1}{2\pi\left(\pi\sigma^2\right)^L}\left(\frac{L\sigma^2}{\beta r^2+L\sigma^2}\right)^L e^{-\frac{1}{\sigma^2}\left(\frac{L\sigma^2}{\beta r^2+L\sigma^2}\right)\left[\|\tilde{\mathbf{y}}\|^2+\|\hat{\boldsymbol{\alpha}}\|^2 r^2\right]}$$

$$\cdot\int_0^{2\pi}e^{\frac{2}{\sigma^2}\left(\frac{L\sigma^2}{\beta r^2+L\sigma^2}\right)r|\tilde{\mathbf{y}}^*\hat{\boldsymbol{\alpha}}|\cos\left(\theta+\arg\left[\tilde{\mathbf{y}}^*\hat{\boldsymbol{\alpha}}\right]\right)}d\theta$$

$$=\frac{1}{\left(\pi\sigma^2\right)^L}\left(\frac{L\sigma^2}{\beta r^2+L\sigma^2}\right)^L e^{-\frac{1}{\sigma^2}\left(\frac{L\sigma^2}{\beta r^2+L\sigma^2}\right)\left[\|\tilde{\mathbf{y}}\|^2+\|\hat{\boldsymbol{\alpha}}\|^2 r^2\right]}I_0\left(\frac{2Lr|\tilde{\mathbf{y}}^*\hat{\boldsymbol{\alpha}}|}{\beta r^2+L\sigma^2}\right).$$

Let $\left\{\hat{\boldsymbol{\alpha}}/\|\hat{\boldsymbol{\alpha}}\|,\mathbf{z}_1,\mathbf{z}_2,\dots,\mathbf{z}_L\right\}$ be an orthonormal basis for $\mathscr{C}^L$, and let

$$\mathbf{Z}=\left[\frac{\hat{\boldsymbol{\alpha}}}{\|\hat{\boldsymbol{\alpha}}\|}\;\middle|\;\mathbf{z}_1\;\middle|\;\mathbf{z}_2\;\middle|\;\cdots\;\middle|\;\mathbf{z}_L\right],\quad\mathbf{w}=\mathbf{Z}^*\tilde{\mathbf{y}}=\begin{bmatrix}w_1\\w_2\\\vdots\\w_L\end{bmatrix},\quad\tilde{\mathbf{w}}=\begin{bmatrix}w_2\\\vdots\\w_L\end{bmatrix}.$$

Then, $\mathbf{Z}$ is unitary, $w_1=\tilde{\mathbf{y}}^*\hat{\boldsymbol{\alpha}}/\|\hat{\boldsymbol{\alpha}}\|$, and

$$p(\mathbf{y})=\frac{1}{\left(\pi\sigma^2\right)^L}\int_{\mathscr{R}^+}\left(\frac{L\sigma^2}{\beta r^2+L\sigma^2}\right)^L I_0\left(\frac{2Lr\|\hat{\boldsymbol{\alpha}}\|w_1}{\beta r^2+L\sigma^2}\right)e^{-\frac{1}{\sigma^2}\left(\frac{L\sigma^2}{\beta r^2+L\sigma^2}\right)\left[\|\mathbf{w}\|^2+\|\hat{\boldsymbol{\alpha}}\|^2 r^2\right]}\mu(dr)$$

$$=\frac{1}{\left(\pi\sigma^2\right)^L}\int_{\mathscr{R}^+}\left(\frac{L\sigma^2}{\beta r^2+L\sigma^2}\right)^L\left[I_0\left(\frac{2Lr\|\hat{\boldsymbol{\alpha}}\|w_1|}{\beta r^2+L\sigma^2}\right)e^{-\frac{1}{\sigma^2}\left(\frac{L\sigma^2}{\beta r^2+L\sigma^2}\right)\left[|w_1|^2+\|\tilde{\mathbf{w}}\|^2+\|\hat{\boldsymbol{\alpha}}\|^2 r^2\right]}\right]\mu(dr),$$



Finally, given that symbol $x$ is transmitted, $w_1$ is independent of $\tilde{\mathbf{w}}$ with $w_1 \sim \mathscr{N}\left(\|\hat{\boldsymbol{\alpha}}\|x, |x|^2 \frac{\beta}{L} + \sigma^2\right)$

and $\tilde{\mathbf{w}} \sim \mathscr{N}\left(0, \left(|x|^2 \frac{\beta}{L} + \sigma^2\right)\mathbf{I}\right)$. Hence, $|w_1|^2 \sim \frac{1}{2}\left(|x|^2 \frac{\beta}{L} + \sigma^2\right)\chi_2^2\left(\frac{2L\|\hat{\boldsymbol{\alpha}}\|^2 |x|^2}{\beta |x|^2 + L\sigma^2}\right)$,

$\tilde{\mathbf{w}} \sim \frac{1}{2}\left(|x|^2 \frac{\beta}{L} + \sigma^2\right)\chi_{2L-2}^2$, and

$$I_{\hat{\boldsymbol{\alpha}}}(\mathbf{x};\mathbf{y}) = \int_{\mathscr{R}^+} \ln\left[\left(\frac{L\sigma^2}{\beta r^2 + L\sigma^2}\right)^L\right]\mu(dr) - L\ln(\pi e) - L\ln(\sigma^2) - \int_{\mathscr{R}^+} E\left\{\ln[p(\mathbf{y})] \mid \|\mathbf{x}\| = r\right\}\mu(dr)$$

$$= \int_{\mathscr{R}^+} \ln\left[\left(\frac{L\sigma^2}{\beta r^2 + L\sigma^2}\right)^L\right]\mu(dr) - L\ln(\pi e) - L\ln(\sigma^2)$$

$$- \int_{\mathscr{R}^+} E\left\{\ln\left[\frac{1}{(\pi\sigma^2)^L}\int_\rho \left(\frac{L\sigma^2}{\beta\rho^2 + L\sigma^2}\right)^L I_0\left(\frac{2L\rho\|\hat{\boldsymbol{\alpha}}\|\|w_1\|}{\beta\rho^2 + L\sigma^2}\right)e^{-\frac{1}{\sigma^2}\left(\frac{L\sigma^2}{\beta\rho^2 + L\sigma^2}\right)\left[|w_1|^2 + \|\tilde{\mathbf{w}}\|^2 + \|\hat{\boldsymbol{\alpha}}\|^2\rho^2\right]}\mu(d\rho)\right] \mid \|\mathbf{x}\| = r\right\}\mu(dr)$$

$$= -L - \int_{\mathscr{R}^+} E\left\{\ln\left[\int_\rho \left(\frac{\beta r^2 + L\sigma^2}{\beta\rho^2 + L\sigma^2}\right)^L I_0\left(\frac{2L\rho\|\hat{\boldsymbol{\alpha}}\|\|w_1\|}{\beta\rho^2 + L\sigma^2}\right)e^{-\frac{1}{\sigma^2}\left(\frac{L\sigma^2}{\beta\rho^2 + L\sigma^2}\right)\left[|w_1|^2 + \|\tilde{\mathbf{w}}\|^2 + \|\hat{\boldsymbol{\alpha}}\|^2\rho^2\right]}\mu(d\rho)\right] \mid \|\mathbf{x}\| = r\right\}\mu(dr)$$

$$= \begin{cases} -L - \int_{\mathscr{R}^+}\left[\int_0^\infty \ln\left[\int_{\mathscr{R}^+} f(x,0,r,\rho,L,\beta,\sigma,\hat{\boldsymbol{\alpha}})\mu(d\rho)\right]\chi_2^2\left(\frac{2Lr^2\|\hat{\boldsymbol{\alpha}}\|^2}{\beta r^2 + L\sigma^2}, x\right)dx\right]\mu(dr), & L = 1, \\ -L - \int_{\mathscr{R}^+}\left[\int_0^\infty \int_0^\infty \ln\left[\int_{\mathscr{R}^+} f(x,y,r,\rho,L,\beta,\sigma,\hat{\boldsymbol{\alpha}})\mu(d\rho)\right]\chi_2^2\left(\frac{2Lr^2\|\hat{\boldsymbol{\alpha}}\|^2}{\beta r^2 + L\sigma^2}, x\right)\chi_{2L-2}^2(y)dxdy\right]\mu(dr), & L > 1, \end{cases}$$

where

$$f\left(x,y,r,\rho,L,\beta,\sigma,\hat{\boldsymbol{\alpha}}\right) = \left(\frac{\beta r^2 + L\sigma^2}{\beta\rho^2 + L\sigma^2}\right)^L I_0\left(2\sqrt{\frac{x}{2}\left(\frac{L\rho^2\|\hat{\boldsymbol{\alpha}}\|^2}{\beta\rho^2 + L\sigma^2}\right)\left(\frac{\beta r^2 + L\sigma^2}{\beta\rho^2 + L\sigma^2}\right)}\right)e^{-\left[\frac{(x+y)(\beta r^2 + L\sigma^2)}{2(\beta\rho^2 + L\sigma^2)} + \frac{L\rho^2\|\hat{\boldsymbol{\alpha}}\|^2}{\beta\rho^2 + L\sigma^2}\right]},$$

as claimed. ∎



**Proof of Corollary 3**. We give the proof for the case $L>1$, which is the more difficult of the two cases. The case $L=1$ is very similar. The result of Lemma 4 implies that to maximize $I_{\hat{\boldsymbol{\alpha}}}(\mathbf{x};\mathbf{y})$, we must solve the constrained minimization problem

$$\min_{\mu} \int_{\mathcal{R}^+}\left[\int_0^\infty\int_0^\infty \ln\left[\int_{\mathcal{R}^+} f(x,y,r,\rho,L,\beta,\sigma,\hat{\boldsymbol{\alpha}})\mu(d\rho)\right]\chi_2^2\left(\frac{2Lr^2\|\hat{\boldsymbol{\alpha}}\|^2}{\beta r^2+L\sigma^2},x\right)\chi_{2L-2}^2(y)dxdy\right]\mu(dr) \tag{A.1}$$

subject to $\int_{\mathcal{R}^+}\mu(dr)\le 1,\quad \int_{\mathcal{R}^+}r^2\mu(dr)\le 1,\quad \mu\ge 0,$

over the space of all signed measures $\mu$ on $\mathcal{R}^+$, where $\mu\ge 0$ implies $\mu$ is a positive measure and

$$f(x,y,r,\rho,L,\beta,\sigma,\hat{\boldsymbol{\alpha}})=\left(\frac{\beta r^2+L\sigma^2}{\beta\rho^2+L\sigma^2}\right)^L I_0\left(2\sqrt{\frac{x}{2}\left(\frac{L\rho^2\|\hat{\boldsymbol{\alpha}}\|^2}{\beta\rho^2+L\sigma^2}\right)\left(\frac{\beta r^2+L\sigma^2}{\beta\rho^2+L\sigma^2}\right)}\right)e^{-\left[\frac{(x+y)(\beta r^2+L\sigma^2)}{2(\beta\rho^2+L\sigma^2)}+\frac{L\rho^2\|\hat{\boldsymbol{\alpha}}\|^2}{\beta\rho^2+L\sigma^2}\right]}.$$

It follows from Theorem 1, §9.4 in [23][4] that a positive measure $\mu^*$ solves (A.1) only if there exist nonnegative constants $\lambda_1,\lambda_2$ and a nonnegative function $g(r)$ such that the Lagrangian

$$J(\mu,\lambda_1,\lambda_2,g)=\int_{\mathcal{R}^+}\left[\int_0^\infty\int_0^\infty \ln\left[\int_{\mathcal{R}^+} f(x,y,r,\rho,L,\beta,\sigma,\hat{\boldsymbol{\alpha}})\mu(d\rho)\right]\chi_2^2\left(\frac{2Lr^2\|\hat{\boldsymbol{\alpha}}\|^2}{\beta r^2+L\sigma^2},x\right)\chi_{2L-2}^2(y)dxdy\right]\mu(dr)$$
$$+\lambda_1\left[\int_{\mathcal{R}^+}r^2\mu(dr)-1\right]+\lambda_2\left[\int_{\mathcal{R}^+}\mu(dr)-1\right]-\int_{\mathcal{R}^+}g(r)\mu(dr)$$

is stationary at $\mu^*$ and

$$\int_{\mathcal{R}^+}\mu^*(dr)=1,\quad \int_{\mathcal{R}^+}r^2\mu^*(dr)=1,\quad \int_{\mathcal{R}^+}g(r)\mu^*(dr)=0.$$

We notice first that by making a suitable change of variables, the Lagrangian can be rewritten as

$$J(\mu,\lambda_1,\lambda_2,g)=\int_{\mathcal{R}^+}\left(\frac{1}{\beta r^2+L\sigma^2}\right)^2\left(\int_0^\infty\int_0^\infty\left[\begin{array}{c}\ln\left[\int_{\mathcal{R}^+}f(u,v,\rho,L,\beta,\sigma,\hat{\boldsymbol{\alpha}})\mu(d\rho)\right]\\ \chi_2^2\left(\frac{2Lr^2\|\hat{\boldsymbol{\alpha}}\|^2}{\beta r^2+L\sigma^2},\frac{u}{\beta r^2+L\sigma^2}\right)\chi_{2L-2}^2\left(\frac{v}{\beta r^2+L\sigma^2}\right)\end{array}\right]dudv\right)\mu(dr)$$
$$+\int_{\mathcal{R}^+}\ln\left[\left(\beta r^2+L\sigma^2\right)^L\right]\mu(dr)+\lambda_1\left[\int_{\mathcal{R}^+}r^2\mu(dr)-1\right]+\lambda_2\left[\int_{\mathcal{R}^+}\mu(dr)-1\right]-\int_{\mathcal{R}^+}g(r)\mu(dr)$$

where

$$f(u,v,\rho,L,\beta,\sigma,\hat{\boldsymbol{\alpha}})=\left(\frac{1}{\beta\rho^2+L\sigma^2}\right)^L I_0\left(\sqrt{\left(\frac{2L\rho^2\|\hat{\boldsymbol{\alpha}}\|^2}{\beta\rho^2+L\sigma^2}\right)\left(\frac{u}{\beta\rho^2+L\sigma^2}\right)}\right)e^{-\frac{1}{2}\left[\frac{u+v}{\beta\rho^2+L\sigma^2}+\frac{2L\rho^2\|\hat{\boldsymbol{\alpha}}\|^2}{\beta\rho^2+L\sigma^2}\right]}.$$

---

[4] Technically, we need to worry about a few details such as the definition of the cost function for signed measures $\mu$ for which $\int_{\mathcal{R}^+}f(x,y,r,\rho,L,\beta,\sigma,\hat{\boldsymbol{\alpha}})\mu(d\rho)\le 0$ and whether or not $\mu^*$ is a "regular point" of the constraint region, but we ignore these technical details here, which cause no problems in our case.

Then using the relationships

$$\chi_2^2(\lambda,x)=\frac{1}{2}e^{-(\lambda+x)/2}I_0\left(\sqrt{\lambda x}\right),\quad \chi_{2L-2}^2(y)=\frac{y^{L-2}e^{-y/2}}{2^{L-1}(L-2)!},$$

we get

$$J(\mu,\lambda_1,\lambda_2,g)=\int_{\mathcal{R}^+}\ln\left[\left(\beta r^2+L\sigma^2\right)^L\right]\mu(dr)+\lambda_1\left[\int_{\mathcal{R}^+}r^2\mu(dr)-1\right]+\lambda_2\left[\int_{\mathcal{R}^+}\mu(dr)-1\right]-\int_{\mathcal{R}^+}g(r)\mu(dr)$$
$$+\frac{1}{2^L(L-2)!}\int_{\mathcal{R}^+}\left[\int_0^\infty\int_0^\infty f(u,v,r,L,\beta,\sigma,\hat{\boldsymbol{\alpha}})\ln\left[\int_{\mathcal{R}^+}f(u,v,\rho,L,\beta,\sigma,\hat{\boldsymbol{\alpha}})\mu(d\rho)\right]dudv\right]\mu(dr).$$

It follows that the Lagrangian is stationary at $\mu^*$ if and only if, for all signed measures $\mu$,

$$0=\lim_{\delta\to0}\frac{1}{\delta}\left[J(\mu^*+\delta\mu,\lambda_1,\lambda_2,g)-J(\mu^*,\lambda_1,\lambda_2,g)\right]$$
$$=\int_{\mathcal{R}^+}\left(\ln\left[\left(\beta r^2+L\sigma^2\right)^L\right]+\lambda_1 r^2+\lambda_2-g(r)\right)\mu(dr)$$
$$+\frac{1}{2^L(L-2)!}\int_{\mathcal{R}^+}\left[\int_0^\infty\int_0^\infty f(u,v,r,L,\beta,\sigma,\hat{\boldsymbol{\alpha}})\left(1+\ln\left[\int_{\mathcal{R}^+}f(u,v,\rho,L,\beta,\sigma,\hat{\boldsymbol{\alpha}})\mu^*(d\rho)\right]\right)dudv\right]\mu(dr),$$

but this can be true for all $\mu$ if and only if

$$0=\ln\left[\left(\beta r^2+L\sigma^2\right)^L\right]+\lambda_1 r^2+\lambda_2-g(r)$$
$$+\frac{1}{2^L(L-2)!}\int_0^\infty\int_0^\infty f(u,v,r,L,\beta,\sigma,\hat{\boldsymbol{\alpha}})\left(1+\ln\left[\int_{\mathcal{R}^+}f(u,v,\rho,L,\beta,\sigma,\hat{\boldsymbol{\alpha}})\mu^*(d\rho)\right]\right)dudv,$$



for all $r \geq 0$. Clearly, this implies that

$$g(r) = \ln\left[\left(\beta r^2 + L\sigma^2\right)^L\right] + \lambda_1 r^2 + \lambda_2$$
$$+ \frac{1}{2^L (L-2)!} \int_0^\infty \int_0^\infty f(u,v,r,L,\beta,\sigma,\hat{\boldsymbol{\alpha}})\left(1 + \ln\left[\int_{\mathcal{R}^+} f(u,v,\rho,L,\beta,\sigma,\hat{\boldsymbol{\alpha}})\mu^*(d\rho)\right]\right) du\, dv,$$

so (A.1) is minimized only if

$$0 \leq \ln\left[\left(\beta r^2 + L\sigma^2\right)^L\right] + \lambda_1 r^2 + \lambda_2$$
$$+ \frac{1}{2^L (L-2)!} \int_0^\infty \int_0^\infty f(u,v,r,L,\beta,\sigma,\hat{\boldsymbol{\alpha}})\left(1 + \ln\left[\int_{\mathcal{R}^+} f(u,v,\rho,L,\beta,\sigma,\hat{\boldsymbol{\alpha}})\mu^*(d\rho)\right]\right) du\, dv,$$

for all $r \geq 0$. Furthermore, we must have

$$0 = \int_{\mathcal{R}^+} g(r)\mu^*(dr)$$
$$= \int_{\mathcal{R}^+}\left(\begin{array}{l}\ln\left[\left(\beta r^2 + L\sigma^2\right)^L\right] + \lambda_1 r^2 + \lambda_2 \\ + \dfrac{1}{2^L (L-2)!} \int_0^\infty \int_0^\infty f(u,v,r,L,\beta,\sigma,\hat{\boldsymbol{\alpha}})\left(1 + \ln\left[\int_{\mathcal{R}^+} f(u,v,\rho,L,\beta,\sigma,\hat{\boldsymbol{\alpha}})\mu^*(d\rho)\right]\right) du\, dv\end{array}\right)\mu^*(dr)$$

which implies that

$$0 = \ln\left[\left(\beta r^2 + L\sigma^2\right)^L\right] + \lambda_1 r^2 + \lambda_2$$
$$+ \frac{1}{2^L (L-2)!} \int_0^\infty \int_0^\infty f(u,v,r,L,\beta,\sigma,\hat{\boldsymbol{\alpha}})\left(1 + \ln\left[\int_{\mathcal{R}^+} f(u,v,\rho,L,\beta,\sigma,\hat{\boldsymbol{\alpha}})\mu^*(d\rho)\right]\right) du\, dv,$$

except on a set of $\mu^*$-measure zero. This establishes the necessity of the conditions in Corollary 3. To see that these conditions are also sufficient, we note that the set of all signed measures $\mu$ that satisfy the three constraints

$$\int_{\mathcal{R}^+} \mu(dr) \leq 1, \quad \int_{\mathcal{R}^+} r^2 \mu(dr) \leq 1, \quad \mu \geq 0,$$

is convex, and the cost function

$$I(\mu) = \int_{\mathcal{R}^+}\left[\int_0^\infty \int_0^\infty \ln\left[\int_{\mathcal{R}^+} f(x,y,r,\rho,L,\beta,\sigma,\hat{\boldsymbol{\alpha}})\mu(d\rho)\right]\chi_2^2\left(\frac{2Lr^2\|\hat{\boldsymbol{\alpha}}\|^2}{\beta r^2 + L\sigma^2}, x\right)\chi_{2L-2}^2(y)dx\, dy\right]\mu(dr)$$



is convex in $\mu$ (since $I_{\hat{\boldsymbol{\alpha}}}(\mathbf{x};\mathbf{y})$ is concave as a function of input distribution). Now suppose that

$$0 \le \ln\left[\left(\beta r^2 + L\sigma^2\right)^L\right] + \lambda_1 r^2 + \lambda_2$$
$$+ \frac{1}{2^L(L-2)!}\int_0^\infty\int_0^\infty f\left(u,v,r,L,\beta,\sigma,\hat{\boldsymbol{\alpha}}\right)\left(1+\ln\left[\int_{\mathcal{R}^+}f\left(u,v,\rho,L,\beta,\sigma,\hat{\boldsymbol{\alpha}}\right)\mu^*(d\rho)\right]\right)dudv,$$

Then for any measure $\bar{\mu}$ that satisfies the constraints, we have

$$I(\bar{\mu}) - I\left(\mu^*\right) = \lim_{\delta\to 0}\frac{1}{\delta}\delta\left[I(\bar{\mu}) - I\left(\mu^*\right)\right]$$
$$\ge \lim_{\delta\to 0}\frac{1}{\delta}\left[I\left(\mu^* + \delta\left[\bar{\mu} - \mu^*\right]\right) - I\left(\mu^*\right)\right]$$
$$= \lim_{\delta\to 0}\frac{1}{\delta}\left[J\left(\mu^* + \delta\left[\bar{\mu} - \mu^*\right],\lambda_1,\lambda_2,0\right) - J\left(\mu^*,\lambda_1,\lambda_2,0\right)\right]$$
$$= \ln\left[\left(\beta r^2 + L\sigma^2\right)^L\right] + \lambda_1 r^2 + \lambda_2$$
$$+ \frac{1}{2^L(L-2)!}\int_0^\infty\int_0^\infty f\left(u,v,r,L,\beta,\sigma,\hat{\boldsymbol{\alpha}}\right)\left(1+\ln\left[\int_{\mathcal{R}^+}f\left(u,v,\rho,L,\beta,\sigma,\hat{\boldsymbol{\alpha}}\right)\mu^*(d\rho)\right]\right)dudv$$
$$\ge 0.$$

Hence, $\mu^*$ minimizes $I(\mu)$ subject to the constraints. This establishes the desired result. ∎